\documentclass[12pt,letterpaper]{article}
\pdfoutput=1
\usepackage[colorlinks=false,
   linkcolor=red, 
   citecolor=blue,
    filecolor=red,
    urlcolor=red,
    linktoc=all, 
    pdfstartview=FitV,
    bookmarksopen=true]{hyperref}
\usepackage[left=2cm,top=1cm,right=3cm,nohead]{geometry}

\usepackage[utf8]{inputenc} 
\usepackage{epsfig,ulem,latexsym,amsfonts,mathtools,amsthm,amssymb,amsbsy,multirow,slashed,color,
mathrsfs,wasysym,textcomp,subfigure,wrapfig,comment,bbold,array,longtable,multirow}
\usepackage{tabularx}
\usepackage{graphicx}
\usepackage{setspace}
\usepackage{subfigure}
\usepackage[bf]{caption}
\usepackage{braket}
\usepackage{comment}
\usepackage{float}
\usepackage{tikz}
\usetikzlibrary{mindmap,trees,shadows}
\colorlet{color1}{gray!25}
\usetikzlibrary{positioning}
\usetikzlibrary{intersections} 

\newlength{\PicScale}
\usepackage[UKenglish]{babel}
\usepackage[toc,page]{appendix}
\usepackage{hhline}
\usepackage{geometry}
\usepackage{cite}
\usepackage{datetime}
\usepackage{bbm} 
\usepackage{bm} 
\hypersetup{
    pdftitle={},
    pdfauthor={},
    pdfsubject={}
} 

\newcommand{\be}{\begin{equation}}
\newcommand{\ee}{\end{equation}}

\newcommand{\beq}{\begin{equation}}
\newcommand{\eeq}{\end{equation}}

\newcommand{\bea}{\begin{eqnarray}}
\newcommand{\eea}{\end{eqnarray}}
\newcommand{\bmat}{\left(\!\!\begin{array}}
\newcommand{\emat}{\end{array}\!\!\right)}

\newcommand{\ZZ}{{\mathbb{Z}}}

\newcommand{\tm}{{\text{-}}}
\newcommand{\mra}{{\mathrm{A}}}
\newcommand{\mrd}{{\mathrm{D}}}
\newcommand{\mre}{{\mathrm{E}}}
\newcommand{\mrc}{{\mathrm{C}}}
\newcommand{\mrb}{{\mathrm{B}}}
\newcommand{\mrf}{{\mathrm{F}}}
\newcommand{\mrg}{{\mathrm{G}}}

\newcommand{\cz}{{\mathcal{Z}}}
\newcommand{\ch}{{\mathcal{H}}}
\newcommand{\cf}{{\mathcal{F}}}

\newcommand{\uno}{{\mathbb 1}}

\newcommand{\rO}{{\mathrm{O}}}
\newcommand{\rG}{{\mathrm{G}}}

\newcommand{\nlc}{{\mathrm{II}_{1,17}}}
\newcommand{\rii}{{\mathrm{II}}}

\newcommand{\uo}{{\mathrm{U}(1)}}

\newcommand{\sug}{{\mathrm{SU}}}
\newcommand{\spg}{{\mathrm{Sp}}}
\newcommand{\sog}{{\mathrm{SO}}}
\newcommand{\sping}{{\mathrm{Spin}}}

\newcommand{\hosp}{{\mathrm{Spin}(32)/\mathbb{Z}_2}}

\newcommand{\RT}{g}

\newcolumntype{M}[1]{>{\centering\arraybackslash}m{#1}}
\newcolumntype{N}{@{}m{0pt}@{}}

\newcommand{\mfn}[1]{\mbox{\footnotesize$#1$}}

\newcommand{\op}{\hspace{1pt}}

\newcommand{\bsmat}{\left(\begin{smallmatrix}}
\newcommand{\esmat}{\end{smallmatrix}\right)}
\newsavebox{\smlmatE}
\savebox{\smlmatE}{$\begin{smallmatrix}E_{11} & E_{12}\\ E_{21} & E_{22}\end{smallmatrix}$}
\newsavebox{\smlmatA}
\savebox{\smlmatA}{$\begin{smallmatrix}a_{1}\\ a_{2}\end{smallmatrix}$}

\numberwithin{equation}{section}

\makeatletter
\def\@cline#1-#2\@nil{
  \omit
  \@multicnt#1
  \advance\@multispan\m@ne
  \ifnum\@multicnt=\@ne\@firstofone{&\omit}\fi
  \@multicnt#2
  \advance\@multicnt-#1
  \advance\@multispan\@ne
  \leaders\hrule\@height\arrayrulewidth\hfill
  \cr
  \noalign{\nobreak\vskip-\arrayrulewidth}}
\makeatother

\graphicspath{{./Images/}}

\begin{document}
\pagestyle{empty}
\begin{center}        
  {\bf\LARGE Exploring the landscape of  \\  CHL strings on $T^d$\\ [3mm]}

\large{ Anamar\'{\i}a Font$^{\sharp\, *}$, Bernardo Fraiman$^{\dag\, **\,\flat}$, Mariana Gra\~na$^{\flat}$, \\[-1mm]
Carmen A. N\'u\~nez$^{\dag\, **}$ and H\'ector Parra De Freitas$^{\flat}$
 \\[2mm]}
{\small  $^\sharp$ Facultad de Ciencias, Universidad Central de Venezuela\\[-2mm] }
{\small\it A.P.20513, Caracas 1020-A,  Venezuela\\ }
{\small  $^*$ Max-Planck-Institut f\"ur Gravitationsphysik, Albert-Einstein-Institut\\ [-2mm]}
{\small\it 14476 Golm, Germany\\ }
{\small  $^\dag$ Instituto de Astronom\'ia y F\'isica del Espacio (IAFE-CONICET-UBA)\\ [-2mm]}
{\small  $**$ Departamento de F\'isica, FCEyN, Universidad de Buenos Aires (UBA) \\ [-2mm]}
{\small\it Ciudad Universitaria, Pabell\'on 1, 1428 Buenos Aires, Argentina\\ }
{\small $\flat$  Institut de Physique Th\'eorique, Universit\'e Paris Saclay, CEA, CNRS\\ [-2mm]}
{\small\it  Orme des Merisiers, 91191 Gif-sur-Yvette CEDEX, France.\\[0.2cm] } 

{\small \verb"afont@fisica.ciens.ucv.ve, bfraiman@iafe.uba.ar, mariana.grana@ipht.fr,"\\[-3mm]}
{\small \verb"carmen@iafe.uba.ar, hector.parradefreitas@ipht.fr"} \\[.2cm]

\small{\bf Abstract} \\[3mm]\end{center}
 Compactifications of the heterotic string on special $T^d/{\mathbb Z}_2$ orbifolds realize a landscape of string models with 16 supercharges and a gauge group  on the left-moving sector of reduced rank $d+8$. The momenta of untwisted and twisted states span a lattice known as the Mikhailov lattice $\rii_{(d)}$, which is not self-dual for $d > 1$. By using computer algorithms which exploit the properties of lattice embeddings, we perform a systematic exploration of the moduli space for $d \le 2$, and give a list of maximally enhanced points where the $U(1)^{d+8}$ enhances to a rank $d+8$ non-Abelian gauge group. For $d = 1$, these groups are simply-laced and simply-connected, and in fact can be obtained from the Dynkin diagram of $\mre_{10}$. For $d = 2$ there are also symplectic  and doubly-connected groups. For the latter we find the precise form of their fundamental groups from embeddings of lattices into the dual of $\rii_{(2)}$. 
 Our results easily generalize to $d > 2$.

\newpage



\setcounter{page}{1}
\pagestyle{plain}
\renewcommand{\thefootnote}{\arabic{footnote}}
\setcounter{footnote}{0}

{\footnotesize{\tableofcontents}}
\newpage

\section{Introduction}
\label{sec:intro}

The rich landscape of string theory can be charted with a high level of rigor in regions where there is a full world-sheet
description. However, even for well understood string vacua with relatively simple geometries for the extra
dimensions, not much is known about the  structure of the moduli spaces and the classification of the possible gauge groups that 
can appear. These features contain a great deal of information about the properties and predictions of the theory. In particular, they are essential ingredients in the tests of duality conjectures, in the attempts to bring string theory closer to observable aspects of fundamental physics, and in applications of the swampland and string lamppost principles.

Ten dimensional string constructions with sixteen supercharges only admit $\mre_8\times \mre_8$ and $\hosp$ gauge groups \cite{Adams:2010zy,Kim:2019vuc} (the latter being locally equivalent to $\sog(32)$). In lower dimensions  many other gauge groups can be realized at special points of the moduli space, but these have only been partially identified. A systematic investigation on the structure of the moduli space  of  toroidal compactifications of the heterotic string was started
in \cite{Fraiman:2018ebo,Font:2020rsk}, where all the maximal gauge symmetry groups and the points in moduli space where they arise were presented for $S^1$ and $T^2$ compactifications.
The landscape of heterotic strings on $T^d$ contains gauge groups of (left) rank $d+16$ \cite{Narain:1985jj}. 
General criteria
to establish whether a gauge group is realized or not on $T^d$ were stated  in \cite{Font:2020rsk}
using lattice
embedding techniques. Modifying
the method of deleting nodes in the extended Dynkin diagram of the Narain lattice ${\rm II}_{1,17}$ \cite{GoddardOlive1985,Cachazo:2000ey}, we developed
more general algorithms to explore the moduli space, and all the maximally enhanced gauge groups, moduli, and other relevant information
about the embeddings in ${\rm II}_{d,d+16}$ were given for  $d \le 2$. In agreement with the duality between the heterotic string on $T^2$ and
F-theory on K3, all possible gauge groups on $T^2$ match all possible ADE types of singular
fibres of elliptic K3 surfaces that were found in \cite{SZ}.

With the motivation to get a better  understanding of the landscape of  string theory,  in this paper we extend the analysis of  \cite{Font:2020rsk}
to compactifications of the $\mre_8 \times \mre_8$ heterotic string on
$T^d/{\mathbb Z}_2$ asymmetric orbifolds which realize the  so-called CHL string \cite{Chaudhuri:1995fk, Chaudhuri:1995bf} (in $10-d$ dimensions with $d\ge 1$). This ${\mathbb Z}_2$ acts by exchanging
the two $\mre_8$ components of the momentum lattice, together with a shift by half a
period along one of the compact directions. One of the effects of this (freely acting)  ${\mathbb Z}_2$ modding is to
remove eight of the U(1) gauge bosons from the spectrum, thereby reducing the
rank of the gauge group by eight. The moduli space of the CHL string in $10-d$ dimensions is locally $\frac{\sog(d,d+8)}{\sog(d) \times \sog(d+8)}$ and world-sheet current algebras  can be realized at level 2 or 1.

The momenta of the physical states of the 9-dimensional CHL string   belong to the Lorentzian even self-dual lattice ${\rm II}_{1,9}$  \cite{Mikhailov:1998si}.   At generic points of the moduli space the (left) gauge symmetry is Abelian, namely $\uo^{9}$. 
In the absence of Wilson lines and for  generic values of the radius, some vector states of the untwisted sector  become massless  and enhance the gauge group  to $\mre_8\times \uo$. At the self-dual orbifold radius $R=\sqrt{2}$ (taking $\alpha'=1$), two twisted states become massless and a further enhancement to $\mre_8 \times \sug(2)$ takes place. Eight other non-Abelian ADE groups of maximal rank 9 can be found at other special points of the moduli space. All of these groups have world-sheet current algebras realized at level 2. In section \ref{sec:review} we list these groups, which can be easily obtained by deleting nodes from the generalized Dynkin diagram for $\rii_{1,9}$ (which is the same as the Dynkin diagram of the group $E_{10}$),  in an analogous way as for the $S^1$ compactification \cite{Fraiman:2018ebo}.

In less than nine  dimensions
the lattice of  momenta of the physical states (the so-called Mikhailov lattice) is even but not self-dual \cite{Mikhailov:1998si}.  This can be understood by noting that there is an asymmetry between the possible winding states along the orbifolded direction and those along the remaining ones, obstructing an automorphism that would make the lattice self-dual. On the other hand, this asymmetry enriches the pattern of gauge symmetries with respect to those found from the Narain lattice, and we find in addition gauge groups of BCF type. 

As in the case of $T^d$
compactifications, there does not seem to exist a Generalized Dynkin Diagram (GDD)  from
which one can extract all possible enhancements for $d > 1$. Although   many
different GDDs can  be constructed, it is uncertain whether they can produce the whole set of enhanced gauge groups.  Hence, we adapt 
the  exploration algorithm that was
introduced in \cite{Font:2020rsk}  for the Narain lattices to the Mikhailov lattices. 
We find that, for $d = 2$, the algorithm generates a list of 61 groups of maximal enhancement. In this case, the CHL string is a realisation of the anomaly free theories with 16 supercharges and rank 10  gauge groups
\cite{Kim:2019ths}.   We find that the ADE groups arise at level 2 while C groups appear at level 1 ($\mra_1$ also appears at level 1, but $\mra_1=\mrc_1$). Taking into account that the exploration algorithm produces all possible maximal enhancements
in  $T^2$ compactifications strongly indicates that these results exhaust all the possibilities. 

Roughly half of the  enhanced gauge groups in the CHL string in eight dimensions are multiply connected, while the rest are simply connected. Importantly, they are seen to satisfy the condition derived in \cite{Cvetic:2020kuw} for anomaly free 8d $\mathcal{N} = 1$ supergravities, a result that can be proven in general also for $T^2$ compactifications \cite{Cvetic:2021sjm}.

We note that our results for the 8d CHL string extend under T-duality to compactifications of the $\hosp$ heterotic string on tori without vector structure \cite{Witten:1997bs}, as discovered in \cite{Lerche:1997rr} and further discussed in \cite{deBoer:2001wca}. These theories are also dual to F-theory on K3 with a frozen singularity \cite{Witten:1997bs, Bhardwaj:2018jgp}, 
hence it should be possible to reproduce our results in that context. Similarly, the $d = 1$ case is dual to M theory on the 
M\"obius strip \cite{Dabholkar:1996pc,Aharony:2007du}. We do not dwell on these dualities here, focusing our attention only on the 
$(\mre_8 \times \mre_8)$ heterotic side. 

The paper is organized as follows. In section \ref{sec:review} we review the construction of the CHL string in nine dimensions  
as an $S^1/{\mathbb Z}_2$ orbifold of the $\mre_8 \times \mre_8$ heterotic string. We then find all the maximal enhancements from 
the Generalized Dynkin diagram and list them in Table \ref{tab:CHLd1}. The T-duality map among the states of the theory is also
checked. 
The more general setting of the CHL string in $D=10-d$ dimensions (with $D\le 9$) is considered in section \ref{sec:9-d}, where the theory 
 is realized as an orbifold of heterotic compactifications on $T^d$. We construct several extended diagrams, each of which gives a different set of enhanced gauge groups. In section \ref{sec:algorithms}
we explain the methods used in the algorithm that searches for maximal enhancement points and
illustrate them with an explicit example. We then present the maximal enhancements generated by  this procedure in the eight dimensional theory and collect the final results in Table \ref{tab:CHLd2}  of section \ref{sec:results}. 
Conclusions are the subject of section \ref{sec:conclusions}. The compactification of the heterotic string in the asymmetric $S^1/\ZZ_2$ orbifold is studied in some detail in Appendix \ref{app:pf}.
Finally, 
the world-sheet realisation of the space-time gauge symmetries  is briefly  discussed  in Appendix \ref{App:CA}.

\section{The nine-dimensional CHL String}
\label{sec:review}

In this section we review the construction of the CHL string in nine dimensions \cite{Chaudhuri:1995fk} as an $S^1/{\mathbb Z}_2$ orbifold of the $\mre_8 \times \mre_8$ heterotic string \cite{Chaudhuri:1995bf} and fix our conventions. We recall  the massless spectrum and study the possible gauge symmetries from the point of view of lattice embeddings. We will see that, as in the case of the heterotic string on $S^1$, this problem is well under control.

\subsection{Constructing the theory from the heterotic string}
\label{ss:construction}
Consider the $\mre_8 \times \mre_8$ heterotic string with the coordinate $x^9$ compactified on a circle of radius $R$. Varying $R$ and turning on the Wilson line $A$ on the compact direction we sweep through the Narain moduli space
\begin{equation}\label{Narainmod}
	\mathcal{M}_\text{Narain} = \frac{\rO(1,17,\mathbb{R})}{\rO(17,\mathbb{R})} \bigg/\rO(1,17,\mathbb{Z}),
\end{equation}
with the discrete T-duality group $\rO(1,17,\mathbb{Z})$ determining its global structure. These compactifications yield theories with gauge group of rank 17 (ignoring the graviphoton). However, the class of nine-dimensional theories with 16 unbroken supercharges also contains reduced rank theories, with gauge groups of rank 9 and 1. Those of rank 9 are realized in the CHL string, and have moduli space
\begin{equation}\label{modCHL}
	\mathcal{M}_{\text{CHL}} = \frac{\rO(1,9,\mathbb{R})}{ \rO(9,\mathbb{R})} \bigg/\rO(1,9,\mathbb{Z}),
\end{equation}
as will be made clear at the end of this section.

For our purposes, it is convenient to construct the CHL string as an orbifold of the $\mre_8 \times \mre_8'$ heterotic string following \cite{Chaudhuri:1995bf}. The orbifold symmetry $\RT=\rm R \rm T$ consists of the outer automorphism R of the $\mre_8 \times \mre_8'$ lattice accompanied by a half turn T around the compactification circle, namely
\begin{equation}\label{RTCHL}
	{\rm R}: ~~~ \Gamma_{\mre_8} \oplus \Gamma_{\mre_8'} \to \Gamma_{\mre_8'} \oplus \Gamma_{\mre_8}, ~~~~~ {\rm T}:  ~~ ~ x^9 \to x^9 + \pi R.
\end{equation}
Since $x^9 \sim x^9 + 2\pi R$  in the parent theory, $\RT^2 = 1$ and  $\RT$ defines a freely-acting $\mathbb{Z}_2$ orbifold.

To find the spectrum of this theory, we start by recalling the components of the internal momentum of the heterotic string in nine dimensions:
\begin{subequations}
	\begin{align}
	p_{R}&= \frac1{\sqrt2 R}\left[n-(R^2+\tfrac12 A^2) m-\Pi \cdot A\right] \, , \label{pR} \\
	p_{L}&=\frac1{\sqrt2 R}\left[n+(R^2-\tfrac12 A^2) m-\Pi\cdot A\right]
	, \label{pL} \\[2mm]
	p^{\hat I} &= \Pi^{\hat I}+A^{\hat I} m \, , \label{pI}
	\end{align}
	\label{momenta}%
\end{subequations}
 where $\hat I =1,....,16$, $n \in \mathbb{Z}$ is the momentum number on the circle, $m\in \mathbb{Z}$ is the winding number and $\Pi\in\Gamma_8 \oplus \Gamma_8$, with $\Gamma_8 \equiv \Gamma_{\mre_8}$.  The momenta form the unique even self-dual Lorentzian lattice $\nlc$ (up to $\sog(1,17)$ boosts given by the moduli), with vectors labeled by the quantum numbers $m, n, \Pi^{\hat I}$. We use the convention $\alpha' = 1$.
 
 On the $S^1/{\mathbb Z}_2$ orbifold, the Wilson lines are restricted to take the form 
 \beq \label{WilsonL}
 A=(a,a) \ , \quad a \in {\mathbb R}^8 . 
 \eeq
 Similarly, it is convenient to decompose the heterotic momenta as
 \begin{equation}\label{pidecomp}
 \Pi = (\pi,\pi'), ~~~~~ \pi, \pi' \in \Gamma_8,
 \end{equation}
and to define the symmetric and antisymmetric combinations
\begin{equation}\label{diagmom}
p^I_+ = \frac{1}{\sqrt{2}}(p^I + p^{I+8}), ~~~~~p^I_- = \frac{1}{\sqrt{2}}(p^I - p^{I+8}), \quad I=1,...,8
\end{equation}
Defining moreover the symmetric combination
\beq\label{rho}
\rho = \pi + \pi' \ \ \in  \Gamma_8 \ ,
\eeq 
the components \eqref{momenta} can be written as
\begin{subequations}
	\begin{align}
	p_{R}&= \frac1{\sqrt2 R}\left[n-R^2 m -a^2 m-\rho \cdot a\right] \, , \label{pRCHL} \\
	p_{L}&=\frac1{\sqrt2 R}\left[n+ R^2 m - a^2 m-\rho\cdot a \right] = p_R + \sqrt{2} R m
	\, , \label{pLCHL} \\[2mm]
	p_+ &= \frac{1}{\sqrt{2}} \left( \rho+ 2 a m \right) \, , \label{pI+} \\
	p_- &= \frac{1}{\sqrt{2}} (\pi-\pi') \, ,\label{pI-} 
	\end{align}
	\label{momentaCHL}%
\end{subequations}
and the total internal momentum vector  is $P=({{\bf p_R}};{\bf p_L})\equiv(p_R;p_L,p_+,p_-)$. 


The orbifold action on the momenta can be written as
\beq 
{\RT} \ |p_R;p_L, p_+,p_-\rangle = e^{2i\pi v\cdot P }|p_R; p_L, p_+,-p_-\rangle\, ,
\eeq
where the inner product is defined with respect to the metric diag$(-1,+1,\dots , +1)$. 
The shift vector $v$ is constrained by the condition that $g$ has order two. Choosing $v_-=0$
implies that $2 v$ belongs to the Narain lattice $\nlc$. Besides, the condition that the shift corresponds
to the geometric translation of $x^9$ by half a period amounts to $e^{2i\pi v\cdot P } = e^{i\pi n}$ and leads to
\beq \label{v}
v=\frac1{2\sqrt2} \left(-R -\frac{ a^2}{R}; R - \frac{a^2}{R},2a,~0\right).
\eeq
Notice that $2v$ equals the Narain lattice vector obtained by substituting $m = 1, n = 0$, and  $\pi = \pi' = 0$ in the formulae \eqref{momentaCHL}.
The lattice vectors can be conveniently traded for states $|m, n,\pi,\pi'\rangle$, which depend on the quantum numbers and transform as
\beq \label{RTCHL2}
{\RT} \  |m, n,\pi,\pi'\rangle = e^{i\pi n}|m, n,\pi',\pi\rangle,
\eeq
for all values of the moduli. 

The action of $g$ on the left-moving bosons living on $\Gamma_{\mre_8} \oplus \Gamma_{\mre_8'}$, 
denoted $Y^I$ and $Y'{}^I=Y^{I+8}$, $I=1,\ldots,8$,
is the exchange $Y^I \leftrightarrow Y'{}^I$, or $Y_\pm^I \to \pm Y_\pm^I$ where
\begin{equation}
\label{Ypmdef}
    Y_\pm^I = \frac{1}{\sqrt{2}}(Y^I \pm Y'{}^I).   
\end{equation}
The action on the space-time coordinates is just the translation in $x^9$. The corresponding oscillators then transform as
\begin{equation}\label{internalg}
g(\alpha^I)=\alpha^{I+8} \, , \quad g(\alpha^{I+8})= \alpha^I \ , \quad g(\alpha^{\mu})=\alpha^\mu ,
\eeq
where $\mu=2,..., 9$ refers to the space-time transverse coordinates.
Notice also that $g(\alpha_{\pm}^I)=\pm  \alpha_{\pm}^I$ for the $Y_\pm^I$ oscillators.

In the untwisted sector, the spectrum consists of states of the parent theory invariant under the orbifold action. 
The invariant states are superpositions of the form
\beq \label{twistedspectrumgen}
\ket{\varphi}_\text{untwisted} = \frac{1}{\sqrt{2}}\bigg( \alpha \ket{m,n,\pi, \pi'}+(-1)^n g(\alpha) \ket{m,n,\pi',\pi} \bigg) \, ,
\eeq
where $\alpha$ denotes any possible combination of oscillators and $g(\alpha)$ its image under $g$, given by \eqref{internalg}. 

In the twisted sector, the internal chiral bosons $Y^{I}$ and $Y'^I$ satisfy the boundary conditions
\begin{equation}\label{twistedcond1}
    Y^I(\sigma + 2\pi) = Y'^I(\sigma) + Q^I,  ~~~~~ Y'^I(\sigma + 2\pi) = Y^I(\sigma) + Q'^I,
\end{equation}
where $Q, Q'$ are arbitrary (fixed) vectors in $\Gamma_8$ which specify the precise way of exchanging $\mre_8 \leftrightarrow \mre_8'$ \cite{Elitzur:1986rs}.
The $Y_\pm^{I}$ then obey
\begin{equation}\label{twistedcondpm}
    Y_\pm^I(\sigma + 2\pi) = \pm Y^I_\pm(\sigma) + \frac{1}{\sqrt{2}}(Q^I \pm Q'^I)\, ,
\end{equation}
and have oscillator expansions
\begin{equation}\label{oscexpand}
\begin{split}
    Y^I_+(\tau+\sigma) &= \tfrac12 y_{+,0}^I +\tfrac{1}{2\pi} p_+^I (\tau+\sigma) +  i\sqrt{\frac{\alpha'}{2}}\sum_{n \neq 0} 
    \frac{\alpha^{I}_{+,\,n}}{n} e^{-in(\tau+\sigma)}\, ,\\
    Y^I_-(\tau+\sigma) &= \tfrac12 y_{-,0}^I +  i\sqrt{\frac{\alpha'}{2}}\sum_{s \, \in \, \ZZ+\tfrac12} \frac{\alpha^{I}_{-,\,s}}{s}
    e^{-i s(\tau+\sigma)}\, ,
\end{split}
\end{equation}
where $p_+^I \equiv \tfrac{1}{\sqrt2}(Q^I+Q'^I)$ and $y_{-,0}^I \equiv \tfrac{1}{\sqrt{2}}(Q^I - Q'^I)$. The boson corresponding to the compact  $x^9$ dimension   satisfies
\begin{equation}\label{twistedcond2}
\begin{split}
X^9(\sigma + 2\pi) &= X^9(\sigma) +  \pi R +  2\pi R \tilde m \equiv X^9(\sigma) + 2\pi R m,
\end{split}
\end{equation}
with $\tilde m \in \mathbb Z$, and hence $m \in \mathbb{Z} + \tfrac12$.

The twisted states have three distinctive features: they have half-integer winding  $m$, 
the occupation numbers of their oscillators can be half-integer or integer valued, and they do not have antisymmetric momentum 
$p_-^I$. We write them as 
\beq
\ket{\varphi}_\text{twisted} = \ket{m,n,\rho} \, , ~~~~~ \  \label{statesT}
\eeq
up to the action of oscillators. Note that upon quantisation the symmetric momentum takes the form 
$p_+ = \tfrac{1}{\sqrt2}(\rho + 2am)$, with $\rho = Q + Q' \in \Gamma_8$, coinciding with the untwisted symmetric momentum in \eqref{diagmom}.
The projection on invariant states in the twisted sector is best deduced from the partition function.
Using results in Appendix \ref{app:pf} it can be shown that all states satisfying level matching do survive the 
orbifold projection.

In the NS sector for the right movers (which gives the space-time bosons), the mass and level matching conditions are 
\bea
 M^2&=&{\bf p_L}^2+{\bf p_R}^2+2(N_L+N_R)+2{\mathfrak a}-1 \ ,\label{chlMass1}\\
0&=&{\bf p_L}^2-{\bf p_R}^2+2(N_L-N_R)+2{\mathfrak a}+1 \label{eqsU} \ ,
\eea 
where the zero point energy ${\mathfrak a}$ is -1 in the untwisted sector, as usual, and $-\tfrac{1}{2}$ in the twisted sector, since the left-moving side part receives contributions from 16 periodic bosons $\{Y^I_+, X^{\mu}\}$ (with $\mu$ labelling the 8 transverse directions) and 8 anti-periodic bosons $\{Y^I_-\}$. Concretely,
\begin{equation}
\mathfrak{a}_\text{twisted} = 16 \times {\mathfrak a}_{\text{periodic}} + 8 \times {\mathfrak a}_{\text{anti-periodic}} = -16 \times\frac{1}{24} + 8 \times \frac{1}{48} = -\frac{1}{2} \ .
\end{equation}

It is convenient to define the modified `oscillator number'
\begin{equation}\label{modosc}
	N_L' = N_L + \delta, ~~~~~~~~~~~~~~~ \delta =
	\begin{cases}  \frac{1}{2}\, p_-^2 &  \text{Untwisted} \\   \ \frac{1}{2} & \text{Twisted} \end{cases}\,,
\end{equation}
where $p_-^2$ is an integer (cf. \eqref{momentaCHL}), and the nine-dimensional momentum 
\begin{equation}
    P_L = (p_L,p_+),
\end{equation}
which allows to rewrite the formulas \eqref{chlMass1} and \eqref{eqsU} in an $\rO(1,9)$ covariant form as
\begin{subequations}
	\begin{align}
	M^2  &= P_L^2+p_R^2+2(N'_L+N_R)-3 \label{chlMass2} \\
	     &= \frac{1}{2} Z^T \mathcal H  Z  + 2(N'_L+N_R)-3 \label{chlMass2'}
	\end{align}
\end{subequations}
\vspace{-0.7in}
\begin{subequations}
	\begin{align}
		~~~0& = P_L^2-p_R^2+ 2(N'_L-N_R) -1  \label{chlMatch2} \\
		&=\frac{1}{2}Z^T \eta Z + 2(N'_L-N_R)-1  \label{chlMatch2'}\ .
	\end{align}
\end{subequations}
Here we have defined the charge vector
\begin{equation}
	Z \equiv \ket{\ell,n;\rho},
\end{equation}
with
\begin{equation}
	\ell \equiv 2m,
\end{equation}
and $\rho$ is defined in \eqref{rho}. Note that $\ell$ is always an integer, and is odd (even) for twisted (untwisted) states. $\mathcal{H}$ is the so-called `generalized metric'
\begin{equation}
	{\cal H} = \frac{1}{R^2}\begin{pmatrix}
	E^2 /2& -a^2 & Ea \\ -a^2 & 2 & -2a \\ Ea^T & -2a^T &  R^2 + 2 a^T a
	\end{pmatrix}\,,
\end{equation}
where $a$ is taken to be a row vector and the lower right $R^2$ term is implicitly multiplied by $\mathbb{1}_8$ so that $\mathcal{H}$ is a $10\times 10$ matrix, and
\begin{equation}\label{Edef}
    E \equiv R^2 + a^2 \, .
\end{equation}
Finally, $\eta$ is the $\rO(1,9)$ metric  
\beq \label{eta19}%
\eta = \begin{pmatrix}0& 1 & 0 \\
	1 & 0 & 0 \\
	0 & 0 & \mathbb{1}_{8}
\end{pmatrix} .
\eeq

The important result
\begin{equation}
	Z^2 \equiv Z^T \eta Z = 2\ell n + \rho^2 \in 2\mathbb{Z}
\end{equation}
implies that the charge vectors $Z$ span the even self-dual Lorentzian lattice $\rii_{1,9} \simeq \rii_{1,1}\oplus \Gamma_8$, since $\ell, n \in \mathbb{Z}$ and $\rho \in \Gamma_8$. 
The  correspondence between the states of the theory and the elements of $\rii_{1,9}$ was first derived in
\cite{Mikhailov:1998si}. Full details of the derivation are presented in Appendix \ref{app:pf}, where we discuss the partition function 
of the $S^1/\ZZ_2$ orbifold.

It can now be seen that the local structure of the moduli space \eqref{modCHL} is $\rO(1,9,\mathbb{R})/\rO(9,\mathbb{R})$ due to the reduction of the Wilson line from 16 to 8 components and the invariance of eqs. \eqref{chlMass2} and \eqref{chlMatch2} under $\rO(9,\mathbb{R})$ rotations of $P_L$. Furthermore, the automorphism group $\rO(1,9,\mathbb{Z})$ of $\rii_{1,9}$ corresponds to the T-duality group of the theory, giving the global structure for $\mathcal{M}_\text{CHL}$. The similarities between $\mathcal{M}_\text{CHL}$ and $\mathcal{M}_\text{Narain}$ (cf. eq. \eqref{Narainmod}) allow to carry out an analysis of the nine-dimensional CHL string mirroring the one performed for $S^1$ compactifications in \cite{Cachazo:2000ey}, namely constructing the fundamental region of the moduli space whose codimension $r \leq 9$ boundaries give enhanced semisimple gauge groups of rank $r$. This ensures that we are able to easily find all possible gauge group enhancements in the theory, as we explain shortly.

\subsection{Massless vectors}
\label{massless}

From equations \eqref{chlMass2} and \eqref{chlMatch2} we see that the NS sector contains massless states with $N_R = \frac{1}{2}$, $p_R = 0$ and 
\begin{equation}
	P_L^2 = 2(1-N_L') ~~~~~\Rightarrow~~~~~ N_L' = 0, 1, \frac{1}{2}.
\end{equation}
Of these, untwisted states can have $N_L = 0, 1$ and twisted states $N_L = 0$ (cf. eq. \eqref{modosc}). For $N_L=1$,  besides the universal gravitational sector, the massless spectrum contains the  left abelian  KK gauge vector
\beq\label{KKvec}
\alpha_{-1}^9\tilde\psi^\mu_{-\frac12}|0\rangle \ , 
\eeq
with $\tilde\psi_{-\frac12}^\mu$ the coefficient of the Laurent expansion of the  right-moving fermions, $\mu=2,...,8$, and
  the 8 symmetric combinations  of the Cartan sector of the heterotic theory  that survive the  R projection 
\be
\frac1{\sqrt2}(\alpha_{-1}^I+\alpha_{-1}^{I+8}) ~\tilde\psi^\mu_{-\frac12}|0\rangle\, ,\label{cartan}
\ee 
implying that the gauge group of the theory has rank 9.

For $N_L=0$, the set of massless states  depends on the point in moduli space. The $p_R = 0$ condition reads (cf. eq. \eqref{pRCHL})  
\begin{equation}
	\tfrac12 E \, \ell-n+a\cdot\rho = 0\, , \label{m03}
\end{equation}
with $E$ defined in \eqref{Edef}, while the level matching condition \eqref{chlMatch2'} becomes a constraint on the norm of $\rii_{1,9}$ vectors:
\begin{equation}\label{m04}
	Z^2 = 2\ell n + \rho^2 = 4(1-N_L') = 4 \text{ or } 2.
\end{equation}
The states with $Z^2 = 4$ correspond to $N'_L=0$, and from the definition of $N'_L$ given in \eqref{modosc} we see that this is only possible in the untwisted sector, with $\pi=\pi'$. From equation \eqref{twistedspectrumgen} we see that such states could only exist with $n$ even. However, substitution in \eqref{m04} gives $2\ell n + \rho^2 = 4q + 4\pi^2 = 4$, with $q$ even, or $\pi^2 = 1 - q = $ odd, which is inconsistent since $\pi \in \Gamma_8$. In compactifications to lower dimensions, such massless states do appear, and correspond to roots of gauge algebras at level 1, being long roots for non-ADE algebras (see section \ref{sec:9-d}). On the other hand, states with $Z^2=2$ are well defined in this case, and correspond to roots of ADE algebras at level 2. They come both from the twisted and untwisted sectors (the latter with $\pi=0$ or $\pi'=0$). We summarise this in Table \ref{tab:massless}.

\begin{table}[htb]
	\begin{center}
		\begin{tabular}{|c|c|c|}\hline
			 & twisted & untwisted   \\ \hline 
			$Z^2$ &         2                    & 2    \\  \hline
			$\ell$   &   odd  &   even             \\ \hline
			$n$   &   integer & integer            \\ \hline
			$\rho$ & $\Gamma_8$      & $\Gamma_8$       \\\hline
		\end{tabular}
		\caption[Caption for LOF]{Quantum numbers of the massless states in the twisted and untwisted sector in nine dimensions. The states must satisfy \eqref{m03} to be massless.}
		\label{tab:massless}
	\end{center}
\end{table}

At a generic point in the moduli space there are no massless states (twisted or untwisted) other than \eqref{KKvec}-\eqref{cartan}, since condition \eqref{m03} can only be satisfied generically for $Z = 0$, and therefore generically the gauge group is $\uo^9$. Enhanced gauge symmetry appears at special points in the moduli space, as we will show.

Let us look at the simple situation where $a=0$. The massless condition \eqref{m03} is trivially satisfied for states with $\ell = n = 0$, and the level matching condition \eqref{m04} with $\rho^2 = 2$, hence we get the massless untwisted sates with charge vectors
\begin{equation}
	Z = \ket{0,0;\rho}, ~~~~~ \rho^2 = 2.
\end{equation}
These are just the 240 roots of the $\mre_8$ arising from the symmetric combination of the two $\mre_8$'s in the parent theory. In the twisted sector, since $\ell$ is odd, equation \eqref{m03} is not satisfied for generic values of  the compactification radius, since $R = \sqrt{E}$ in this case. The surviving gauge group for $a=0$ and generic  $R$ is then $\mre_8 \times \uo$. Interestingly enough, taking $R=1$ when $a=0$ does not lead to additional states that enhance the $\uo$ to $\sug(2)$,
as occurs in the $S^1$ compactification. For this enhancement to occur we must actually take $R = \sqrt{2}$, i.e. $E = 2$, so that equations \eqref{m03} and \eqref{m04} are solved by 
\begin{equation}
	Z = \pm \ket{1,1;0},
\end{equation}
corresponding to two \textit{twisted} states with winding number $m = \pm \tfrac12$. 

As we review in Appendix \ref{App:CA},  in this example the  world-sheet  realisation of the $\mre_8 \times  \sug(2)$ space-time gauge symmetry is provided by a  Kac-Moody  algebra at level $k=2$. It is interesting to compare the radius that gives this enhancement  in the orbifold theory with the self-dual radius $R_k$ in the standard $S^1$ compactification where the enhancement occurs at $R_k=1$ and the gauge group is realized at level $1$. They are related as $R = {\sqrt{k}} R_k$. For generic Wilson lines this enhancement occurs at
\beq \label{Ek}
 E_k={k}^{-1} E \, = 1.
\eeq
In the following section we show that this is a generic feature: while maximal enhancement  in the heterotic string on $S^1$ occurs at $E=1$ and the Kac-Moody  algebra is realized at $k=1$,  in the nine-dimensional orbifold theory they occur at $E=2$ and $k=2$, i.e. both enhancements occur at $E_k = 1$. 
 This is actually expected from T-duality. We will shortly explain that in the orbifold theory the self-dual point is $E=2$.

\subsection{Maximal enhancements from the Generalized Dynkin diagram}
\label{sec:enhancements9d}

As we have commented in section \ref{ss:construction}, the structure of the moduli space of the nine-dimensional CHL string, $\mathcal{M}_\text{CHL}$, is similar to that of $S^1$ compactifications of the heterotic string,  $\mathcal{M}_\text{Narain}$. In particular, its global structure is given by $\rO(1,9,\mathbb{Z})$, the group of automorphisms of a Lorentzian even self-dual lattice. This group is reflexive, meaning that it can be generated by a finite set of Weyl reflections on the moduli space cover $\rO(1,9,\mathbb{R})/\rO(9,\mathbb{R})$, each of which fixes an hyperplane at the boundary of the fundamental domain. Each one of these reflections is uniquely associated to a short root quantum state that becomes massless on its fixed hyperplane, such that all possible enhanced semisimple gauge groups of rank $r$ may be found at their $r$-fold intersections  (for  details see \cite{Cachazo:2000ey}).

The upshot is that given the set of 10 roots corresponding to the boundaries of $\mathcal{M}_\text{CHL}$, we may simply impose that some of them satisfy the massless condition \eqref{m03} (condition \eqref{m04} is satisfied by construction), so that they become the simple roots of some simply laced gauge algebra. This can be done neatly by introducing the Generalized Dynkin Diagram (GDD) \cite{GoddardOlive1985} for the lattice $\rii_{1,9}$ shown in Figure \ref{edd1,9}, which is  the over-extended Dynkin Diagram for $\mre_8$, usually denoted $\mre_{10}$. The roots 1 through 8 are the simple roots of $\mre_8$, and we take them to have the following embedding in $\rii_{1,9}$
\beq
Z_i = \ket{0,0;\alpha_i} \ , \quad i=1,...,8,
\eeq
where the $\alpha_i$ are listed in Table \ref{tab:e8}. The root 0 corresponds to the lowest root of $\mre_8$ with the additional property that it has $n = -1$, i.e.
\begin{equation}\label{z0}
	Z_0 = \ket{0,-1;\alpha_0}.
\end{equation}
Finally, the root $\texttt{C}$ lies in the hyperbolic sublattice $\rii_{1,1}$ and reads
\begin{equation}
	Z_\texttt{C} = \ket{1,1;0}.
\end{equation}

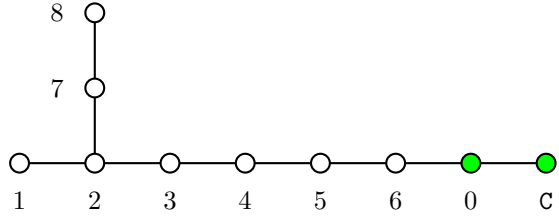
\begin{figure}[htb]
	\begin{center}
		\begin{tikzpicture}[scale=.25]
		\draw[thick] (0 cm,0) circle (5 mm) node [shift={(0.0,-0.5)}] {\mfn{1}} ;
		\draw[thick] (.5 cm, 0) -- (3.5 cm,0);    
		\draw[thick] (4 cm,0) circle (5 mm) node [shift={(0.0,-0.5)}] {\mfn{2}};
		\draw[thick] (4.5 cm, 0) -- (7.5 cm,0); 
		\draw[thick] (8 cm,0) circle (5 mm) node [shift={(0.0,-0.5)}] {\mfn{3}};
		\draw[thick] (8.5 cm, 0) -- (11.5 cm,0);     
		\draw[thick] (12 cm,0) circle (5 mm) node [shift={(0.0,-0.5)}] {\mfn{4}};
		\draw[thick] (12.5 cm, 0) -- (15.5 cm,0);     
		\draw[thick] (16 cm,0) circle (5 mm) node [shift={(0.0,-0.5)}] {\mfn{5}};
		\draw[thick] (16.5 cm, 0) -- (19.5 cm,0);     
		\draw[thick] (20cm,0) circle (5 mm) node [shift={(0.0,-0.5)}] {\mfn{6}};
		\draw[thick] (4 cm, 0.5cm) -- (4 cm, 3.5cm);     
		\draw[thick] (4cm,4cm) circle (5 mm) node [shift={(-0.5,0.0)}] {\mfn{7}};
		\draw[thick] (4 cm, 4.5cm) -- (4 cm, 7.5cm);     
		\draw[thick] (4cm,8cm) circle (5 mm) node [shift={(-0.5,0.0)}] {\mfn{8}};
		\draw[thick] (20.5 cm, 0) -- (23.5 cm,0);     
		\draw[thick, fill = green] (24cm,0) circle (5 mm) node [shift={(0.0,-0.5)}] {\mfn{0}};
		\draw[thick] (24.5 cm, 0) -- (27.5 cm,0);     
		\draw[thick, fill=green] (28cm,0) circle (5 mm) node [shift={(0.0,-0.5)}] {\footnotesize{$\texttt{C}$}};
		\end{tikzpicture}
	\end{center}
	\caption{Generalized Dynkin Diagram for the lattice $\text{II}_{1,9}$. The coloring of the nodes 0 and \texttt{C} reflects the fact the the associated states have nonzero momentum and/or winding, as opposed to the white nodes.}\label{edd1,9}
\end{figure}

\begin{table}[h!]\begin{center}
\renewcommand{\arraystretch}{1.5}
{\footnotesize
	\begin{tabular}{|c|c!{\vrule width 1.2pt}c|c|}
		\hline
		$i$ &$\kappa_i$ & $\alpha_i$ & $w_i$ \\[3pt]
		\hline\hline
		1 & 3 & $(1,\tm1,0,0,0,0,0,0)$ & $\tm(\tm\frac12,\frac12,\frac12,\frac12,\frac12,\frac12,\frac12,\tm\frac52)$  \\[3pt]\hline
		2 & 6&  $(0,1,\tm1,0,0,0,0,0)$ & $\tm(0,0,1,1,1,1,1,\tm5)$   \\[3pt]\hline
		3 & 5&  $(0,0,1,\tm1,0,0,0,0)$ & $\tm(0,0,0,1,1,1,1,\tm4)$   \\[3pt]\hline
		4 & 4&  $(0,0,0,1,\tm1,0,0,0)$ & $\tm(0,0,0,0,1,1,1,\tm3)$   \\[3pt]\hline
		5 &  3&  $(0,0,0,0,1,\tm1,0,0)$ & $\tm(0,0,0,0,0,1,1,\tm2)$   \\[3pt]\hline
		6 &  2&  $(0,0,0,0,0,1,\tm1,0)$ & $(0,0,0,0,0,0,\tm1,1)$   \\[3pt]\hline
		7 & 4& $\tm(1,1,0,0,0,0,0,0)$ & $\tm(\frac12,\frac12,\frac12,\frac12,\frac12,\frac12,\frac12,\tm\frac72)$   \\[3pt]\hline 
		8 & 2& $(\frac12,\frac12,\frac12,\frac12,\frac12,\frac12,\frac12,\frac12)$ & $(0,0,0,0,0,0,0,2)$   \\[3pt]\hline
		0 & 1 & $(0,0,0,0,0,0,1,\tm1)$ & $(0,0,0,0,0,0,0,0)$ \\ \hline 
	\end{tabular}
}
\caption{Simple roots $\alpha_i$, Kac marks $\kappa_i$ and fundamental weights $w_i$ of $\mre_8$.}
\label{tab:e8}\end{center}\end{table}  

Maximally enhanced (rank 9) non-Abelian gauge groups are then found by deleting one node in the GDD such that the remaining nodes form the Dynkin diagram of an ADE algebra. Imposing the condition \eqref{m03} on the roots associated to the remaining nodes gives rise to 9 constraints on the moduli and defines a singular point $(E,a)$ at the boundary of the fundamental domain with maximally enhanced gauge group. More generally, deleting $s$ nodes defines a subvariety 
of dimension $s-1$ with generic semisimple gauge group of rank $10-s$, given by the remaining Dynkin diagram.

Note that for maximal enhancements the node \texttt{C} cannot be broken, since the remaining diagram corresponds to the infinite dimensional algebra $\mre_9$. This means that all maximal enhancements must contain this node, and from equation \eqref{m03} this implies that $E = 2$. The massless condition then reduces to
\begin{equation}\label{m03eff}
	a \cdot \rho = \ell - n.
\end{equation}
Deletion of the $i$th node, $i = 0,...,8$, corresponds to the Wilson line
\begin{equation}
	a = \frac{1}{\kappa_i}w_i,
\end{equation}
with no sum over $i$, where $w_i$ and $\kappa_i$ are respectively the fundamental weight and Kac mark listed in Table \ref{tab:e8}. It is easy to show that this prescription exactly solves equation \eqref{m03eff} for the remaining roots $Z_{j \neq i}$, while violating the one for $Z_i$ since $w_i\cdot\alpha_i / \kappa_i \notin \mathbb{Z}$, $i \neq 0$. In fact, these values for the Wilson line correspond to those for a shift vector breaking $\mre_8$ to a maximal regular subgroup \cite{kac:1969}.

The maximal enhancements are listed in Table \ref{tab:CHLd1}, where the subindex on the gauge group indicates that the world-sheet Kac-Moody algebra is realized at level 2. This is explained in detail in Appendix \ref{App:CA}. Moreover, note that the relation \eqref{Ek} is satisfied in all cases, since $E = 2$.

\begin{table}[htb]
\begin{center}
\renewcommand{\arraystretch}{1.5}
\begin{tabular}{|c!{\vrule width 1.2pt}c!{\vrule width 1.2pt}c|c|} \hline
	$i$ & Gauge group root lattice & $E$ & $-a$ \\ \hline \hline
	1 & $\mra_9$ & 2 & $(-\tfrac16,\tfrac16,\tfrac16,\tfrac16,\tfrac16,\tfrac16,\tfrac16, -\tfrac56)$ \\ \hline
	2 & $\mra_1 + \mra_2 + \mra_6$ & 2 & $(0,0,\tfrac16, \tfrac16, \tfrac16, \tfrac16, \tfrac16, -\tfrac56)$ \\ \hline
	3 & $\mra_4 + \mra_5$ & 2 & $(0,0,0,\tfrac15, \tfrac15, \tfrac15, \tfrac15, -\tfrac45)$ \\ \hline
	4 & $\mrd_5 + \mra_4$ & 2 & $(0,0,0,0,\tfrac14, \tfrac14, \tfrac14, -\tfrac34)$ \\ \hline
	5 & $\mre_6 + \mra_3$ & 2 & $(0,0,0,0,0,\tfrac13, \tfrac13, -\tfrac23)$ \\ \hline
	6 & $\mre_7 + \mra_2$ & 2 & $(0,0,0,0,0,0,\tfrac12,-\tfrac12)$ \\ \hline
	7 & $\mra_1 + \mra_8$ & 2 & $(\tfrac18,\tfrac18,\tfrac18,\tfrac18,\tfrac18,\tfrac18,\tfrac18,-\tfrac78)$ \\ \hline
	8 & $\mrd_9$ & 2 & $(0,0,0,0,0,0,0,-1)$ \\ \hline
	0 & $\mre_8 + \mra_1$ & 2 & $(0,0,0,0,0,0,0,0)$ \\ \hline
\end{tabular} 
\caption{Maximal enhancements in the nine-dimensional theory, obtained by deleting the $i$th node in the GDD shown in Figure \ref{edd1,9}. All groups arise at level 2. The Wilson line is always of the form $a = w_i/\kappa_i$ (cf. Table \ref{tab:e8}).}
\label{tab:CHLd1}\end{center}\end{table}

\subsection{T-duality}
\label{ss:tduality}

The T-duality group of the nine-dimensional CHL string is $\rO(1,9,\mathbb{Z})$, the automorphism group of $ \rii_{1,9}$. Of particular interest is the Weyl reflection, say $T$, generated by the root $Z_\texttt{C}$, whose action on the moduli and the quantum numbers $\ell, n, \rho$ is
\begin{equation}
	T:~~~~~E \leftrightarrow \frac{4}{E}, ~~~~~ a \leftrightarrow \frac{2a}{E}, ~~~~~ \ell \leftrightarrow n,  ~~~~~ \rho \leftrightarrow -\rho
\end{equation}
while $N'_L$ is invariant. Note that this transformation is not inherited from the T-duality group of the parent theory on $S^1$, although it is analogous to the transformation $E \to 1/E$ found there. In fact, in the $S^1/\ZZ_2$ orbifold
some states in the untwisted sector are transformed under $T$ to states in the twisted sector.  
Twisted states with $\ell$ odd and $n$ even are mapped to untwisted states with $\ell$ even and $n$ odd (cf. Table \ref{tab:massless}), and vice versa. This mixing of the two sectors under T-duality was originally noted in \cite{Mikhailov:1998si}.

In Appendix \ref{app:pf}, we show that the partition function of the $S^1/\ZZ_2$ orbifold is invariant under $T$.
One can also see explicitly how the mixing of untwisted and twisted states occurs at the level of the Hilbert space by taking 
into account the difference in the ground states and internal oscillators of the two sectors. As a simple example consider the twisted state with $\ell = 1$, $n = 0$, $\rho = r$, with $r$ a root of $\mre_8$, and no left oscillators. Since T-duality preserves the norms of the momenta $p_R^2$ and $P_L^2$, it should also preserve the value of $N_L'$ to leave the mass \eqref{chlMass2} unaffected. In this case, $N_L' = \tfrac12$, and so the transformed untwisted state must have $p_-^2 = 1$ (cf. eq. \eqref{modosc}). It is not hard to see that it should take the form
\begin{equation}
    \frac{1}{\sqrt{2}}\left( \ket{0,1;r,0} - \ket{0,1;0,r}\right),
\end{equation}
where the notation is that of Eq. \eqref{twistedspectrumgen}. 

The mapping is more complicated when oscillators are involved. Consider for instance the set of twisted states with charge vector 
$Z = \ket{1,0;0}$ and $N_L' = 2$, i.e. $N_L = \tfrac32$. The allowed combinations of oscillators along the eight directions $I$ that 
can act on $Z$ are
\begin{equation}
\alpha^I_{-,-\frac12} \alpha^J_{-,-\frac12} \alpha^K_{-,-\frac12}, ~~~ \alpha^I_{+,-1}\alpha^J_{-,-\frac12}, 
~~~ \alpha^I_{-,-\frac32}, \quad I,J,K=1,\ldots,8,
\end{equation}
giving $120 + 64 + 8 = 192$ states. 
Their T-dual untwisted states, labelled by $\ket{\ell,n;\pi,\pi'}$, must have $\ell=0$, $n=1$, $\pi'=-\pi$
since $\rho = 0$, and they must also add up to 192 states. For the first 120 twisted states the T-duality is 
\begin{equation}
\alpha^I_{-,-\frac12} \alpha^J_{-,-\frac12} \alpha^K_{-,-\frac12}, \ket{1,0;0,0} \quad \leftrightarrow  \quad  
\frac{1}{\sqrt{2}}\left(\ket{0,1;r,-r} - \ket{0,1;-r,r}\right),
\end{equation}
where $r$ is any of the 120 positive roots of $\mre_8$ (the other 120 give the same states up to an overall irrelevant sign). We see that $p_-^2 = 2r^2 = 4$, hence $N_L' = 2$ as required. 

For the remaining states the mapping reads
\begin{equation}
\begin{split}
\alpha^I_{+,-1}\alpha^J_{-,-\frac12} \ket{1,0;0,0} \quad &\leftrightarrow  \quad   
\alpha^I_{+,-1}\alpha^J_{-,-1} \ket{0,1;0,0}, \\[2mm]
\alpha^I_{-,-\frac32},  \ket{1,0;0,0} \quad& \leftrightarrow  \quad   \alpha^I_{-,-2}\ket{0,1;0,0}. 
\end{split}
\end{equation}
Here we have used that in the untwisted sector the $\alpha^I_{-}$ oscillators have integer occupation number and 
under the orbifold action pick up a minus sign so that the full states are invariant.

\section{The CHL string in $D$ dimensions } \label{sec:9-d}
We now consider the more general setting of the CHL string in $D$ external dimensions, with $D\le 9$. It is realized as an orbifold of heterotic compactifications on $T^d$ (with $d=10-D$), where the orbifold symmetry is again $g = \text{RT}$ (cf. eq. \eqref{RTCHL}), with T a half-turn around one of the cycles of $T^d$. We will choose this cycle to be  along  $x^9$, while the others remain unaffected.

\subsection{Extending the nine-dimensional construction}

The moduli of the $\mre_8\times \mre_8$ heterotic string on $T^d$  are the torus metric $G_{ij}$, the antisymmetric tensor $B_{ij}$ and the Wilson lines $A_i$, where $i,j = 1,...,d$. Again, the Wilson lines have to be invariant under the R  symmetry, which implies that they are of the form $A_i = (a_i,a_i)$. Generalizing \eqref{Edef}, we define the moduli
\begin{equation}
    E_{ij} = G_{ij} + B_{ij} + a_i\cdot a_j,
\end{equation}
and  the quantum numbers
\begin{equation}
    \ell^i \equiv 2m^i, ~~~~~n_i, ~~~~~ \rho^I \equiv \pi^I + \pi'{}^I,
\end{equation}
where $m^i$ and $n_i$ are the winding  and momentum numbers along the $i$th direction and $\pi^I, \pi'{}^I$ are the same as in \eqref{pidecomp}. The momenta \eqref{momentaCHL} are then generalized to
\begin{subequations}
	\begin{align}
	p_{R}&= \frac{1}{\sqrt{2}} \left(n_i -\tfrac12 E_{ij} \ell^j - a_i \cdot \rho \right) \hat e^i, \label{pRCHLD} \\
	p_{L}&=\frac{1}{\sqrt{2}}\left(n_i +(G_{ij} -\tfrac12 E_{ij}) \ell^j - a_i \cdot \rho \right) \hat e^i = p_R + \frac{1}{\sqrt{2}} \ell^i e_i
	\, , \label{pLCHLD} \\[2mm]
	p_+ &= \frac{1}{\sqrt{2}} \left( \rho+ \ell^ia_i \right) \, , \label{pI+D} \\
	p_- &= \frac{1}{\sqrt{2}} (\pi-\pi') \, ,\label{pI-D} 
	\end{align}
	\label{momentaCHLD}%
\end{subequations}
 where $e_i$ is the vielbein for the torus metric, i.e. $e_i \cdot e_j = G_{ij}$, and $\hat e^i$ its inverse.
 
The construction of the spectrum in section \ref{sec:review} carries over with some differences. Basically, the $i = 1$ direction behaves as in the nine-dimensional case, while the other directions $i \geq 2$ behave as in the usual $T^d$ compactification. In particular, the charge vectors
\begin{equation}
    Z \equiv \ket{\ell^1,...,\ell^d,n_1,...,n_d;\rho}
\end{equation}
have $\ell^1$ odd (even) for twisted (untwisted) states, but $\ell^2,...,\ell^d$ are always even, while in general, $n_1,...,n_d \in \mathbb{Z}$ and $\rho \in \Gamma_8$. 

The Lorentzian metric \eqref{eta19} generalizes to
\begin{equation}\label{etaD}
    \eta = 
    \begin{pmatrix}
    0  & \mathbb{1}_d & 0\\
    \mathbb{1}_d  & 0 & 0\\
    0 & 0 & \mathbb{1}_8
    \end{pmatrix}
\end{equation}
and, together with the allowed values for the quantum numbers, already suggests that the vectors $Z$ span the lattice
\begin{equation}\label{Mikh}
    \rii_{(d)} \simeq \rii_{d-1,d-1}(2) \oplus \rii_{1,9}.
\end{equation}
The $(2)$ at the right of $\rii_{d-1,d-1} \simeq \overset{d-1}{\bigoplus} \rii_{1,1}$ means that the norm squared of its vectors is scaled by a factor of 2, in this case due to $\ell^2,...,\ell^d$ always being even. This is in agreement with \cite{Mikhailov:1998si}, where these lattices were initially introduced. We therefore refer to $\rii_{(d)}$ in this context as the Mikhailov lattice. This is the analog of the Narain lattice $\rii_{d,d+16}$, but with the important difference that it is not self-dual (except for the $d=1$ case reviewed in section \ref{sec:review}). More details of these lattices can be found  in Appendix \ref{app:pf}, and here we present the spectrum as originally worked out in \cite{Mikhailov:1998si}.

The left moving sector of the theory now includes $d$ abelian KK gauge vectors like \eqref{KKvec}, so that the gauge group is of rank $8+d$. A generic point in the moduli space has gauge group $\uo^{d+8}$, but at special points this group is enhanced. The novel feature for $d > 1$ compactifications is that states with $Z^2 = 4$ can become massless and certain enhanced gauge groups are not simply laced, as we now show.

The zero mass and level matching conditions generalizing \eqref{m03} and \eqref{m04} are
\begin{equation}\label{m03D}
	\frac{1}{2}E_{ij}\ell^j - n_i + a_i \cdot \rho = 0, ~~~~~ i = 1,...,d \, , 
\end{equation}
\vspace{-0.3 in}
\begin{equation}\label{m04D}
	Z^2 = 2\ell^i n_i + \rho^2 = 4 \text{ or } 2 \, .
\end{equation}
Let us take for the moment $d = 2$. An untwisted state with $Z^2 = 4$ has $n_1$ even and $\rho^I = 2\pi^I$. Substituting in \eqref{m04D} gives $2\ell^1 n_1 + 2\ell^2 n_2 + \rho^2 = 2\ell^2 n_2 + 4q + 4\pi^2 = 4$, with $q$ even, but in contrast  to the situation in $d = 1$, it can be solved by an appropriate choice of $\ell^2$ and $n_2$. Indeed, the product $\ell^2n_2$ can be any even number, say $2p$ with $p\in \mathbb{Z}$. Then \eqref{m04D} reduces to $\pi^2 = 1 - q - p$, which admits solutions in $\rii_{(2)}$ if $p$ is odd. 
These states give rise to $\mrc_n$ gauge algebras at level 1, where they play the role of long roots when $n \geq 2$ ($\mrc_1 = \mra_1$). For $d \geq 3$ there are more possibilities such as $\mrb_n$ and $\mrf_4$ algebras. In Table \ref{tab:massless2} we record the values of the quantum numbers that massless states can have for $d \geq 2$, together with the squared length $Z^2$ of the charge vector.

\begin{table}[htb]
	\begin{center}
		\begin{tabular}{|c|c|c|c|}\hline
			& twisted & \multicolumn{2}{c|}{untwisted}    \\ \hline 
			$Z^2$ &         2                    & 2 &    4    \\  \hline
			$\ell^1$   &   odd  &   even             & even \\ \hline
			$n_1$   &   integer & integer  &      even           \\ \hline
			$\ell^i$   &   even  &   even             & even \\ \hline
			$n_i$   &   integer & integer  &      integer           \\ \hline
			$\rho$ & $\Gamma_8$      & $\Gamma_8$       & $2\Gamma_8$ \\\hline
		\end{tabular}
		\caption[Caption for LOF]{Quantum numbers of the massless states in the twisted and untwisted sector. The index $i > 1$ corresponds to further compactifications of the nine-dimensional theory. States with $Z^2 = 4$ can only be massless in $D<9$ dimensions. The states must satisfy \eqref{m03D} to be massless.}
		\label{tab:massless2}
	\end{center}
\end{table}

\subsection{Generalized Dynkin Diagrams}
\label{ss:gdd}
As in $T^d$ compactifications of the heterotic string, there does not seem to exist a GDD for $d > 1$ from which one can extract all possible enhancements. One obstruction to obtaining such a GDD is that the group of automorphisms for the Mikhailov lattice, similarly to the Narain lattice, is not generated by simple reflections when $d > 1$. In \cite{Font:2020rsk} it was found that indeed there are many possible GDDs in $T^{2}$ heterotic compactifications, each of which provides partial information about the possible gauge symmetry ehnancements in the theory. Here we extend this construction to the CHL string, where the story is similar.

The simplest kind of GDD that can be constructed is just the GDD of $\rii_{1,9} \subset \rii_{(d)}$ together with some nodes representing vectors with $Z^2 = 4$ in its orthogonal complement $\rii_{d-1,d-1}(2)$. The maximal enhancements that can be read from these diagrams are those of the form 
\begin{equation}
	(\rG_9)_2 + (\hat \rG_{d-1})_1 \, ,
\end{equation}
where $\rG_9$ is any maximal rank algebra of the $d = 1$ theory at level 2, and $\hat \rG_{d-1}$ is any rank $d-1$ algebra of ADE type at level 1. The roots of $\rG_9$ are the same as before, with $\ell^2 = ... = \ell^d = n_2 =... = n_d = 0$, and the roots of $\hat \rG_{d-1}$ have $\ell^1 = n_1 = \rho^I = 0$. 

For $d = 2$, there is only one such diagram, shown in Figure \ref{edd101a}, giving maximal enhancements of the form $(\rG_9)_2 + (\mra_1)_1$. The extra vector is
\begin{equation}
	Z_{\texttt{C}_2} = \ket{0,2,0,1;0},
\end{equation}
and the moduli for the enhancements are found by constraining them with equation \eqref{m03D}. The results are straightforward generalizations of those in Table \ref{tab:CHLd1}, with $E_{ij} = \text{diag}(2,1)$, $a_1 = a$, $a_2 = 0$, and an extra $(\mra_1)_1$ factor in every algebra. 

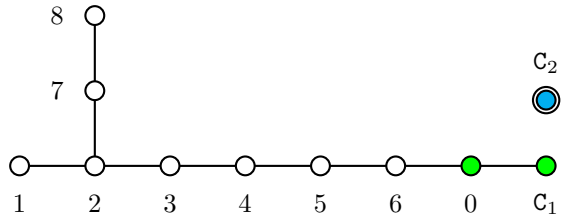
\begin{figure}[htb]
	\begin{center}
		\begin{tikzpicture}[scale=.25]
		\draw[thick] (0 cm,0) circle (5 mm) node [shift={(0.0,-0.5)}] {\mfn{1}} ;
		\draw[thick] (.5 cm, 0) -- (3.5 cm,0);    
		\draw[thick] (4 cm,0) circle (5 mm) node [shift={(0.0,-0.5)}] {\mfn{2}};
		\draw[thick] (4.5 cm, 0) -- (7.5 cm,0); 
		\draw[thick] (8 cm,0) circle (5 mm) node [shift={(0.0,-0.5)}] {\mfn{3}};
		\draw[thick] (8.5 cm, 0) -- (11.5 cm,0);     
		\draw[thick] (12 cm,0) circle (5 mm) node [shift={(0.0,-0.5)}] {\mfn{4}};
		\draw[thick] (12.5 cm, 0) -- (15.5 cm,0);     
		\draw[thick] (16 cm,0) circle (5 mm) node [shift={(0.0,-0.5)}] {\mfn{5}};
		\draw[thick] (16.5 cm, 0) -- (19.5 cm,0);     
		\draw[thick] (20cm,0) circle (5 mm) node [shift={(0.0,-0.5)}] {\mfn{6}};
		\draw[thick] (4 cm, 0.5cm) -- (4 cm, 3.5cm);     
		\draw[thick] (4cm,4cm) circle (5 mm) node [shift={(-0.5,0.0)}] {\mfn{7}};
		\draw[thick] (4 cm, 4.5cm) -- (4 cm, 7.5cm);     
		\draw[thick] (4cm,8cm) circle (5 mm) node [shift={(-0.5,0.0)}] {\mfn{8}};
		\draw[thick] (20.5 cm, 0) -- (23.5 cm,0);     
		\draw[thick, fill = green] (24cm,0) circle (5 mm) node [shift={(0.0,-0.5)}] {\mfn{0}};
		\draw[thick] (24.5 cm, 0) -- (27.5 cm,0);     
		\draw[thick, fill=green] (28cm,0) circle (5 mm) node [shift={(0.0,-0.5)}] {\footnotesize{$\texttt{C}_1$}};
		\draw[fill=cyan, thick] (28cm,3.5cm) circle (5 mm) node [shift={(0.0,0.5)}] {\footnotesize{$\texttt{C}_2$}};
		\draw[thick] (28cm,3.5cm) circle (7 mm);
		\end{tikzpicture}
		\caption{Generalized Dynkin diagram giving enhancements $(\rG_9)_2 + (\mra_1)_1$. Green (blue) coloring means that the state has nontrivial momentum number and/or winding only along direction 1 (2). The double border of the $\texttt{C}_2$ node indicates that it corresponds to a vector with $Z^2 = 4$. }\label{edd101a}
	\end{center}
\end{figure}

Although the GDD just constructed may seem trivial, it serves as a starting point for a much more interesting one. Simply make the replacement
\begin{equation}\label{c2c2'}
	Z_{\texttt{C}_2} \to Z_{\texttt{C}_2'} = \ket{2,2,0,1;0}.
\end{equation}
This new vector has nonzero inner product with $Z_{\texttt{C}_1}$ such that the resulting Dynkin diagram is the one shown in Figure \ref{edd101b}. The first thing to note is that for maximal enhancements, neither of the nodes $\texttt{C}_1$ and $\texttt{C}_2'$ can be removed. This means that there will always be some non-ADE factor $\mrc_n$ in the resulting gauge algebra, and that $E_{ij}$ will always take the value
\begin{equation}
	E_{ij} = 
	\begin{pmatrix}
	2 & -2 \\ 0 & 1
	\end{pmatrix} \, ,
\end{equation}
as can be seen by substituting $Z_{\texttt{C}_1}$ and $Z_{\texttt{C}_2'}$ in \eqref{m03D}. Moreover note that removing node 8 does not lead to a valid Dynkin diagram, i.e. a finite-dimensional semisimple Lie algebra. This is consistent with the fact that the predicted moduli for such a point has torus metric with negative determinant, as one can easily check. The valid maximal enhancements given by this diagram can be read off from Table \ref{tab:CHLd1} ($i \neq 8$) by taking $E_{ij} = \bigl(\begin{smallmatrix}2&-2\\0&1\end{smallmatrix}\bigr)$, $a_1 = a$, $a_2 = 0$ and replacing the rightmost $(\mra_n)_2$ factor by $(\mrc_{n+1})_1$. 

\begin{figure}[htb]
	\begin{center}
		\begin{tikzpicture}[scale=.25]
		\draw[thick] (0 cm,0) circle (5 mm) node [shift={(0.0,-0.5)}] {\mfn{1}} ;
		\draw[thick] (.5 cm, 0) -- (3.5 cm,0);    
		\draw[thick] (4 cm,0) circle (5 mm) node [shift={(0.0,-0.5)}] {\mfn{2}};
		\draw[thick] (4.5 cm, 0) -- (7.5 cm,0); 
		\draw[thick] (8 cm,0) circle (5 mm) node [shift={(0.0,-0.5)}] {\mfn{3}};
		\draw[thick] (8.5 cm, 0) -- (11.5 cm,0);     
		\draw[thick] (12 cm,0) circle (5 mm) node [shift={(0.0,-0.5)}] {\mfn{4}};
		\draw[thick] (12.5 cm, 0) -- (15.5 cm,0);     
		\draw[thick] (16 cm,0) circle (5 mm) node [shift={(0.0,-0.5)}] {\mfn{5}};
		\draw[thick] (16.5 cm, 0) -- (19.5 cm,0);     
		\draw[thick] (20cm,0) circle (5 mm) node [shift={(0.0,-0.5)}] {\mfn{6}};
		\draw[thick] (4 cm, 0.5cm) -- (4 cm, 3.5cm);     
		\draw[thick] (4cm,4cm) circle (5 mm) node [shift={(-0.5,0.0)}] {\mfn{7}};
		\draw[thick] (4 cm, 4.5cm) -- (4 cm, 7.5cm);     
		\draw[thick] (4cm,8cm) circle (5 mm) node [shift={(-0.5,0.0)}] {\mfn{8}};
		\draw[thick] (20.5 cm, 0) -- (23.5 cm,0);     
		\draw[thick, fill = green] (24cm,0) circle (5 mm) node [shift={(0.0,-0.5)}] {\mfn{0}};
		\draw[thick] (24.5 cm, 0) -- (27.5 cm,0);     
		\draw[thick] (28.2cm, 0) -- (28.2cm ,3.5cm);
		\draw[thick] (27.8cm, 0) -- (27.8cm ,3.5cm); 
		\draw[thick] (27 cm,  2.5cm) -- (28cm, 1.5cm) --  (29 cm,  2.5cm);
		\draw[thick, fill=green] (28cm,0) circle (5 mm) node [shift={(0.0,-0.5)}] {\footnotesize{$\texttt{C}_1$}};
		\draw[thick, fill=yellow] (28cm,3.5cm) circle (5 mm) node [shift={(0.0,0.5)}] {\footnotesize{$\texttt{C}_2'$}};
		\draw[thick] (28cm,3.5cm) circle (7 mm);
		\end{tikzpicture}
		\caption{Generalized Dynkin diagram for $d = 2$ theories with enhancements to algebras with $\mrc_n$ factor. It is obtained from the GDD in figure \ref{edd101a} by replacing the node $\texttt{C}_2$ with $\texttt{C}_2'$ as shown in \eqref{c2c2'}. Yellow coloring means that the state has nontrivial momentum number and/or winding along directions 1 and 2.}
		\label{edd101b} 
	\end{center}
\end{figure}
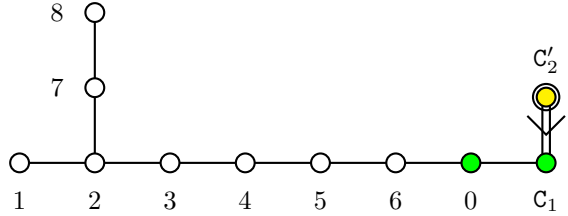

Another interesting possibility is to change the winding of the state associated to node $0$,
\begin{equation}\label{00'}
	Z_0 = \ket{0,0,-1,0;\alpha_0} \to Z_{0'} = \ket{0,0,0,-2;\alpha_0},
\end{equation}
and switch back $Z_{\texttt{C}_2'} \to Z_{\texttt{C}_2}$. This results in the diagram shown in Figure \ref{edd103}, and the corresponding maximal enhancements can be read from Table \ref{tab:CHLd1} ($i \neq 8$) by taking $E_{ij} = \text{diag}(2,1)$, $a = 0$, $a = a_2$, and replacing the rightmost $(\mra_n)_2$ factor by $(\mra_1)_2 + (\mrc_n)_1$. We see that the breaking of $\mre_8$ is now done by the second Wilson line $a_2$ and not $a_1$. These enhancements are complementary to those of the GDD in Figure \ref{edd101b}, as they are still limited to having a $\mrc_n$ factor (except for the trivial case with $a_i = 0$). In fact, to obtain other non-ADE factors such as $\mrf_4$ or $\mrb_n$ with $n> 2$ (recall that $\mrb_2 \simeq \mrc_2$), we must go to compactifications to dimensions lower than 8. 

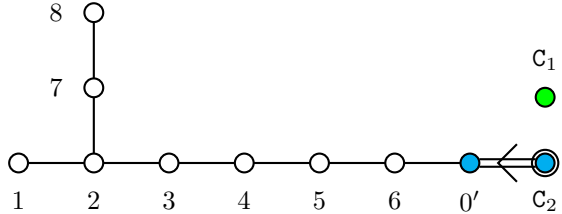
\begin{figure}[htb]
	\begin{center}
		\begin{tikzpicture}[scale=.25]
		\draw[thick] (0 cm,0) circle (5 mm) node [shift={(0.0,-0.5)}] {\mfn{1}} ;
		\draw[thick] (.5 cm, 0) -- (3.5 cm,0);    
		\draw[thick] (4 cm,0) circle (5 mm) node [shift={(0.0,-0.5)}] {\mfn{2}};
		\draw[thick] (4.5 cm, 0) -- (7.5 cm,0); 
		\draw[thick] (8 cm,0) circle (5 mm) node [shift={(0.0,-0.5)}] {\mfn{3}};
		\draw[thick] (8.5 cm, 0) -- (11.5 cm,0);     
		\draw[thick] (12 cm,0) circle (5 mm) node [shift={(0.0,-0.5)}] {\mfn{4}};
		\draw[thick] (12.5 cm, 0) -- (15.5 cm,0);     
		\draw[thick] (16 cm,0) circle (5 mm) node [shift={(0.0,-0.5)}] {\mfn{5}};
		\draw[thick] (16.5 cm, 0) -- (19.5 cm,0);     
		\draw[thick] (20cm,0) circle (5 mm) node [shift={(0.0,-0.5)}] {\mfn{6}};
		\draw[thick] (4 cm, 0.5cm) -- (4 cm, 3.5cm);     
		\draw[thick] (4cm,4cm) circle (5 mm) node [shift={(-0.5,0.0)}] {\mfn{7}};
		\draw[thick] (4 cm, 4.5cm) -- (4 cm, 7.5cm);     
		\draw[thick] (4cm,8cm) circle (5 mm) node [shift={(-0.5,0.0)}] {\mfn{8}};
		\draw[thick] (20.5 cm, 0) -- (23.5 cm,0);     
		\draw[thick, fill = cyan] (24cm,0) circle (5 mm) node [shift={(0.0,-0.5)}] {\mfn{0'}};
		\draw[thick] (24.5 cm, 0.2cm) -- (27.5 cm,0.2cm);
		\draw[thick] (24.5 cm, -0.2cm) -- (27.5 cm,-0.2cm);
		\draw[thick] (26.5 cm,  1cm) -- (25.5cm, 0cm) --  (26.5 cm,  -1cm);				     
		\draw[thick, fill=cyan] (28cm,0) circle (5 mm) node [shift={(0.0,-0.5)}] {\footnotesize{$\texttt{C}_2$}};
		\draw[thick] (28cm,0) circle (7 mm);
		\draw[thick, fill=green] (28cm,3.5cm) circle (5 mm) node [shift={(0.0,0.5)}] {\footnotesize{$\texttt{C}_1$}};
		\end{tikzpicture}
		\caption{Generalized Dynkin diagram for $d = 2$ theories, obtained by replacing the node $0$ by $0'$ as shown in \eqref{00'}.}\label{edd103}
	\end{center}
\end{figure}

To obtain $\mrf_4$ and $\mrb_3$, for example, simply go to 7 dimensions and add one node $\texttt{C}_3$ to the diagram in Figure \ref{edd101b}, such that it has one link with $\texttt{C}_2$. Deleting node $6$ then yields $(\mre_7)_2 + (\mrf_4)_1$, while deleting node $0$ yields $(\mre_8)_2 + (\mrb_3)_1$. Note that the algebra $\mrg_2$ is absent in the theory, since there are no massless states with $Z^2 = 6$ regardless of the number of compactified dimensions.

It is possible to construct many other GDDs, as we have done for $T^2$ heterotic compactifications in \cite{Font:2020rsk}. However, it is not guaranteed that doing so will produce  all possible enhancements. Indeed, we lack a complete understanding of the significance of these diagrams yet. To get a more exhaustive list of enhancements we turn to the so-called exploration algorithm, which we present in the next section. 

\section{Exploring the moduli space}
\label{sec:algorithms}

In a previous work \cite{Font:2020rsk} we developed an algorithm for $T^d$ compactifications which, starting from a point $p_0$ of the moduli space corresponding to a (semisimple) gauge group of maximal rank $r_{max} = d+16$, gives a set of new points of maximal enhancement. Heuristically, it searches for maximal enhancement points which are connected to $p_0$ through some variety with generic gauge group of rank $r_{max} - 1$. In the case of $S^1$ and $T^2$ compactifications, this algorithm was proven to be exhaustive by comparing with previous results \cite{Cachazo:2000ey,SZ}.

For the present investigation we have modified this algorithm in order to apply it to the CHL compactifications. This is required by the technicalities of working with Mikhailov lattices as opposed to Narain lattices, specially for compactifications to spacetime dimensions lower than nine, where non-ADE root lattices appear. 

In the following section we explain the methods used in our algorithm and illustrate them with an explicit example. We then present the maximal enhancements generated by iterating this procedure, collecting the final results in table \ref{tab:CHLd2}  of section \ref{sec:results}.

\subsection{Exploration algorithm}
\label{ss:exploration}

The purpose of our algorithm is to take as input some point $p_0$ of maximal enhancement and return a list of other such points $p_k$ related to $p_0$ in some specific, controllable way. To this end, it is best to specify $p_0$ not by its moduli, but by its root lattice $L_0$ via some generating matrix (in general, by generating matrix we mean a matrix whose rows are a basis for some lattice) of simple roots embedded in the Mikhailov lattice. Both sets of data are equivalent as one can recover one from the other using equations \eqref{m03D} and \eqref{m04D}. However, the lattice $L_0$ is more amenable to discrete operations, which we now describe.

Consider the $(10-d)$-dimensional ($d \geq 1$) CHL string at a point $p_0$ in moduli space specified by a set of $d+8$ simple roots with quantum numbers  $\ell^i, n_i$ and  $\rho$. Substituting each one of them in \eqref{m03D} gives $d$ real constraints on the $d \times (d+8)$ moduli. It follows that deletion of some simple root $r_0$ defines a $d$-dimensional subvariety in moduli space which contains $p_0$. Generically, this subvariety contains many more maximal enhancement points $p_k$, each one corresponding to a distinct simple root $r_k$ replacing $r_0$, $r_k \neq r_0$. It is in this sense that  the $p_k$ are neighbors of $p_0$. To generate such a root $r_k$ we solve a system of equations stating that $r_k$ must have inner product $0, -1$ or $-2$ with all other roots, its squared length must be $2$ or $4$ and it must be embedded in the Mikhailov lattice $\rii_{(d)}$ in accordance with Table \ref{tab:massless2}.

In order to make sure that the root lattice obtained by replacing $r_0 \to r_k$ corresponds to the gauge group $G_k$ at $p_k$, we have to take care of an ambiguity in the relation between the moduli of $p_k$ and the root lattice $L_k \equiv L$ of $G_k$. Even though the embedding of $L$ in $\rii_{(d)}$ specifies the moduli via the constraints mentioned above, it is also true that any sublattice $L' \subseteq L$ with $\text{rank}(L') = \text{rank}(L)$ will give the same moduli. When we replace $r_0 \to r_k$ there is therefore the possibility that the lattice obtained will not be $L$ but some $L'$. This ambiguity is eliminated if we implement a procedure, which we explain below, that takes $L'$ and returns $L$ by adding the missing roots. This adding of roots will be referred to as a \textit{saturation} of $L'$ to $L$.

To saturate $L'$ we recall that all of its even overlattices are contained in the dual lattice ${L'}^*$, so that in particular $L' \subseteq L \subseteq {L'}^*$. It suffices then to compute the vectors dual to $L'$, select those which correspond to roots embedded in $\rii_{(d)}$ and add them to $L'$. In practice this is done by iterating an algorithm which replaces one root vector in the generating matrix for $L'$ such that $\det L'$ gets smaller (indicating that $L'$ has been extended) and is still embedded in  $\rii_{(d)}$. When all attempts to do this leave the determinant of the lattice invariant, $L'$ has been saturated to the true root lattice $L$ at $p_k$.

\subsection{Example}
\label{ss:example}
To illustrate this procedure we first  consider an exploration of the neighborhood of the point in moduli space corresponding to eight dimensional CHL with gauge algebra $(\mra_1 + \mra_3 + \mrd_6)_2$ given by the moduli
\begin{equation}
E_{ij} = 
\begin{pmatrix}
2 & 0 \\ 0 & 1
\end{pmatrix},
~~~~~
a_1 = (0^7,1), ~~~~~ a_2 = (0^3,-\tfrac{1}{2}^4,\tfrac{1}{2}).
\end{equation}
The root lattice $L_0$ is generated by the rows $(\ell^1,\ell^2,n_1,n_2; \rho)$ of the $10 \times 12$ matrix
\begin{equation}
\setcounter{MaxMatrixCols}{20}
\mathcal{G}_0 = 
\begin{footnotesize}\begin{bmatrix}
1 & 2 & -1 & -1 & 0 & 0 & 0 & 1 & 1 & 1 & 1 & -2\\
0 & 0 & 0 & 0 & 1 & -1 & 0 & 0 & 0 & 0 & 0 & 0\\
0 & 0 & 0 & 0 & 0 & 1 & -1 & 0 & 0 & 0 & 0 & 0\\
0 & 0 & 0 & 0 & -1 & -1 & 0 & 0 & 0 & 0 & 0 & 0\\
1 & 0 & 1 & 0 & 0 & 0 & 0 & 0 & 0 & 0 & 0 & 0\\
-1 & 0 & 0 & 0 & 0 & 0 & 0 & 0 & 0 & 0 & 1 & 1\\
0 & 0 & 0 & 0 & 0 & 0 & 0 & 0 & 0 & 1 & -1 & 0\\
0 & 0 & 0 & 0 & 0 & 0 & 0 & 0 & 1 & -1 & 0 & 0\\
0 & 0 & 0 & 1 & 0 & 0 & 0 & -1 & -1 & 0 & 0 & 0\\
1 & 0 & -1 & -1 & 0 & 0 & 0 & 0 & 0 & 0 & 0 & -2\\
\end{bmatrix}\end{footnotesize},
\end{equation}
from which the gauge algebra is read by computing its Cartan matrix $\mathcal{G}_0 \eta \mathcal{G}_0^T$, with $\eta$ given in \eqref{etaD}. Note also that $\mathcal G_0$ is not a square matrix due to the fact that it gives an embedding of a rank 10 lattice into the rank 12 lattice $\rii_{(2)}$. We have chosen this particular vacuum because, as we explain below, it neighbors another vacuum with globally non-trivial gauge group. To obtain it we have applied the algorithm described here to another vacuum which can be obtained from the GDD construction explained in section \ref{ss:gdd}. 

Starting from $\mathcal{G}_0$, one of the paths that our algorithm will follow is to remove, for example, the 6th row. This breaks $(\mrd_6)_2 \to (2 \mra_1 + \mra_3)_2$ and eliminates two real constraints on the moduli (cf. eq. \eqref{m03D}), which taking into account the remaining $20-2 = 18$ constraints read
\begin{equation}
E_{ij} = 
\begin{pmatrix}
2 & x \\
0 & y 
\end{pmatrix},
~~~~~
a_1 = (0_3, x, (-x)_3, 1), ~~~~~ a_2 = (0_3, y - \tfrac{3}{2}, (\tfrac{1}{2} - y)_3, \tfrac{1}{2}) \, ,
\end{equation}
with the subindex 3 meaning that the quantity is repeated 3 times. In other words, the moduli are now constrained to a plane $(x,y)$ with generic gauge algebra $(3 \mra_1 + 2 \mra_3)_2$. Our algorithm will now generate a new simple root $\alpha$ by picking out a solution to the set of equations
\begin{equation}
\begin{cases}
\mathcal{G}_{0,mn}\alpha_n = k_m, & k_m \in \{0, -1, -2\}, ~~~ m \neq 6\\
\alpha^2 = N,&  N \in \{2,4\},
\end{cases}
\end{equation}
where $\alpha = (\ell^i,n_i;\rho)$ is constrained to lie in $\rii_{(2)}$, meaning that $\ell^1, n_1, n_2 \in \mathbb{Z}$, $\ell^2 \in 2\mathbb{Z}$ and $\rho \in \Gamma_8$. One possible solution with $N=4$  is
\begin{equation}
\setcounter{MaxMatrixCols}{20}
\alpha = 
\begin{bmatrix}
0 & 2 & -2 & -3 & 0 & 0 & 0 & 0 & 2 & 2 & 2 & -2
\end{bmatrix}.
\end{equation}
The new matrix $\mathcal{G}_1$ resulting from this exchange of roots ($\alpha$ is now in the 6th row) is seen to generate the root lattice $L_1$ corresponding to the gauge algebra $(2 \mra_1 + 2 \mra_3)_2 + (\mrc_2)_1$ and the moduli are fixed to
\begin{equation}\label{moduliex}
E_{ij} = 
\begin{pmatrix}
2 & 0 \\ 0 & \tfrac{5}{4}
\end{pmatrix}, ~~~~~
a_1 = (0_7,1), ~~~~~ a_2 = (0_3, -\tfrac{1}{4}, (-\tfrac{3}{4})_3, \tfrac{1}{2}).
\end{equation}

To check that $L_1$ contains all the solutions to equations $\eqref{m03D}$ and $\eqref{m04D}$, our algorithm calculates the generating matrix $\mathcal{G}^*_1$ for the dual lattice $L^*_1$:

\begin{equation}
\setcounter{MaxMatrixCols}{20}
\mathcal{G}^*_1 = 
\begin{footnotesize}\begin{bmatrix}
\tfrac{1}{2} & 1 & -\tfrac{1}{2} & -\tfrac{1}{2} & 0 & 0 & 0 & \tfrac{1}{2} & \tfrac{1}{2} & \tfrac{1}{2} & \tfrac{1}{2} & -1\\
0 & 0 & 0 & 0 & \tfrac{1}{2} & -\tfrac{1}{2} & -\tfrac{1}{2} & 0 & 0 & 0 & 0 & 0\\
0 & 0 & 0 & 0 & 0 & 0 & -1 & 0 & 0 & 0 & 0 & 0\\
0 & 0 & 0 & 0 & -\tfrac{1}{2} & -\tfrac{1}{2} & -\tfrac{1}{2} & 0 & 0 & 0 & 0 & 0\\
1 & 1 & 0 & -\tfrac{3}{2} & 0 & 0 & 0 & 0 & 1 & 1 & 1 & -1\\
\tfrac{1}{2} & 1 & -\tfrac{1}{2} & -\tfrac{3}{2} & 0 & 0 & 0 & 0 & 1 & 1 & 1 & -1\\
0 & 0 & 0 & \tfrac{1}{4} & 0 & 0 & 0 & -\tfrac{1}{4} & \tfrac{1}{4} & \tfrac{1}{4} & -\tfrac{3}{4} & 0\\
0 & 0 & 0 & \tfrac{1}{2} & 0 & 0 & 0 & -\tfrac{1}{2} & \tfrac{1}{2} & -\tfrac{1}{2} & -\tfrac{1}{2} & 0\\
0 & 0 & \tfrac{3}{4} & 0 & 0 & 0 & 0 & -\tfrac{3}{4} & -\tfrac{1}{4} & -\tfrac{1}{4} & -\tfrac{1}{4} & 0\\
\tfrac{1}{2} & 0 & -\tfrac{1}{2} & -\tfrac{1}{2} & 0 & 0 & 0 & 0 & 0 & 0 & 0 & -1\\
\end{bmatrix}\end{footnotesize}.
\end{equation}
It then constructs generic integer linear combinations of the rows corresponding to roots lying in $\rii_{(2)}$ and adds them to $L_1$ by replacing one of the rows of $\mathcal{G}_1$. This is done in an exhaustive way, but in this particular case no such replacement decreases the determinant of $\mathcal{G}_1$, hence $L_1$ is saturated. This means that the gauge algebra at this point in moduli space is indeed  $(2 \mra_1 + 2 \mra_3)_2 + (\mrc_2)_1$. 

\subsection{Matter states and global data}

There are two other sets of data of importance that can be obtained by our methods, namely the matter states in the lowest massive level associated to fundamental representations of the gauge group $G$, and the global structure of $G$, i.e. the fundamental group $\pi_1(G)$. Both of these problems involve finding overlattices of root lattices which are primitively embedded in the momentum lattice $\rii_{(d)}$ or its dual $\rii_{(d)}^*$, as we now explain.

\subsubsection{Computing the overlattice}

\label{ss:overlattice}
By primitively embedded overlattice we mean the intersection of the real span of the root lattice, $L \otimes \mathbb{R}$, and the momentum lattice $\rii_{(d)}$ in the ambient space $\mathbb{R}^{d+8,d}$. In terms of the momenta $p_{L,R}$ this means all vectors which satisfy the constraint $p_R = 0$ but $p_L$ is unconstrained. Generally such an overlattice $M$ corresponds to an extension of $L$ by a set of fundamental weights $\{\mu,\mu',...\}$, and the quotient $M/L$ can be put in correspondence with a subgroup $K$ of the center of the universal cover $\tilde G$ of $G$, denoted $Z(\tilde G)$ (cf. Table \ref{tab:Z}). It follows that the overlattice data can be encoded in the generators $\{k,k',...\}$ of $K$. 

\begin{table}[H] 
	\begin{center}
		\footnotesize
		\def\arraystretch{1}
		\begin{tabular}
			{|@{\hskip 0.1cm}>{$}c<{$}@{\hskip 0.1cm}|@{\hskip 0.1cm}>{$}c<{$}@{\hskip 0.1cm}|}
			\hline
			\tilde G &  Z(\tilde G)   \\ \hline
			\sug(n+1)   & \mathbb{Z}_{n+1}\\ \hline
			\sping(2n+1),\,\spg(2n),\,\mre_7 &\mathbb{Z}_2   \\ \hline
			\mre_6 &\mathbb{Z}_3  \\ \hline
			\sping(4n+2) & \mathbb{Z}_4 \\ \hline
			\sping(4n)     & \mathbb{Z}_2 \times \mathbb{Z}_2 \\ \hline
			\mre_8,\,\mrf_4,\, \mrg_2  &\mathbb{1}     \\ \hline
		\end{tabular}
		\caption{Center $Z(\tilde G)$ of compact connected simple groups $\tilde G$.}
		\label{tab:Z}
			\end{center}
\end{table}

Computing the weight vectors $\mu_i$ can be done by a slight generalization of the saturation algorithm described at the end of section \ref{ss:exploration}. Indeed, what it basically does is a computation of an overlattice of $L$ which is also a root lattice. By relaxing this last constraint, the same algorithm can be used to compute $M$. Returning to the example of section \ref{ss:example}, we apply this algorithm and find that $L$ can be extended to an overlattice $M$ in $\rii_{(2)}$ by adding the weight vector
\begin{equation}\label{weightex}
\mu = \ket{2,2,-1,-2;0,0,-1,0,2,1,1,-3} \, .
\end{equation}
In other words, the vector $\mu$ satisfies $p_R = 0$ (cf. eq. \eqref{m03D}) with the moduli given in \eqref{moduliex}, but is not in $L$. Determining the precise $K \subset Z(\tilde G)$ now amounts to determining the element in $Z(\tilde G)$ to which $\mu$ corresponds. To do this we recall that
\begin{equation}
Z(\tilde G) = \Lambda_\text{weight}/\Lambda_\text{root} \,
\end{equation}
where $\Lambda_\text{weight}$ is the weight lattice, which in particular contains $M$, and $\Lambda_\text{root} = L$. The weight $\mu$ together with all its $L$-translations constitutes an equivalence class $[\mu] \in Z(\tilde G)$. 

In general, for $\tilde G$ a semisimple group with $s$ simple factors, $Z(\tilde G)$ is a product of $s+t$ cyclic groups,
\begin{equation}
Z(\tilde G) = \mathbb{Z}_{p_1} \times \cdots \times \mathbb{Z}_{p_{s+t}}\, ,
\end{equation}
where $t$ is the number of $\mrd_{2n}$ factors since they contribute each a $\mathbb{Z}_2 \times \mathbb{Z}_2$ group (see Table \ref{tab:Z}). Any element of $Z(\tilde G)$ can therefore be written as a tuple
\begin{equation}
k = (k_1,...,k_{s+t}) \, ,
\end{equation}
where $k_i \sim k_i + p_i$, and the ordering of the $k_i$'s is appropriately specified in each case. In our example, we have
\beq \label{tildeG}
\tilde G = \sug(2)^2 \times \sug(4)^2 \times \spg(2) , \quad \quad Z(\tilde G) = \mathbb{Z}_2^2 \times \mathbb{Z}_4^2 \times \mathbb{Z}_2 \, ,
\eeq
and each central element is of the form
\begin{equation}
k = (k_1,k_2,k_3,k_4,k_5) \mod (2,2,4,4,2) \, .
\end{equation}
To determine which equivalence class $k$ contains the weight vector $\mu$, we first note that each possible $k$ can be put in correspondence with a combination of fundamental weights of $\tilde G$. If for example one looks at the fundamental weights $w_i$ of $\sug(n)$,  one finds that $[w_i] = i \in \mathbb{Z}_n$ (up to the outer automorphism of $\sug(n)$ which maps $i \to -i \mod n$). For $\spg(2)$, the only non trivial element of the center contains the weight corresponding to the short simple root (or equivalently the spinor class in $\sping(5) = \spg(2)$). Using these facts one finds that the $\mu$ given in \eqref{weightex} is contained in  
\begin{equation}
k = (1,1,2,2,1) \, .
\end{equation}
To verify this, one can compute the fundamental weights (labeled by $i$) $w_{j,i}$ of each simple factor (labeled by $j$) and check that the vector 
\begin{equation}
w_{1,1} + w_{2,1} + w_{3,2} + w_{4,2} + w_{5,1}
\end{equation}
can be translated by roots in $L$ to the given $\mu$. Keep in mind that these calculations are performed with respect to the particular embedding of $L$ and $M$ in $\rii_{(2)}$. 

%

Having determined the explicit form of $k = [\mu] \in Z(\tilde G)$, we immediately find that $K = \mathbb{Z}_2$, since $2k = (2,2,4,4,2) = (0,0,0,0,0)$, i.e. $k$ is an order 2 element. Moreover, it is uniquely in correspondence with the fundamental representation $(\mathbf{2,2,6,6,4})$ of $\tilde G$. Indeed, one can explicitly find all the states which form this representation with mass $M^2 = 4$. It suffices to construct such a state from the weight vector \eqref{weightex} and act on it with the Weyl group of the enhanced gauge group, which is a subset of the subgroup of T-dualities that leave the moduli invariant. In this way all the states forming the corresponding representation of $\tilde G$ are obtained.  

\subsubsection{Computing the fundamental group}
\label{ss:fund}
As explained in \cite{Cvetic:2021sjm} (see also \cite{Mikhailov:1998si}) the fundamental group of $G$ can be computed as the quotient $M^\vee / L^\vee$, where $L^\vee$ and $M^\vee$ are respectively the coroot lattice and the cocharacter lattice of $G$ \footnote{The computation of the fundamental group has been corrected from v1 of this paper.}. For every $G$, $L^\vee$ is embedded in the dual Mikhailov lattice $\rii_{(d)}^*(2)$, where the $(2)$ means that it is also rescaled by a factor of $\sqrt{2}$ to make it even, and $M^\vee$ corresponds to its overlattice primitively embedded in $\rii_{(d)}^*(2)$. In practice this means that to compute the fundamental groups we need to find embeddings of the lattices $L^\vee$ in the dual Mikhailov lattice and then apply the procedure explained before to get the respective $M^\vee$. 
	
Even though the exploration algorithm was designed to find points of maximal symmetry enhancement in moduli space, it can be considered on its own as an algorithm for finding embeddings of lattices into other lattices. For this reason it can be used also to compute all possible root lattices in $\rii_{(d)}^*(2)$. This is due to the fact that the data that we manipulate through this algorithm corresponds to the lattice vectors themselves and not the moduli or the momenta. A point that has to be made clear however is that the condition for a vector in the lattice to be a root is that it is of norm 2, or that it is of norm 4 and furthermore has even inner product with all other vectors in the lattice. This is the statement which generalizes the conditions for massless states shown in Table \ref{tab:massless2} to any basis for the momentum lattice that we choose. It applies both to $\rii_{(d)}$ and $\rii_{(d)}^*(2)$.

In eight dimensions, for example, we have
\begin{equation}
	\rii_{(2)} = \rii_{1,1}(2)\oplus \rii_{1,1} \oplus \mre_8  ~~\Rightarrow~~ \rii_{(2)}^*(2) = \rii_{1,1} \oplus \rii_{1,1}(2) \oplus \mre_8 (2)\,.
\end{equation}
We can take as a starting point for the exploration the root lattice of, say, $\mrb_{10}$, which can be constructed by hand and is expected to embed into $\rii_{(2)}^*(2)$ since it is the coroot lattice of $\mrc_{10}$ which embeds into $\rii_{(2)}$. After a few steps, the algorithm produces a list of root lattices which correspond exactly to the coroot lattices of the gauge algebras found by exploring the original lattice $\rii_{(2)}$. In particular, we find the root 	lattice 
\begin{equation}\label{rootsex}
	L = 2 \mra_1 (2) \oplus  2 \mra_3 (2) \oplus \mrb_2,
\end{equation}
which corresponds to the coroot lattice $L^\vee$ of the model used in the examples of Sections \ref{ss:example} and \ref{ss:overlattice}. One may apply exactly the same procedure of the last section to compute its overlattice and the subgroup
of $Z(\tilde G^\vee)$ to which it corresponds, where $\tilde G^\vee$ is the simply connected gauge group with root lattice in \eqref{rootsex}. Since this subgroup coincides with $M^\vee / L^\vee$, its generators $k_i$ give precisely the fundamental group $\pi_1 (G) \subset Z(\tilde G) \simeq Z(\tilde G^\vee)$, which we refer to as $H$, i.e. $G = \tilde G / H$. In this case, we find two generators 
\begin{equation}
	k = (0,1,0,2,1)\,, ~~~~~ k' = (1,0,2,0,1)\,
\end{equation}
of order 2, so that $H = \mathbb{Z}_2 \times \mathbb{Z}_2$, and the gauge group is 
\begin{equation}
	G = \frac{\sug(2)^2 \times \sug(4)^2 \times \sping(5)}{\mathbb{Z}_2 \times \mathbb{Z}_2}\,.
\end{equation}
This result is in agreement with that of \cite{Cvetic:2021sjm}.

\subsubsection{Anomaly for center symmetries}

It has been shown in \cite{Cvetic:2020kuw} that in order for  an 8d $\mathcal{N} = 1$ supergravity theory with global gauge group $G = \tilde G / H$ to be consistent, the following condition must be satisfied:
\begin{equation}\label{consistency8d}
\sum_{i = 1}^s \alpha_{\tilde G_i}  m_i  k_i^2 = 0 \mod 1\, ,
\end{equation}
where $\tilde G_i$ are the $s$ simple factors in $\tilde G$, $\alpha_{\tilde G_i}$ are the conformal dimensions of the Kac-Moody representations which generate the center  \cite{Cordova:2019uob}, 
$m_i$ are free parameters in the supergravity theory and $k = (k_1,...,k_s)$ is the generator of $H \in Z(\tilde G)$. This condition ensures that the $H$ center symmetry is free of anomalies. In the string theory whose low energy limit corresponds to this supergravity theory, $m_i$ are the levels of the world-sheet current algebra of  $\tilde G_i$. It can be shown in general that \eqref{consistency8d} is satisfied by construction for all $G = \tilde G / H$ obtained from the heterotic string on $T^2$ and the 8d CHL string \cite{Cvetic:2021sjm}. Here we give a brief alternative proof for this fact in the $T^2$ case, and comment briefly on the CHL case.

The fact that the gauge groups that arise from the Narain lattice $\rii_{2,18}$ satisfy \eqref{consistency8d} by construction is relatively easy to see. For this we recall that the conformal dimension $\alpha_{\tilde G_i}$ can be written as
\begin{equation}
	\alpha_{\tilde G_i} = \frac{w_i^2}{\alpha_\ell^2}\,,
\end{equation}
where $w_i$ is the fundamental weight that generates the center of the group $\tilde G_i$ and $\alpha_\ell$ is the highest root, which is a long root. In this case, all possible gauge groups are of ADE type, so that $\alpha_\ell^2 = 2$, and have $m_i = 1$. We can therefore rewrite \eqref{consistency8d} as
\begin{equation}
	\sum_{i = 1}^{s} (w_i k_i)^2 = 0 \mod 2\,,
\end{equation}
which is the statement that the weight vector $\sum_{i = 1}^sw_i k_i$ is even. For ADE groups, the root and coroot lattices are the same, and since the Narain lattice is also self-dual, the global structure is given by the overlattice $M$ which embeds primitively into $\rii_{2,18}$ and is given by precisely this weight vector (cf. Sections \ref{ss:overlattice} and \ref{ss:fund}). It is of course possible that there is more than one weight vectors involved, in which case the situation is analogous. Since the Narian lattice is even, all overlattices $M$ must be also even and so the condition \eqref{consistency8d} is satisfied by construction. 

For the CHL string the situation is more subtle since the Mikhailov lattice is not self dual and there are symplectic groups. One can understand why groups occuring in this case should satisfy \eqref{consistency8d} by noting that all of them can be constructed from groups arising from the Narain lattice by a suitable projection \cite{Cvetic:2021sjm}, and so they must also preserve condition \eqref{consistency8d}. It is straightforward to verify the that this is the case given the $H$ generators displayed in Table \ref{tab:CHLd2subgroups}.

\subsubsection{Globally non-trivial groups of lower rank}
\label{ss:breaking}

So far we have discussed maximally enhanced gauge groups. For non-Abelian groups of lower rank there are of course many more possibilities. In particular, the list of all possible gauge groups arising in $T^2$ compactifications of the heterotic string is 5366, of which only 336 are of maximal rank; this was determined by Shimada in \cite{Shimada2000OnEK} from the point of view of elliptic K3 surfaces, and in principle applies to the heterotic string on $T^2$ in light of its duality with F-theory on K3. 

An important fact that was noticed in \cite{Shimada2000OnEK} is that all possible gauge groups of rank lower than 18 (the maximal rank in $T^2$ compactifications) which are simply-connected can be obtained from those of rank 18 which are also simply-connected by deleting nodes in the corresponding Dynkin diagram (e.g. $\mra_{m+n+1} \to \mra_m + \mra_{n}$). For groups with non-trivial fundamental group $H$, this is not necessarily true. For example, the gauge group $\sping(8)^4 / (\mathbb{Z}_2 \times \mathbb{Z}_2)$ cannot be enhanced to a higher rank group, so that, conversely, it cannot be found by deleting a node  as just described. We note that Shimada has given a set of rules for obtaining such gauge groups (see theorems 2.4-2.7 of \cite{Shimada2000OnEK}), but they do not correspond to arbitrary node deletion and are rather involved.

Here we will not attempt to repeat this analysis for the CHL string, but instead ask the following question: what gauge groups with non-trivial $H$ can be obtained by breaking maximally enhanced groups via node deletion? Given that all maximal enhancements in 9d have trivial $H$ (cf. Table \ref{tab:CHLd1}), we will restrict ourselves to the 8d theory. In this case, there are 29 such groups, 24 with $H = \mathbb{Z}_2$ and 5 with $H = \mathbb{Z}_2 \times \mathbb{Z}_2$ (cf. Table \ref{tab:CHLd2}).We record them with their corresponding $k$'s in Table \ref{tab:CHLd2subgroups}.

It is easiest to find the answer to our question by brute force. Just delete one of the simple roots in the embedding of the rank 10 root lattice $L$ into the Mikhailov lattice $\rii_{(2)}$ and check if the resulting rank 9 lattice $L' \subset L$ still has a nontrivial weight overlattice $W' \subset W$. This will give rank 9 semisimple gauge groups with $H = \mathbb{Z}_2$ or $\mathbb{Z}_2\times\mathbb{Z}_2$ (as there are no other possibilities). Repeating the same procedure gives groups of rank 8 with the same $H$, and so on.

There is only one non-simply-connected gauge group of rank 4, namely $SU(2)^4/\mathbb{Z}_2$, and there are none for rank $\leq 3$. On the other hand, all of the 29 rank 10 groups can be broken to the rank 4 one. Analogously, $SU(2)^7/(\mathbb{Z}_2\times\mathbb{Z}_2)$ is the only one gauge group of rank $7$ with $H=\mathbb{Z}_2\times\mathbb{Z}_2$. There are no groups with that $H$ for rank $\leq 6$ and all of the five rank 10 groups with that fundamental group can be broken to the rank 7 one. In Figure \ref{grafo} we present a graph which encodes the breaking patterns that preserve the $\mathbb{Z_2}\times\mathbb{Z_2}$. Graphs of this type were studied in \cite{Chabrol:2019bvn} at the level of the algebra for the heterotic string on $T^2$ from the point of view of F-theory.


\begin{table}[H] 
	\begin{scriptsize}
		\begin{center}
			\def\arraystretch{1.0}
			\begin{tabular}
				{!{\vrule width 1.2pt}@{\hskip 0.1cm}>{$}c<{$}@{\hskip 0.1cm}|@{\hskip 0.2cm}>{$}l<{$}@{\hskip 0.05cm}|@{\hskip 0.1cm}>{$}c<{$}@{\hskip 0.1cm}|@{\hskip 0.0cm}>{$}c<{$}@{\hskip 0.0cm}!{\vrule width 1.2pt}}
				\hline
			\# & L & H & k   \\
				\hline
1& 2 \text{A}_2+2 \text{A}_3 & \mathbb{Z}_2 & 
\begin{array}{cccc}
  0  &  0  &  2  &  2  \\
\end{array}
 \\
\hline 2& 2 \text{A}_5 & \mathbb{Z}_2 & 
\begin{array}{cc}
  3  &  3  \\
\end{array}
 \\
\hline 3& 2 \text{A}_1+\text{A}_3+\text{A}_5 & \mathbb{Z}_2 \times \mathbb{Z}_2 & 
\begin{array}{cccc}
  0  &  1  &  0  &  3  \\
  1  &  0  &  2  &  3  \\
\end{array}
 \\
\hline 4& \text{A}_1+\text{A}_4+\text{A}_5 & \mathbb{Z}_2 & 
\begin{array}{ccc}
  1  &  0  &  3  \\
\end{array}
 \\
\hline 6&\text{A}_1+\text{A}_2+\text{A}_7 & \mathbb{Z}_2 & 
\begin{array}{ccc}
  0  &  0  &  4  \\
\end{array}
 \\
\hline 7& \text{A}_1+\text{A}_9 & \mathbb{Z}_2 & 
\begin{array}{cc}
  1  &  5  \\
\end{array}
 \\
\hline 9&\text{A}_1+\text{A}_3+\text{A}_5+\text{C}_1 & \mathbb{Z}_2 & 
\begin{array}{cccc}
  1  &  2  &  3  &  0  \\
\end{array}
 \\
\hline 12&2 \text{A}_1+\text{A}_7+\text{C}_1 & \mathbb{Z}_2 & 
\begin{array}{cccc}
  1  &  1  &  4  &  0  \\
\end{array}
 \\
\hline 15&2 \text{A}_1+2 \text{A}_3+\text{C}_2 & \mathbb{Z}_2 \times \mathbb{Z}_2 & 
\begin{array}{ccccc}
  0  &  1  &  0  &  2  &  1  \\
  1  &  0  &  2  &  0  &  1  \\
\end{array}
 \\
\hline 16&\text{A}_1+\text{A}_3+\text{A}_4+\text{C}_2 & \mathbb{Z}_2 & 
\begin{array}{cccc}
  1  &  2  &  0  &  1  \\
\end{array}
 \\
\hline 18&3 \text{A}_1+\text{A}_5+\text{C}_2 & \mathbb{Z}_2 \times \mathbb{Z}_2 & 
\begin{array}{ccccc}
  0  &  0  &  0  &  3  &  1  \\
  1  &  1  &  1  &  0  &  1  \\
\end{array}
 \\
\hline 19&\text{A}_3+\text{A}_5+\text{C}_2 & \mathbb{Z}_2 & 
\begin{array}{ccc}
  0  &  3  &  1  \\
\end{array}
 \\
\hline 26&2 \text{A}_1+2 \text{A}_2+\text{C}_4 & \mathbb{Z}_2 & 
\begin{array}{ccccc}
  1  &  1  &  0  &  0  &  1  \\
\end{array}
 \\
\hline 27&\text{A}_1+\text{A}_2+\text{A}_3+\text{C}_4 & \mathbb{Z}_2 & 
\begin{array}{cccc}
  0  &  0  &  2  &  1  \\
\end{array}
 \\
\hline
\end{tabular}
\hskip 0.8cm
\begin{tabular}
				{!{\vrule width 1.2pt}@{\hskip 0.1cm}>{$}c<{$}@{\hskip 0.1cm}|@{\hskip 0.2cm}>{$}l<{$}@{\hskip 0.05cm}|@{\hskip 0.1cm}>{$}c<{$}@{\hskip 0.1cm}|@{\hskip 0.0cm}>{$}c<{$}@{\hskip 0.0cm}!{\vrule width 1.2pt}}
				\hline
			\# & L & H & k   \\
			\hline 28&2 \text{A}_1+\text{A}_4+\text{C}_4 & \mathbb{Z}_2 & 
\begin{array}{cccc}
  1  &  1  &  0  &  1  \\
\end{array}
 \\
\hline 31&2 \text{A}_1+\text{A}_2+\text{C}_6 & \mathbb{Z}_2 & 
\begin{array}{cccc}
  0  &  1  &  0  &  1  \\
\end{array}
 \\
				\hline
33& \text{A}_1+\text{A}_3+\text{C}_6 & \mathbb{Z}_2 & 
\begin{array}{ccc}
  1  &  0  &  1  \\
\end{array}
 \\
\hline 36&2 \text{A}_1+\text{C}_8 & \mathbb{Z}_2 & 
\begin{array}{ccc}
  0  &  0  &  1  \\
\end{array}
 \\
\hline 37&\text{A}_2+\text{C}_8 & \mathbb{Z}_2 & 
\begin{array}{cc}
  0  &  1  \\
\end{array}
 \\
\hline 40&2 \text{D}_5 & \mathbb{Z}_2 & 
\begin{array}{cc}
  2  &  2  \\
\end{array}
 \\
\hline 42&\text{A}_1+\text{A}_2+\text{C}_2+\text{D}_5 & \mathbb{Z}_2 & 
\begin{array}{cccc}
  1  &  0  &  1  &  2  \\
\end{array}
 \\
\hline 43&\text{A}_1+\text{C}_4+\text{D}_5 & \mathbb{Z}_2 & 
\begin{array}{ccc}
  0  &  1  &  2  \\
\end{array}
 \\
\hline 45&\text{A}_1+\text{A}_3+\text{D}_6 & \mathbb{Z}_2 \times \mathbb{Z}_2 & 
\begin{array}{ccc}
  0  &  2  &  (1,1)  \\
  1  &  0  &  (0,1)  \\
\end{array}
 \\
\hline 46&\text{A}_2+\text{C}_2+\text{D}_6 & \mathbb{Z}_2 & 
\begin{array}{ccc}
  0  &  1  &  (1,0)  \\
\end{array}
 \\
\hline 47&\text{C}_4+\text{D}_6 & \mathbb{Z}_2 & 
\begin{array}{cc}
  1  &  (1,1)  \\
\end{array}
 \\
\hline 48&\text{A}_1+\text{C}_2+\text{D}_7 & \mathbb{Z}_2 & 
\begin{array}{ccc}
  1  &  1  &  2  \\
\end{array}
 \\
\hline 49&2 \text{A}_1+\text{D}_8 & \mathbb{Z}_2 \times \mathbb{Z}_2 & 
\begin{array}{ccc}
  0  &  0  &  (0,1)  \\
  1  &  1  &  (1,0)  \\
\end{array}
 \\
\hline 56&\text{A}_1+\text{A}_2+\text{E}_7 & \mathbb{Z}_2 & 
\begin{array}{ccc}
  1  &  0  &  1  \\
\end{array}
 \\
\hline 58&\text{A}_1+\text{C}_2+\text{E}_7 & \mathbb{Z}_2 & 
\begin{array}{ccc}
  0  &  1  &  1  \\
\end{array}
 \\ \hline
			\end{tabular}
			\caption{Maximal enhancement groups with non-trivial global structure for the  8-dimensional CHL string. The $k$'s are  the generator of $H$. All ADE groups arise at level $2$ while C groups arise at level $1$.}
			\label{tab:CHLd2subgroups}
		\end{center}
	\end{scriptsize}
\end{table}
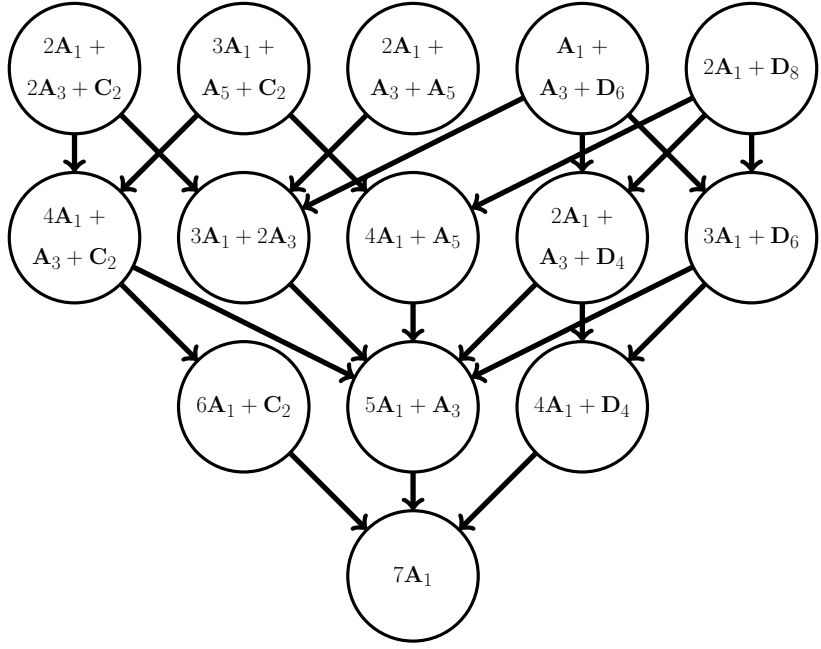
\begin{figure}[H]
	\begin{center}
		\begin{tikzpicture}[scale=2.25][ every annotation/.style = {draw, fill = white, font = \large}]
		\tikzset{concept/.append style={fill={none}}}
		\tikzset{every concept/.style={minimum size=12em, text width=12em}}
		\path[mindmap,concept color=black,text=black, every node/.style=
		{concept},
		root/.style    = {concept color=black, font=\huge\bfseries,text width=10em, scale=0.35},
		level 1 concept/.append style={font=\normalsize\bfseries,text width=3em,level distance=0em,inner sep=0pt},
		level 2 concept/.append style={font=\normalsize\bfseries,text width=3.3em,level distance=6em},
		level 3 concept/.append style={text=black,font=\normalsize\bfseries,text width=4em,level distance=6em},
		]
		node[root] at (2,0) (7A1){$7\text{A}_1$}
		node[root] at (1,1) (6A1+C2){$6\text{A}_1+\text{C}_2$}
		node[root] at (2,1) (5A1+A3){$5\text{A}_1 + \text{A}_3$}
		node[root] at (3,1) (4A1+D4){$4\text{A}_1 + \text{D}_4$}
		node[root] at (0,2) (4A1+A3+C2){$4\text{A}_1 + \text{A}_3 + \text{C}_2$}
		node[root] at (1,2) (3A1+2A3){$3\text{A}_1 + 2\text{A}_3$}
		node[root] at (2,2) (4A1+A5){$4\text{A}_1 + \text{A}_5$}
		node[root] at (3,2) (2A1+A3+D4){$2\text{A}_1 + \text{A}_3 + \text{D}_4$}
		node[root] at (4,2) (3A1+D6){$3\text{A}_1 + \text{D}_6$}
		node[root] at (0,3) (2A1+2A3+C2){$2\text{A}_1 + 2\text{A}_3 + \text{C}_2$}
		node[root] at (1,3) (3A1+A5+C2){$3\text{A}_1 + \text{A}_5 + \text{C}_2$}
		node[root] at (2,3) (2A1+A3+A5){$2\text{A}_1 + \text{A}_3 + \text{A}_5$}
		node[root] at (3,3) (A1+A3+D6){$\text{A}_1 + \text{A}_3+\text{D}_6$}
		node[root] at (4,3) (2A1+D8){$2\text{A}_1 + \text{D}_8$};
		\draw [<-, line width=2pt](7A1) to (6A1+C2);
		\draw [<-, line width=2pt](7A1) to (5A1+A3);
		\draw [<-, line width=2pt](7A1) to  (4A1+D4);
		\draw [<-, line width=2pt](6A1+C2) to 	(4A1+A3+C2);
		\draw [<-, line width=2pt](4A1+A3+C2) to (2A1+2A3+C2);
		\draw [<-, line width=2pt](4A1+A3+C2) to (3A1+A5+C2);
		\draw [<-, line width=2pt](5A1+A3) to 		(4A1+A3+C2);
		\draw [<-, line width=2pt](5A1+A3) to 		(3A1+2A3);
		\draw [<-, line width=2pt](5A1+A3) to 		(4A1+A5);
		\draw [<-, line width=2pt](5A1+A3) to 		(2A1+A3+D4);
		\draw [<-, line width=2pt](5A1+A3) to 		(3A1+D6);
		\draw [<-, line width=2pt](4A1+D4) to  	(2A1+A3+D4);
		\draw [<-, line width=2pt](4A1+D4) to  	(3A1+D6);
		\draw [<-, line width=2pt](3A1+2A3) to  	(2A1+2A3+C2);
		\draw [<-, line width=2pt](3A1+2A3) to  	(2A1+A3+A5);
		\draw [<-, line width=2pt](3A1+2A3) to  	(A1+A3+D6);
		\draw [<-, line width=2pt](4A1+A5) to  		(3A1+A5+C2);
		\draw [<-, line width=2pt](4A1+A5) to  		(2A1+D8);
		\draw [<-, line width=2pt](2A1+A3+D4) to  		(A1+A3+D6);
		\draw [<-, line width=2pt](2A1+A3+D4) to  		(2A1+D8);
		\draw [<-, line width=2pt](3A1+D6) to  		(A1+A3+D6);
		\draw [<-, line width=2pt](3A1+D6) to  		(2A1+D8);
		\end{tikzpicture}
		\caption{Scheme of how deleting nodes in the Dynkin diagrams of maximally enhanced groups with $H = \mathbb{Z}_2 \times \mathbb{Z}_2$ lead to gauge groups with lower rank and also with $H =  \mathbb{Z}_2 \times \mathbb{Z}_2$.} 
		\label{grafo}
	\end{center}
\end{figure}

\subsection{Results}
\label{sec:results}

We collect in Table \ref{tab:CHLd2} the 61 maximally enhanced groups $G = \tilde G/H$ that are realized in the eight-dimensional CHL string, and give the point in moduli space where they arise. ADE groups are realized at level 2 of the Kac-Moody algebra, as explained in Appendix \ref{App:CA}, while C groups arise at level 1. 

Our results for the algebras are  in  complete agreement of these results at the level with those obtained in \cite{Hamada:2021bbz} from F-theory, which appeared while the present paper was being written. 

There are 32 simply connected groups. The rest are of the form $\tilde G /H$ with $H=\mathbb{Z}_2$ or $\mathbb{Z}_2 \times \mathbb{Z}_2$. The fundamental group $H$ is generated in each case by the elements $k \in Z(\tilde G)$ shown in Table \ref{tab:CHLd2subgroups}. Our results are in perfect agreement with those in \cite{Cvetic:2021sjm}. 


Most of the groups shown lie in the subspace of moduli space given by $E_{ij} = \text{diag}(2,1)$, 
and it can actually be shown that the remaining ones can be mapped to this subspace by applying T-dualities. This is analogous to the situation  in the heterotic string on $T^2$ with $E_{ij} = \text{diag}(1,1)$ \cite{Font:2020rsk}. By performing the necessary T-dualities  to realise  the enhancement groups at such $E_{ij}$, however, the Wilson lines get much more complicated, and difficult to handle.

The central charge $c$ of the Kac-Moody algebras of the 9- and 8-dimensional models listed in Tables
\ref{tab:CHLd1} and  \ref{tab:CHLd2} can be easily calculated as explained in Appendix \ref{App:CA}.
A consistency check is that when the difference $(16+d- c)$ is less than one, it is always equal to the central 
charge of a unitary minimal model.

\newpage

\begin{center}
	\def\arraystretch{1.2}
	\begin{scriptsize}
		\begin{tabular}
			{
				!{\vrule width 1.2pt}@{\hskip 0.1cm}>{$}c<{$}@{\hskip 0.1cm}
				|>{$}l<{$}@{\hskip 0.05cm}
				|@{\hskip 0.05cm}>{$}c<{$}@{\hskip 0.05cm}
				!{\vrule width 1.2pt}@{\hskip 0.05cm}
				>{\begin{tiny}$}c<{$\end{tiny}}@{\hskip 0.01cm}
				>{\begin{tiny}$}c<{$\end{tiny}}@{\hskip 0.01cm}
				>{\begin{tiny}$}c<{$\end{tiny}}@{\hskip 0.01cm}
				>{\begin{tiny}$}c<{$\end{tiny}}@{\hskip 0.05cm}
				|@{}>{\begin{tiny}$}c<{$\end{tiny}}@{}|@{}>{\begin{tiny}$}c<{$\end{tiny}}@{}!{\vrule width 1.2pt}
			}\hline
			\# & L & H & E_{11} & E_{21} & E_{22} & E_{12} &  a_1 & a_2 \\
			\hline
			1 & 2 \text{A}_2+2 \text{A}_3 & \mathbb{Z}_2  & 2 &0& 3 & 1 & \frac{{{w}}_3}{4} & \frac{{{w}}_3}{4}-\frac{{{w}}_2}{3}    \\
			\hline
			2& 2 \text{A}_5 & \mathbb{Z}_2  & 2 &0&1& -1 & \frac{{{w}}_6}{2} & \frac{{{w}}_7}{3}-\frac{2 {{w}}_6}{3}    \\
			\hline
			3& 2 \text{A}_1+\text{A}_3+\text{A}_5 & \mathbb{Z}_2^2  & 2 &0& 2& 2 &0& \frac{{{w}}_7}{3}-\frac{{{w}}_6}{6}    \\
			\hline
			4& \text{A}_1+\text{A}_4+\text{A}_5 & \mathbb{Z}_2  & 2 &0& 2& 2 &0& \frac{2 {{w}}_7}{5}-\frac{{{w}}_5}{5}   \\
			\hline
			5& 2 \text{A}_2+\text{A}_6 & \mathbb{1}  & 2 &0&3& \frac{7}{2} & \frac{{{w}}_2}{6} &0\\
			\hline
			6& \text{A}_1+\text{A}_2+\text{A}_7 & \mathbb{Z}_2  & 2 &0& \frac{3}{2}& 2 &0& \frac{{{w}}_2}{6}    \\
			\hline
			7& \text{A}_1+\text{A}_9 & \mathbb{Z}_2  & 2 &0&3& -2 &0& \frac{{{w}}_3}{3}    \\
			\hline
			8& \text{A}_1+2 \text{A}_2+\text{A}_4+\text{C}_1 & \mathbb{1}  & 2 &0& 1&0& \frac{{{w}}_2}{6}   & \frac{{{w}}_2}{3}-\frac{2 {{w}}_5}{3}   \\
			\hline
			9& \text{A}_1+\text{A}_3+\text{A}_5+\text{C}_1 & \mathbb{Z}_2 & 2 &0& 1&0& \frac{{{w}}_8}{2}& \frac{5 {{w}}_8}{6}-\frac{{{w}}_3}{3}    \\
			\hline
			10& \text{A}_4+\text{A}_5+\text{C}_1 & \mathbb{1}  & 2 &0& 1&0& \frac{{{w}}_3}{5} &0\\
			\hline 
			11& \text{A}_1+\text{A}_2+\text{A}_6+\text{C}_1 & \mathbb{1}  & 2 &0& 1&0& \frac{{{w}}_2}{6}&0\\
			\hline
			12& 2 \text{A}_1+\text{A}_7+\text{C}_1 & \mathbb{Z}_2 & 2 &0& 1&0& \frac{{{w}}_8}{2}  & \frac{{{w}}_2}{3}-\frac{2 {{w}}_8}{3}    \\
			\hline
			13& \text{A}_1+\text{A}_8+\text{C}_1 & \mathbb{1}  & 2 &0&1&0& \frac{{{w}}_7}{4} &0\\
			\hline
			14& \text{A}_9+\text{C}_1 & \mathbb{1}  & 2 &0& 1&0& \frac{{{w}}_1}{3} &0\\
			\hline
			15& 2 \text{A}_1+2 \text{A}_3+\text{C}_2 & \mathbb{Z}_2^2 & 2 &0& 1&0& \frac{{{w}}_8}{2} & \frac{{{w}}_3}{4}-\frac{{{w}}_8}{4}   \\
			\hline
			16& \text{A}_1+\text{A}_3+\text{A}_4+\text{C}_2 & \mathbb{Z}_2  & 2 &0& 1&0& \frac{{{w}}_2}{5}& \frac{{{w}}_2}{2}-{{w}}_1    \\
			\hline
			17& 2 \text{A}_4+\text{C}_2 & \mathbb{1}  & 2 &0&1&0& {{w}}_4-\frac{2 {{w}}_3}{3}& \frac{{{w}}_3}{5}    \\
			\hline
			18& 3 \text{A}_1+\text{A}_5+\text{C}_2 & \mathbb{Z}_2^2 & 2 &0& 1&0& \frac{{{w}}_8}{2}& \frac{{{w}}_2}{4}-\frac{3 {{w}}_8}{8}    \\
			\hline
			19& \text{A}_3+\text{A}_5+\text{C}_2 & \mathbb{Z}_2  & 2 &0& 1&2 &0& \frac{{{w}}_3}{8}   \\
			\hline
			20& \text{A}_2+\text{A}_6+\text{C}_2 & \mathbb{1}  & 2 &0& 1&2 &0& \frac{{{w}}_7}{7}    \\
			\hline
			21& \text{A}_8+\text{C}_2 & \mathbb{1}  & 2 &0&1& -2 &0& \frac{{{w}}_3}{7}    \\
			\hline
			22& 2 \text{A}_2+\text{A}_3+\text{C}_3 & \mathbb{1}  & 2 &0& 1&0& \frac{{{w}}_3}{4}& \frac{{{w}}_3}{2}-\frac{{{w}}_2}{3}   \\
			\hline
			23& \text{A}_1+\text{A}_2+\text{A}_4+\text{C}_3 & \mathbb{1}  & 2 &0& 1&0& \frac{{{w}}_2}{10}  & -\frac{{{w}}_2}{6}    \\
			\hline
			24& \text{A}_2+\text{A}_5+\text{C}_3 & \mathbb{1}  & 2 &0& 1&0& \frac{{{w}}_6}{2}  & \frac{{{w}}_7}{3}-\frac{2 {{w}}_6}{3}    \\
			\hline
			25& \text{A}_1+\text{A}_6+\text{C}_3 & \mathbb{1}  & 2 &0& 1&0& \frac{{{w}}_2}{6}  & {{w}}_8-\frac{{{w}}_2}{3}    \\
			\hline
			26& 2 \text{A}_1+2 \text{A}_2+\text{C}_4 & \mathbb{Z}_2  & 2 &0& 1&0& \frac{{{w}}_4}{6}   & \frac{{{w}}_2}{6}    \\
			\hline
			27& \text{A}_1+\text{A}_2+\text{A}_3+\text{C}_4 & \mathbb{Z}_2  & 2 &0& 1&0& \frac{{{w}}_7}{2}-\frac{{{w}}_4}{3} & \frac{{{w}}_4}{4}    \\
			\hline
			28& 2 \text{A}_1+\text{A}_4+\text{C}_4 & \mathbb{Z}_2  & 2 &0& 1&0& \frac{{{w}}_2}{5}  & \frac{{{w}}_2}{10}    \\
			\hline
			29& \text{A}_1+\text{A}_4+\text{C}_5 & \mathbb{1}  & 2 &0& 1&0&0& \frac{{{w}}_3}{5}    \\
			\hline
			30& \text{A}_5+\text{C}_5 & \mathbb{1}  & 2 &0& 1&0& \frac{{{w}}_7}{3}-\frac{{{w}}_6}{6}   & \frac{{{w}}_6}{2}    \\
			\hline
			31& 2 \text{A}_1+\text{A}_2+\text{C}_6 & \mathbb{Z}_2  & 2 &0& 1&0&0& \frac{{{w}}_2}{6}    \\
			\hline
		\end{tabular}
		\hskip 0.8cm
		\begin{tabular}
			{
				!{\vrule width 1.2pt}@{\hskip 0.1cm}>{$}c<{$}@{\hskip 0.1cm}
				|>{$}l<{$}@{\hskip 0.05cm}
				|@{\hskip 0.05cm}>{$}c<{$}@{\hskip 0.05cm}
				!{\vrule width 1.2pt}@{\hskip 0.05cm}
				>{\begin{tiny}$}c<{$\end{tiny}}@{\hskip 0.01cm}
				>{\begin{tiny}$}c<{$\end{tiny}}@{\hskip 0.01cm}
				>{\begin{tiny}$}c<{$\end{tiny}}@{\hskip 0.01cm}
				>{\begin{tiny}$}c<{$\end{tiny}}@{\hskip 0.05cm}
				|@{}>{\begin{tiny}$}c<{$\end{tiny}}@{}|@{}>{\begin{tiny}$}c<{$\end{tiny}}@{}!{\vrule width 1.2pt}
			}\hline
			\# & L & H & E_{11} & E_{21} & E_{22} & E_{12} &  a_1 & a_2 \\
			\hline
			32& 2 \text{A}_2+\text{C}_6 & \mathbb{1}  & 2 &0& 1&0& \frac{{{w}}_2}{2}-{{w}}_1  & \frac{{{w}}_2}{6}    \\
			\hline
			33& \text{A}_1+\text{A}_3+\text{C}_6 & \mathbb{Z}_2  & 2 &0& 1&0& \frac{{{w}}_3}{4} & \frac{{{w}}_3}{8}    \\
			\hline
			34& \text{A}_4+\text{C}_6 & \mathbb{1}  & 2 &0& 1&0& \frac{{{w}}_6}{2}  & \frac{{{w}}_3}{5}    \\
			\hline
			35& \text{A}_1+\text{A}_2+\text{C}_7 & \mathbb{1}  & 2 &0&1& -2 & \frac{{{w}}_2}{6}  &0\\
			\hline
			36& 2 \text{A}_1+\text{C}_8 & \mathbb{Z}_2  & 2 &0& 1&0&0& \frac{{{w}}_7}{4}   \\
			\hline
			37& \text{A}_2+\text{C}_8 & \mathbb{Z}_2  & 2 &-1 & 1 &0&0& \frac{{{w}}_2}{6}    \\
			\hline
			38& \text{A}_1+\text{C}_9 & \mathbb{1}  & 2 &0& 1&0&0& \frac{{{w}}_1}{3}   \\
			\hline
			39& \text{C}_{10} & \mathbb{1}  & 2 &0& 1&-2 & \frac{{{w}}_1}{3}  &0\\
			\hline
			40& 2 \text{D}_5 & \mathbb{Z}_2  & 2 &0& 1&-1 &0& \frac{{{w}}_4}{4}    \\
			\hline
			41& \text{A}_4+\text{C}_1+\text{D}_5 & \mathbb{1}  & 2 &0& 1&0& \frac{{{w}}_4}{4} &0\\
			\hline
			42& \text{A}_1+\text{A}_2+\text{C}_2+\text{D}_5 & \mathbb{Z}_2  & 2 &0& 1&0& \frac{{{w}}_4}{6}  & \frac{{{w}}_4}{4}   \\
			\hline
			43& \text{A}_1+\text{C}_4+\text{D}_5 & \mathbb{Z}_2  & 2 &0& 1&0&0& \frac{{{w}}_4}{4}   \\
			\hline
			44& \text{C}_5+\text{D}_5 & \mathbb{1}  & 2 &0& 1& -2 & \frac{{{w}}_4}{4} &0\\
			\hline
			45& \text{A}_1+\text{A}_3+\text{D}_6 & \mathbb{Z}_2^2  & 2 &0& 2&2 &0& \frac{{{w}}_3}{4}  \\
			\hline
			46& \text{A}_2+\text{C}_2+\text{D}_6 & \mathbb{Z}_2  & 2 &0& 1&2 &0& \frac{{{w}}_4}{6}    \\
			\hline
			47& \text{C}_4+\text{D}_6 & \mathbb{Z}_2  & 2 &0& 1&0& \frac{{{w}}_8}{2} & \frac{{{w}}_6}{2}   \\
			\hline
			48& \text{A}_1+\text{C}_2+\text{D}_7 & \mathbb{Z}_2  & 2 &0& 1&0& \frac{{{w}}_8}{2} & \frac{{{w}}_8}{4}    \\
			\hline
			49& 2 \text{A}_1+\text{D}_8 & \mathbb{Z}_2^2  & 2 &0& 1&-1 & \frac{{{w}}_4}{2}-{{w}}_6& \frac{{{w}}_6}{2}    \\
			\hline
			50& \text{C}_1+\text{D}_9 & \mathbb{1}  & 2 &0& 1&0& \frac{{{w}}_8}{2}     &0\\
			\hline
			51& \text{A}_4+\text{E}_6 & \mathbb{1}  & 2 &0& 1& -1 &0& \frac{{{w}}_5}{3}   \\
			\hline
			52& \text{A}_1+\text{A}_2+\text{C}_1+\text{E}_6 & \mathbb{1}  & 2 &0& 1&0& \frac{{{w}}_5}{6}  & \frac{{{w}}_5}{3}    \\
			\hline
			53& \text{A}_3+\text{C}_1+\text{E}_6 & \mathbb{1}  & 2 &0& 1&0& \frac{{{w}}_5}{3}  &0\\
			\hline
			54& \text{A}_1+\text{C}_3+\text{E}_6 & \mathbb{1}  & 2 &0& 1&0&0& \frac{{{w}}_5}{3}   \\
			\hline
			55& \text{C}_4+\text{E}_6 & \mathbb{1}  & 2 &0& 1&-2 & \frac{{{w}}_5}{3} &0\\
			\hline
			56& \text{A}_1+\text{A}_2+\text{E}_7 & \mathbb{Z}_2  & 2 &0&1& -1 &0& \frac{{{w}}_6}{2}    \\
			\hline
			57& \text{A}_2+\text{C}_1+\text{E}_7 & \mathbb{1}  & 2 &0& 1&0& \frac{{{w}}_6}{2}  &0\\
			\hline
			58& \text{A}_1+\text{C}_2+\text{E}_7 & \mathbb{Z}_2  & 2 &0& 1&0&0& \frac{{{w}}_6}{2}    \\
			\hline
			59& \text{C}_3+\text{E}_7 & \mathbb{1}  & 2 &0& 1&0& \frac{{{w}}_6}{2}& \frac{{{w}}_6}{2}    \\
			\hline
			60& \text{A}_1+\text{C}_1+\text{E}_8 & \mathbb{1}  & 2 &0& 1&0&0&0\\
			\hline
			61& \text{C}_2+\text{E}_8 & \mathbb{1}  & 2 &0&1& -2 &0&0\\
			\hline
			\multicolumn{2}{c}{}\\
		\end{tabular}
		\begin{tiny}
			\begin{table}[H]
				\begin{center}
					\caption{All groups of maximal enhancement in the 8-dimensional CHL string. The Wilson lines are given in terms of the fundamental weights of $\mre_8$, see Table \ref{tab:e8}.
						ADE groups arise at level $2$ 
						and C groups at level $1$. }
					\label{tab:CHLd2}
				\end{center}
			\end{table}
		\end{tiny}
	\end{scriptsize}
\end{center}

\section{Conclusions}
\label{sec:conclusions}

In this work we have studied heterotic string compactifications that realize the CHL branch of 
superstring vacua with 16 supersymmetries in $(10-d)$ dimensions, $d \ge 1$. Such vacua, characterized by
left-moving gauge group of rank $d+8$, were first obtained in the context of type I strings \cite{Bianchi:1991eu}
and later derived in heterotic strings both in the fermionic \cite{Chaudhuri:1995fk} and bosonic formalism \cite{Chaudhuri:1995bf}.
We have followed the latter approach, based on compactification of the $\mre_8 \times \mre_8$ heterotic string on an
asymmetric orbifold $T^d/\ZZ_2$, which
enables a description at any point of the moduli space.
In particular, we have focused on the question of which non-Abelian groups of maximal rank can appear.
We have given a complete answer in $d=1,2$ in the form of a list of allowed groups and the corresponding moduli.
We believe that this list is exhaustive. This claim is supported by our
previous work on compactifications of the heterotic string on $T^d$ \cite{Fraiman:2018ebo,Font:2020rsk} where using the same 
algorithms in $d=1$ we found all groups dictated by the Generalised Dynkin diagram, and in $d=2$ we reproduced the results of \cite{SZ} based on the dual realisation in F-theory on elliptic K3 surfaces. More generally, non-exhaustivity of the exploration algorithm (for arbitrarily many iterations) would imply the existence of special points with maximal enhancement in moduli space which cannot be connected to others by moving along hypersurfaces as illustrated in Section \ref{ss:example}, which does not happen for $d = 1,2$. Moreover, we have verified that starting with different points of maximal enhancing we always obtain the same set of groups. Thus, the algorithm appears to be exhaustive for all $d$.  

Our analysis relies on the Mikhailov lattice $\rii_{(d)}$ underlying the $T^d/\ZZ_2$ asymmetric orbifold.
In analogy with the Narain lattice $\rii_{d,d+16}$ associated to heterotic compactification on $T^d$,
the momenta of all states in the orbifold spectrum lie in $\rii_{(d)}$ and symmetries of the spectrum
correspond to automorphisms of the lattice \cite{Mikhailov:1998si}.
For our purposes an essential fact is that the root lattice of the resulting non-Abelian groups must admit an embedding
in $\rii_{(d)}$, which is even but not self-dual for $d >1$. This last property leads to both simply-laced and non simply-laced groups realized at Kac-Moody level 2 and 1 respectively. The embedding condition gives a systematic prescription to determine
the groups that can arise or not. Moreover, studying embeddings of the coroot and cocharacter lattices in the dual Mikhailov lattice  allows to
determine the global structure of the gauge group \cite{Cvetic:2021sjm}. In this way we have proven that
for $d = 1$ the groups are simply-laced and simply-connected whereas for $d = 2$ there are also symplectic and doubly-connected groups. 

Our results for the global groups exactly match those obtained in \cite{Cvetic:2021sjm}, where they were shown to satisfy 
the condition for anomaly-free one-form center symmetries \cite{Cvetic:2020kuw}.
It would be interesting to check if these results are also 
consistent with constraints imposed by triviality of cobordism classes \cite{Montero:2020icj}. A partial check was carried out in \cite{Cvetic:2021sjm}. 

As mentioned above, the automorphisms of the Mikhailov lattice are T-dualities of the theory. 
As such they restrict the moduli space, and fixed points of discrete transformations are expected to display gauge symmetry 
enhancement. Indeed, we have shown that this is the case in $d=1$. A striking feature of the T-duality in $d=1$ is that
it mixes untwisted and twisted states. This can be demonstrated at the level of the partition function and it is expected
to occur for $d >1$.  It would be interesting to examine T-duality transformations and fixed points for $d>1$.

The methods developed to analyze lattice embeddings can also be applied to find out the non-Abelian groups that actually occur
in the CHL branch when $d\ge 3$. Already considering simple GDDs shows that factors $\mrf_4$ and $\mathrm{SO}(2r+1)$
can appear in agreement with earlier results \cite{Chaudhuri:1995bf, Mikhailov:1998si, deBoer:2001wca}. 
It is further known that for $d \ge 3$, superstring vacua with 16 supercharges exhibit a broader
pattern of rank reduction. In particular, in $d=3$ there are components with rank 7 or 5 for which an underlying
momentum lattice analogous to $\rii_{(d)}$ has been constructed \cite{deBoer:2001wca}. For these components it would be 
interesting to explore the associated moduli spaces, their duality symmetries, and the points of maximal enhancement.
One could also consider toroidal compactifications of the 7-dimensional theories with rank 7 or 5. 

The 8-dimensional CHL string is known to have a dual F-theory description in terms 
of compactification on a K3 surface with frozen singularities \cite{Witten:1997bs, Bhardwaj:2018jgp}.
The gauge groups arising in F-theory on such K3 surfaces were worked out very recently in \cite{Hamada:2021bbz}, and 
agree perfectly with the heterotic groups of maximal enhancing given in Table \ref{tab:CHLd2}, giving yet more support to the exhaustiveness of our algorithm.

To conclude, let us stress that perturbative heterotic compactifications with 16 supercharges are endowed with rich 
structures that allow a detailed exploration of their moduli spaces. We believe that in nine and eight dimensions the compactifications 
of the heterotic and CHL string provide the full landscape of gauge groups of half-maximal supergravities. In lower dimensions there 
remain many open questions that deserve further investigation.

\subsection*{Acknowledgements}
We are grateful to Lilian Chabrol, Silvia Georgescu, Ling Lin, Ruben Minasian, Cheng Peng, Stefan Theisen, and Cumrun Vafa  for interesting comments and valuable insights, and specially to Savdeep Sethi and Miguel Montero for stimulating our interest in this problem. 
A.~Font thanks B. Acharya, G. Aldazabal, K. Narain and I. Zadeh for discussions and collaboration in related subjects.
This work was partially supported
by the ERC Consolidator Grant 772408-Stringlandscape, PIP-CONICET- 11220150100559CO, UBACyT 2018-2021,  ANPCyT- PICT-2016-1358 (2017-2020) and NSF under Grant No. PHY-1748958. 
M.~Gra\~na would like to thank KITP, Santa Barbara, for 
hospitality.

\appendix

\section{Partition function}
\label{app:pf} 

In this appendix we study in some detail the compactification of the heterotic string in the asymmetric $S^1/\ZZ_2$ orbifold, 
where $\ZZ_2$ is the action defined in \eqref{RTCHL}.
We are mostly interested in writing down the partition function in order to derive properties of the resulting spectrum and 
to reproduce the results of \cite{Mikhailov:1998si}. An early discussion of the partition function was given in \cite{Bianchi:1997rf}.

\subsection{Notation and conventions}
\label{ss:prelim}

We will use the bosonic formulation in which there are 16 internal left-moving 
bosonic fields denoted $Y^{\hat I}$, $\hat I=1,\ldots, 16$. Modular invariance requires the 
$Y^{\hat I}$ to be compactified on an even self-dual lattice which in the $\mre_8 \times \mre_8$ heterotic string 
is $\Gamma_8 \oplus \Gamma_8$. As in the text, we will also use the combinations $Y_\pm$ 
in which the exchange of the two $\mre_8$ factors is diagonal, i.e. $Y^I_\pm \to \pm Y^I_\pm$.

To fix notation and conventions we recall the partition function of the 10-dimensional $\mre_8 \times \mre_8$ heterotic string, discussed for instance in \cite{Blumenhagen:2013fgp}.
Up to normalization, 
\be
\cz_{10} (\tau,\bar \tau) = \frac1{\left(\sqrt{\tau_2} \eta(\tau) \eta(\bar\tau)\right)^{8}} \ \cz_{\Gamma_8 \oplus \Gamma_8}(\tau)  \ \cz_\psi(\bar\tau)  \, .
\label{pf10}
\ee
The first factor is due to transverse worldsheet bosons in the lightcone. 
The $\cz_{\Gamma_8 \oplus \Gamma_8}(\tau)$ piece, arising from the 16 internal left-moving bosons, can be written as
\be
\cz_{\Gamma_8 \oplus \Gamma_8}(\tau)=
\frac1{\eta(\tau)^{16}}\, 
 \sum_{p\, \in \, \Gamma_8 \oplus \Gamma_8} q^{\frac12 p^2} 
\, ,
\label{zlatticedef}
\ee
where $q=e^{2\pi i \tau}$. The right-moving world sheet fermions give rise to $\cz_\psi(\bar\tau)$, which is given by
\begin{equation}
\cz_\psi(\bar\tau) = \frac1{2\op \eta(\bar\tau)^4}\left[\vartheta_3^4(\bar\tau) - \vartheta_4^4(\bar\tau) - \vartheta_2^4(\bar\tau) + 
\vartheta_1^4(\bar\tau) \right]\ .
\label{zpsidef}
\end{equation}
For the Jacobi $\vartheta$-functions we use the conventions of \cite{Blumenhagen:2013fgp}.
Recall that $\vartheta_1(\tau)=0$, but it is convenient to introduce it with a sign such as to explicitly have the same GSO projection in 
both NS and R sectors. By virtue of Jacobi's abtruse identity, $\cz_\psi(\bar\tau)$ vanishes identically. Hence, there are equal number of 
spacetime bosons and fermions at every mass level and the theory is supersymmetric. 

\subsection{Compactification on $S^1$}
\label{ss:s1}

The partition function of the heterotic string compactified on a $d$-dimensional torus is discussed
e.g. in \cite{Blumenhagen:2013fgp}. In the $S^1$ case it takes the form 
\begin{equation}
\cz_{S^1} (\tau,\bar \tau) =  \frac1{\left(\sqrt{\tau_2} \eta(\tau) \eta(\bar\tau)\right)^{7}} \ \cz_\psi (\bar\tau)  \  \cz_{\rii}(\tau, \bar\tau) \, .
\label{pfs1}
\end{equation}
The first term is due to the uncompactified worldsheet bosons. The $\cz_\psi(\bar\tau)$, 
coming from the worldsheet fermions, is again given by \eqref{zpsidef}. The last term 
$\cz_{\rii}(\tau, \bar\tau)$  is the contribution of the 16 internal left-moving bosons and the worldsheet boson compactified on $S^1$.
More precisely,
\begin{equation}
\cz_{\rii}(\tau, \bar\tau) = \frac1{\eta(\bar\tau) \eta(\tau)^{17}} 
\sum_{(p_R \op ; p_L, p) \op \in \op \nlc} \hspace*{-5mm}
\bar q^{\frac12 p_R^2} \,  q^{\frac12 p_L^2 + \frac12 p^2}  \ ,
\label{pflat}
\end{equation}
where $\nlc$ is the even, self-dual Narain lattice \cite{Narain:1985jj}. The moduli dependent momenta are given in \eqref{momenta}.

\subsection{Compactification on asymmetric $S^1/\ZZ_2$  orbifold}
\label{ss:s1z2}

The $\ZZ_2$ action consists of a translation by half a period along the circle, i.e. $x^9 \to x^9 + \pi R$, 
together with an exchange of the two $\mre_8$ factors in the lattice $\Gamma_8 \oplus \Gamma_8$,
which amounts to   $Y^I_\pm \to \pm Y^I_\pm$.
Since the $Y^I_\pm$ are purely left-moving, this orbifold can be more accurately described in the formalism of asymmetric 
orbifolds  \cite{Narain:1986qm}, as done in \cite{Chaudhuri:1995bf},  and in more generality in \cite{deBoer:2001wca}.

In an asymmetric orbifold the action of the orbifold generator, denoted $g$, is defined in terms of the
momenta in the Narain lattice, $\nlc$ in the case at hand. Specifically,
\be
\label{gaction}
g\ket{p_R; p_L, p}  = e^{2\pi i(p_L v_L + p\cdot V - p_R v_R)}  \ket{\theta_R \op p_R; \theta_L \op p_L, \theta \op p}
\ee
We are interested in the case with unbroken supersymmetry in which $\theta_L=\theta_R=\uno$. We consider a $\ZZ_2$
orbifold with $g^2 = \uno$. Thus, $\theta$ must be an automorphism of order two acting on $\Gamma_8 \oplus \Gamma_8$.
Without loss of generality we can assume that $\theta V=V$. It then follows that
the shift vector $v=(v_R ; v_L, V)$ satisfies $2 v \in \nlc$. 

We want the effect of the shift $v$ to correspond to the geometric translation of $x^9$ by half a period.
This means that 
\be
\label{shiftcond}
e^{2\pi i(p_L v_L + p\cdot V - p_R v_R)}= e^{i\pi n} \, ,
\ee
where $n$ is the quantized momentum along $S^1$. Using the momenta in \eqref{momenta} yields
\be
(v_R; v_L, V) =\frac12 \left(-\frac{R^2+\frac12 A^2}{\sqrt2\, R}; \frac{R^2 -\frac12 A^2}{\sqrt2\, R}, A \right) \, .
\label{vshift}
\ee
Observe that $v^2=0$.

Let us briefly examine the possibilities for the order two automorphism $\theta$ acting on the lattice $\Gamma_8 \oplus \Gamma_8$. 
In principle, $\theta$ might involve changing the sign of a number of lattice coordinates, but as we will shortly
show, modular invariance requires that the number of $-1$ eigenvalues of $\theta$ be a multiple of eight.
For example, we could take $Y^I \to - Y^I$, $I=1,\ldots,8$. However, this is an inner automorphism of the
$\mre_8$ lattice that can be realized by a shift in the lattice, as explained in the Appendix  of \cite{Dixon:1986jc}.
In other words, such inner automorphisms can be realized by setting $\theta=\uno$ and including an
additional phase $e^{2\pi i p\cdot \delta}$, with $2\delta \in \Gamma_8 \oplus \Gamma_8$, in the right-hand side of \eqref{shiftcond}.
Now, such action could not reduce the rank of the gauge group because all Cartan generators  would be invariant.
Luckily, for the $\mre_8 \times \mre_8$ heterotic theory there is the option of taking $\theta$ to be the exchange of
the two $\mre_8$ factors which is actually an outer automorphism of the lattice $\Gamma_8 \oplus \Gamma_8$.
The corresponding $\theta$ precisely has eight $-1$ eigenvalues so that it is allowed by modular invariance.
Moreover, the rank of the gauge group will be reduced by eight because the action of this $\theta$ will eliminate
eight Cartan generators. 
The upshot is that the appropriate automorphism is the exchange of the two $\mre_8$ factors
and we make this choice in the following. This means that $\theta$ is taken to be ${\rm R}$ defined in \eqref{RTCHL}.

We now proceed as in section \ref{sec:review}, writing $\Pi \in \Gamma_8 \oplus \Gamma_8$ as $\Pi=(\pi, \pi')$, with 
$\pi, \pi' \in \Gamma_8$.
The Wilson line $A$ is split in a similar way. In fact, the condition $\theta V=V$ requires $A=(a,a)$.
We then switch to a basis for $\nlc$ with momenta $(p_R; p_L, p_+, p_-)$ written in \eqref{momentaCHL}. 
The shift $v$ in \eqref{vshift} has components $(v_R;v_L, v_+, v_-)$ given in \eqref{v}.
In this basis the $\ZZ_2$ generator $g$ acts as
\be
\label{gaction2}
g\ket{p_R; p_L, p_+, p_-}  = e^{i \pi n}  \ket{p_R; p_L, p_+, - p_-} \, ,
\ee 
where we substituted \eqref{shiftcond}. Besides, $g$ acts on the left-moving coordinates as $g Y^I_{\pm} = \pm Y^I_{\pm}$.

We also need to describe the sublattices $I$, $I^*$ and $\tilde I$ of $\nlc$ which will enter in the partition function of the 
asymmetric $S^1/\ZZ_2$ orbifold.
The invariant lattice, denoted $I$, is the sublattice of $\nlc$ left invariant by $\theta$ \cite{Narain:1986qm}.
Since in our case $\theta$ is the exchange of the two $\mre_8$ factors, this sublattice corresponds to $\pi=\pi'$, i.e to $p_-=0$, whereas $p_R$, $p_L$ and $p_+$ reduce to
\begin{subequations}
	\label{momI}
	\begin{align}
	p_{R}&= 
	\frac1{\sqrt2 R}\left[n- (R^2 + \tfrac12 \hat a^2)m- \hat \pi \cdot \hat a\right], \label{pRin} \\
	p_{L}&=
	\frac1{\sqrt2 R}\left[n+ (R^2 - \tfrac12 \hat a^2)m- \hat\pi \cdot \hat a \right], \label{pLin} \\[2mm]
	p_+ &= \sqrt2 \left( \pi + a m \right)=\hat\pi + m \hat a,  \label{p+in}
\end{align}
\end{subequations}
where we defined $\hat\pi= \sqrt2 \pi$ and $\hat a= \sqrt2 a$. Notice that $\hat \pi \in \sqrt2 \Gamma_8$.
Rewriting the momenta in terms of the hat quantities is intended to highlight the structure of $I$.
In fact, comparing for instance \eqref{momI} with \eqref{momenta} we see that the invariant lattice is given by
\be
I \simeq \rii_{1,1} \oplus \Gamma_8(2) \, ,
\label{invlat}
\ee
where $\rii_{1,1}$ is the even self-dual lattice of signature $(1,1)$ and $\Gamma_8(2)$ is the lattice whose
Gram matrix is the Gram matrix of $\mre_8$ multiplied by 2 (i.e. the basis is multiplied by $\sqrt2$).
In \eqref{invlat} the symbol $\simeq$ is used because $I$  equals the given lattice at a particular point
in its moduli space, concretely at $\hat a=0$. As we will see shortly, $I$ has discriminant group $A_I=\ZZ_2^8$. 
As shown by Kneser and Nikulin (see Theorem A1 quoted in Appendix A of \cite{Mikhailov:1998si}), $I$
is unique up to a $SO(1,9)$ transformation, parameterised by the 9 moduli $\hat a$ and $R$.

The dual lattice $I^*$ plays an important role in asymmetric orbifolds \cite{Narain:1986qm}.
In the example at hand  we find
\be
I^* \simeq \rii_{1,1} \oplus \Gamma_8(\tfrac12) \, ,
\label{dualinvlat}
\ee
where $\Gamma_8(\frac12)$ is the lattice whose
Gram matrix is the Gram matrix of $\mre_8$ divided by 2 (i.e. the basis is divided by $\sqrt2$).
It follows that
\be
\label{dualdata}
A_I\equiv I^*/I = \ZZ_2^8, \qquad \left|I^*/I\right|= 2^8 \, .
\ee
The results for $I$ and $I^*$ agree with those of \cite{Chaudhuri:1995bf} where $\theta$ was also taken to be 
the exchange of the two $\mre_8$.

Another relevant sublattice is $\tilde I$, defined to be the projection of the Narain lattice $\nlc$ into 
$I$ \cite{Narain:1986qm}.
The elements of $\tilde I$ only have components $\ket{p_R; p_L, p_+}$, given in \eqref{momentaCHL}. To identify $\tilde I$
we recast these components as
\begin{subequations}
	\label{momtilde}
	\begin{align}
	p_{R}&= \frac1{\sqrt2 R}\left[n- (R^2+\tfrac12 \hat a^2) m- \frac{\rho}{\sqrt2} \cdot \hat a\right], \label{pRIt} \\
	p_{L}&=\frac1{\sqrt2 R}\left[n+ (R^2-\tfrac12 \hat a^2) m- \frac{\rho}{\sqrt2} \cdot \hat a \right], \label{pLIt} \\[2mm]
	p_+ &= \frac{\rho}{\sqrt2} +  m \hat a ,  \label{p+It} 	
	\end{align}
\end{subequations}
where $\hat a=\sqrt{2} a$ as before, and $\rho=\pi + \pi'$.
Since $\rho \in \Gamma_8$, $\frac{\rho}{\sqrt2} \in \Gamma_8(\frac12)$.
From the form of the component momenta in \eqref{momtilde} we then conclude that $\tilde I$ is given by
\be
\tilde I \simeq \rii_{1,1} \oplus \Gamma_8(\tfrac12) \, .
\label{tildelat}
\ee
Hence, $\tilde I=I^*$, in agreement with the general result shown in Appendix A of \cite{Narain:1986qm}.

By construction, the invariant lattice $I$ is a sublattice of $\tilde I$, which is consistent with the property $I \subset I^*$.
In fact, $I$ is the sublattice given by $\pi=\pi'$, which implies $\rho \in 2\Gamma_8$.
More explicitly, for $\pi=\pi'$, $\frac{\rho}{\sqrt2} = \sqrt 2 \pi =  \hat \pi$. Therefore, \eqref{momtilde} reproduces 
\eqref{momI} when $\pi=\pi'$.

After describing the general setup we move on to compute the partition function for the asymmetric $S^1/\ZZ_2$ orbifold. 
In a $\ZZ_2$ orbifold with Abelian generator $g$, the partition function 
is a sum of contributions $\cz(g^j, g^k)$, $j,k=0,1$. The first and second entries refer to boundary conditions along
the worldsheet $\sigma$ and $t$ directions.  In operator language, 
$\cz(g^j, g^k) =  \text{Tr}_{\ch_j}\!\! \left(g^k \, q^{L_0} \, \bar q^{\bar L_0}\right)$,
where $\ch_j$ is the $g^j$-twisted Hilbert space, meaning that worldsheet fields are periodic in $\sigma$ up to
a transformation by $g^j$. The sum in $j$ is over twisted sectors whereas the sum in $k$ enforces the projection on states invariant 
under $g$. The double sum is required by modular invariance \cite{Dixon:1985jw,Dixon:1986jc}. 

As we have seen, $g$ does not act on the worldsheet fermions and acts only on the worldsheet $X^9$ by a translation.
This implies that $\cz(g^j, g^k)$ takes the form
\be
\label{genzlk}
\cz(g^j, g^k) = \frac1{\left(\sqrt{\tau_2} \eta(\tau) \eta(\bar\tau)\right)^{7}} \ \cz_\psi(\bar\tau)   \  \cz_{\rii}(g^j, g^k) \, ,
\ee
where $\cz_\psi(\bar\tau)$ is given in \eqref{zpsidef}.
The presence of $\cz_\psi(\tau)$, $\forall j, k$, indicates that supersymmetry is unbroken. 

The full partition function can be expressed as
\be
\label{zz2full}
\cz_{S^1/\ZZ_2} = 
\frac1{\left(\sqrt{\tau_2}\op \eta\op \bar\eta\right)^{\!7}}\ \cz_\psi \op \cz_{\rii_{\text{o}}} \, .
\ee
From now on we will drop the dependence on the modular parameter $\tau$,
and use abbreviations $\eta=\eta(\tau)$, $\bar\eta=\eta(\bar\tau)$, $\vartheta_2=\vartheta_2(\tau)$, and so on.
In turn the full orbifold lattice sum $\cz_{\rii_{\text{o}}}$ is 
\be
\label{zz2fulllat}
\cz_{\rii_{\text{o}}}= \cz_\rii(\uno)  + \cz_\rii(g)  \, ,
\ee
where the untwisted and twisted projected lattice sums are defined as
\begin{subequations}
\label{lato}
\begin{align}
\cz_\rii(\uno) &= \frac12\left[ \cz_\rii(\uno, \uno) + \cz_\rii(\uno, g) \right] \, , \label{untlat}\\
\cz_\rii(g) &= \frac12\left[ \cz_\rii(g, \uno) + \cz_\rii(g, g) \right] \, . \label{twlat}
\end{align}
\end{subequations}
It remains to determine the  lattice pieces $\cz_{\rii}(g^j, g^k)$.  
Below we will analyze the untwisted and twisted sectors separately.

\subsubsection{Untwisted sector}
\label{sss:untwisted}

The $\cz(\uno,\uno)$ term is nothing but the partition function in $S^1$ discussed previously.
From \eqref{pfs1} we see that
\be
\cz_\rii(\uno, \uno) = \frac1{\bar\eta \eta^{17}} 
\sum_{(p_R \op ; p_L, p_+, p_-) \op \in \op \nlc} \hspace*{-5mm}
\bar  q^{\frac12 p_R^2} \,  q^{\frac12 p_L^2 + \frac12 p_+^2 + \frac12 p_-^2}  \, .
\label{z11}
\ee
We stress that the lattice sum in $\cz_\rii(\uno, \uno)$ is over the full $\nlc$ for which we use the basis \eqref{momentaCHL}.

The next step is to obtain $\cz_\rii(\uno,g)$ taking into account the action of $g$.
It turns out that
\be
\cz_\rii(\uno, g) = \frac1{\bar\eta  \eta^{17}} \left(\frac{2 \eta^3}{\vartheta_2}\right)^{\!\!4}
\sum_{(p_R \op ; p_L, p_+) \op \in \op I} \hspace*{-5mm}
\bar  q^{\frac12 p_R^2} \,  q^{\frac12 p_L^2 + \frac12 p_+^2}  \, e^{i\pi n} \, . 
\label{z1g}
\ee
Notice that now the lattice that enters is the invariant lattice $I$ in which $p_-=0$ and the remaining momenta are
given in \eqref{momI}.
The appearance of $I$ in $\cz(\uno, g)$ is a well known result \cite{Narain:1986qm}.
The reason is that $\cz(\uno, g) =  \text{Tr}_{\ch_0}\!\! \left(g \, q^{L_0} \, \bar q^{\bar L_0}\right)$, 
and the insertion of $g$ in the trace removes the non-invariant subspace. 
To explain the prefactor, notice first that $1/\eta^{16}$ in \eqref{z11} is due to the oscillators
in the expansion of the 16 left-moving coordinates $Y^{\hat I}$. Next, in the diagonal basis 
$g Y^I_\pm= \pm Y^I_\pm$, $I=1,\ldots,8$. Using properties of Jacobi functions we then find that 
instead of $1/\eta^{16}$, the $Y^I_\pm$ oscillators contribute
\be
\frac{1}{\eta^{16-\mathfrak{n}}} \left(\frac{2 \eta}{\vartheta_2}\right)^{\!\!\frac{\mathfrak{n}}2} = 
\frac{1}{\eta^{16}} \left(\frac{2 \eta^3}{\vartheta_2}\right)^{\!\!\frac{\mathfrak{n}}2} \, ,
\label{osc1g}
\ee
with $\mathfrak{n}=8$. The parameter $\mathfrak{n}$ counts the number of $-1$ eigenvalues of $g$ acting on the $Y^{\hat I}$. 

\subsubsection{Twisted sector}
\label{sss:twisted}

The partition function in the twisted sector is obtained by the chain of modular transformations
\be
\cz(\uno, g) \xrightarrow{\tau \to -1/\tau} \cz(g,\uno) \xrightarrow{\tau \to \tau + 1} \cz(g,g) \, .
\label{chainmod}
\ee
The modular transformations of $\eta$ and $\vartheta$ functions are standard.
Concerning the lattice sum, the $\tau \to -1/\tau$ transformation of the lattice sum involves Poisson resummation 
while $\tau \to \tau +1$ is elementary.

In the first step we obtain
\be
\cz_\rii(g, \uno) = \frac1{\bar\eta \eta^{17}} \left(\frac{\eta^3}{\vartheta_4}\right)^{\!\!4}
\sum_{(p_R \op ; p_L, p_+) \op \in \op I^*} \hspace*{-5mm}
\bar q^{\frac12 (p_R+v_R)^2} \,  q^{\frac12 (p_L+v_L)^2 + \frac12 (p_++v_+)^2}  \, . 
\label{zg1}
\ee
The dual lattice arises from Poisson resummation. The shifts in the momenta emerge rewriting $e^{i \pi n}$ as in
\eqref{shiftcond}.
The factor $\text{vol}(I)$ equals $\sqrt{\left|I^*/I\right|}=2^4$, cf.
\eqref{dualdata}. It cancels against the original $2^4$ in \eqref{z1g} which actually corresponds to $\sqrt{\det'(1-\theta)}$.
In other words, the degeneracy of the twisted sector is one. In general this degeneracy is given by
${\mathcal D} = \sqrt{\frac{\det'(1-\theta)}{\left|I^*/I\right|}}$, where $\det'$ is the determinant over the eigenvalues of 
$\theta$ different from one \cite{Narain:1986qm}.

The components of $(p_R \op ; p_L, p_+) \op \in \op I^*$ are written in \eqref{momtilde}, while
the shift $v$ has $v_-=0$ and $(v_R \op ; v_L, v_+)$ given in \eqref{v}. 
Therefore, in the twisted sector the momenta have the form
\begin{subequations}
\label{momtwisted}
\begin{align}
p_{R} + v_R &= \frac1{\sqrt2 R}\left[n- (R^2+\tfrac12 \hat a^2)(m+\tfrac12)- \frac{\rho}{\sqrt2} \cdot \hat a\right], \label{pRtw} \\
p_{L} + v_L &=\frac1{\sqrt2 R}\left[n+ (R^2-\tfrac12 \hat a^2)(m+\tfrac12)- \frac{\rho}{\sqrt2} \cdot \hat a \right], \label{pLtw} \\[2mm]
p_+ + v_+&= \frac{\rho}{\sqrt2} +  (m+\tfrac12) \hat a .  \label{p+tw} 	
	\end{align}
\end{subequations}
Comparing with \eqref{momtilde} shows that in the twisted sector the winding number $m$ is shifted by $\frac12$.
This result is expected because in the twisted sector the bosonic field  $X^9$ satisfies the boundary condition
$X^9(t,\sigma + 2\pi)=X^9(t,\sigma) + 2\pi m R + \pi R$.  

Performing a $\tau \to \tau + 1$ transformation we find
\be
\cz_\rii(g, g) = \frac1{\bar\eta \eta^{17}} \left(\frac{e^{\frac{i\pi}4} \eta^3}{\vartheta_3}\right)^{\!\!4}
\sum_{(p_R \op ; p_L, p_+) \op \in \op I^*} \hspace*{-5mm}
\bar q^{\frac12 (p_R+v_R)^2} \, q^{\frac12 (p_L+v_L)^2 + \frac12 (p_++v_+)^2}  \, e^{i\pi (n + \frac{\rho^2}2)} \, . 
\label{zgg}
\ee
Here we used again \eqref{shiftcond} and $v^2=0$.

A further $\tau \to \tau + 1$ transformation gives $\cz(g,g^2)$, but $g^2=\uno$. Therefore, it must be that
\be
 \cz(g,\uno) \xrightarrow{\tau \to \tau + 2} \cz(g,\uno) \, .
\label{modinv1} 
\ee
This is the necessary and sufficient condition for modular invariance at one loop and it is equivalent to level matching
\cite{Dixon:1985jw, Dixon:1986jc, Vafa:1986wx}.
From the above results it is not difficult to check that this condition is satisfied. 
We might as well consider a more general $\ZZ_2$ action such that $\theta$ has $\mathfrak{n}$ negative eigenvalues as in \eqref{osc1g},
$2v \in I$ but $v^2\not=0$. In this case the modular invariance condition \eqref{modinv1} leads to
\be
 \left( e^{\frac{i\pi}2}\right)^{\!\!\frac{\mathfrak{n}}2} e^{2 i\pi v^2} = 1 \ \Rightarrow \  \frac{\mathfrak{n}}8 + v^2 \in \ZZ \, .
\label{modinv2}
\ee
Thus, modular invariance is verified in our case in which $\mathfrak{n}=8$ and $v^2=0$.

Using identities such as $(2/\eta(q)\vartheta_2(q))^4= 1/\eta^8(q^2)$ and $(1/\eta(q)\vartheta_4(q))^4= 1/\eta^8(q^{\frac12})$,
one can check that our results in the untwisted and twisted sector agree with those in 
\cite{Mikhailov:1998si} and \cite{Bianchi:1997rf}.

\subsection{Underlying Mikhailov lattice in the  $S^1/\ZZ_2$ asymmetric orbifold}
\label{ss:z2mik}

So far we know that physical states are characterized by momenta $(p_R;p_L,p_+)$
belonging to an integer lattice of signature $(1,9)$. In the untwisted sector $(p_R;p_L,p_+) \in I^*$, whereas
in the twisted sector $(p_R;p_L,p_+) \in I^* + v$. The untwisted sector has the additional feature that some
momenta belong to the invariant lattice $I$. In \cite{Mikhailov:1998si} it is argued that the full lattice is 
\be
\label{mik1}
\rii_{(1)} = \rii_{1,9} \, ,
\ee
where  $\rii_{1,9}$ is the even self-dual lattice of signature $(1,9)$ which is unique up to $SO(1,9)$ transformations.
In fact, $\rii_{1,9} \simeq \rii_{1,1} \oplus \Gamma_8$.
Below we will deduce this result by mimicking the reasoning in \cite{Mikhailov:1998si}. 
We will refer to $\rii_{(1)}$ as the Mikhailov lattice. 

\subsubsection{A basis for $\rii_{(1)}$}
\label{sss:mikbasis}

To simplify the arguments we set the Wilson line $a$ to zero. Restoring $a$ in the end is straightforward.
More crucially, we introduce the Mikhailov radius $R_M$ according to
\be
\label{rm}
R= \sqrt2 R_M \, ,
\ee
where $R$ is the $S^1$ radius. The components of $(p_R;p_L,p_+) \in I^*$, cf. \eqref{momtilde}, can be recast as
\begin{subequations}
	\label{momp}
	\begin{align}
	p_{R}&= \frac1{2 \op R_M}\left[n- \ell R_M^2\right], \label{pRp} \\
	p_{L}&=\frac1{2 \op R_M}\left[n+ \ell R_M^2\right], \label{pLp} \\
	p_+ &= \frac{\rho}{\sqrt2},  \label{ppp} 	
	\end{align}
\end{subequations}
where we defined $\ell = 2 m$ as in the text. The advantage of redefining the radius and the winding is that now the
momenta in $I^* + v$ take the same form as in \eqref{momp}, but with $\ell=2m+1$, as seen from \eqref{momtwisted}.
Recall that $\rho \in \Gamma_8$. 

A further essential advantage of redefining the winding and the radius is that it allows to identify the momenta of the
Mikhailov lattice $\rii_{(1)}$, which we dub $(\wp_R;\wp_L, \wp)$. Concretely,
\begin{subequations}
	\label{mommik}
	\begin{align}
	\wp_{R}&= \frac1{\sqrt2 \op R_M}\left[n- \ell R_M^2\right], \label{pRmik} \\
	\wp_{L}&=\frac1{\sqrt2 \op R_M}\left[n+ \ell R_M^2\right], \label{pLmik} \\
	\wp &= \rho,  \label{ppmik} 	
	\end{align}
\end{subequations}
where $n, \ell \in \ZZ$ and $\rho \in \Gamma_8$. Notice that formally 
\be
\label{mikistar}
(\wp_R;\wp_L, \wp)=\sqrt2 (p_R;p_L,p_+) \, ,
\ee
but now $\ell$ can be even or odd. Recall that for $(p_R;p_L,p_+)$ the redefined winding $\ell$ is even or odd depending
on whether it belongs to $I^*$ or to $I^*+v$.

The points $(\wp_R;\wp_L, \wp)$ clearly lie in $\rii_{(1)}\simeq \rii_{1,1} \oplus \Gamma_8$.
In particular,
\be
\wp^2 + \wp_L^2 - \wp_R^2 = \rho^2 + 2 \ell n  \in  2\ZZ \, .
\ee 
With fixed $R_M$, a point in $\rii_{(1)}$ is specified by $n$, $\ell$ and $\rho$. 

These quantum numbers allow a neat characterisation of the points in $\rii_{(1)}$ 
based on the simple fact that $\frac12 \rho^2 + \ell n$ can be even or odd. 
Indeed, as noted in \cite{Mikhailov:1998si} there are 3 types of points $(\ell, n, \rho) \in \rii_{(1)}$ given by
\be
\label{mikpoints}
\begin{tabular}{l@{\hspace{3mm}}l}
{\bf 1.}  & $\frac12 \rho^2 + \ell n \in 2\ZZ$, $(\ell, n,\rho) \in 2\rii_{(1)}$, \\[2mm]
{\bf 2.}  & $\frac12 \rho^2 + \ell n \in 2\ZZ$, but $(\ell, n, \rho) \notin 2\rii_{(1)}$, \\[2mm]
{\bf 3.}  & $\frac12 \rho^2 + \ell n \in 2\ZZ+1$.
\end{tabular}
\ee
We will soon see how these points show up in the orbifold spectrum.   

It is interesting to consider the  duality symmetries of the $\rii_{(1)}$ lattice. With zero Wilson line there is just
the T-duality $R_M \to 1/R_M$, $n \leftrightarrow \ell$. The self-dual radius is $R_M=1$, which corresponds to
$R=\sqrt 2$. This explains why enhancement occurs at this value of $R$. Restoring the Wilson line, enhancing takes place
when $E_M=R_M^2 +\frac12 a^2=1$, which translates into $E=R^2+a^2=2$.
We will shortly prove that the partition function enjoys T-duality.

\subsubsection{Rewriting the partition function}
\label{sss:mikpf}

To continue we need to rewrite the partition function in a way more appropriate to reveal 
the underlying lattice. The strategy is to unify the different lattices, namely $I^*$, $I$ and $I^*+v$, into one bigger 
structure. The main outcome will be the  full orbifold lattice sum $\cz_{\rii_{\text{o}}}$, cf. \eqref{zz2fulllat}, expressed 
in a form that shows the correspondence with the Mikhailov lattice. Below we proceed in order.

\bigskip
\noindent
\underline{Untwisted sector}
\medskip

The untwisted projected sum $\cz_\rii(\uno)$ is complicated because the two terms involve different lattices. 
In the term $\cz_\rii(\uno, g)$, given in \eqref{z1g}, the relevant lattice is $I$. 
In the invariant lattice the momenta are again given by \eqref{momp} but with $\rho \in 2 \Gamma_8$. 
We then have
\be
\cz_\rii(\uno, g) = \frac1{\bar\eta \eta^{17}} \, \left(\frac{2\op \eta^3}{\vartheta_2}\right)^{\!\!4} 
\sum_{n\in \ZZ, \, \ell \in 2\ZZ} \hspace*{-3mm}
\bar q^{\frac12 p_R^2} \,  q^{\frac12 p_L^2} e^{i\pi n} \,  \sum_{\rho \op \in 2\Gamma_8} q^{\frac14 \rho^2}\, .
\label{z1gmik}
\ee
The momenta $p_R$ and $p_L$ are given in \eqref{momp}.

The next task is to express $\cz(\uno,\uno)$ in a similar way involving sums over $n$, $\ell$ and $\rho$. This is
more difficult because in $\cz(\uno, \uno)$ the lattice sum is over the whole $\nlc$. In other words, 
the $(p_R;p_L,p_+)$ are in $I^*$, but $p_-$ also appears. In \cite{Mikhailov:1998si} the clever way to deal with this is to
distinguish whether or not $\rho \in \Gamma_8$ also belongs to $2\Gamma_8$. There are two possibilities
\begin{itemize}
\item[{\bf A.}]\ $\rho \in 2\Gamma_8$, $\ell \in 2 \ZZ$. In this case one can write
\be
\pi = \frac{\rho}2 + S, \quad \pi '= \frac{\rho}2 - S, 
\ee
where $S \in \Gamma_8$.
\item[{\bf B.}]\ $\rho \in \Gamma_8/2\Gamma_8$, $\ell \in 2 \ZZ$. In this case one can instead take
\be
\pi = \rho + L, \quad \pi' = -L, 
\ee
where $L \in \Gamma_8$.
\end{itemize} 
For future purposes we have stressed that $\ell \in 2\ZZ$, which is always the case in the untwisted sector.
 
Dividing the states in $\nlc$ into classes {\bf A} and {\bf B} allows to rewrite the sum over $\pi$ and $\pi'$ as
\be
\sum_{\pi \in \Gamma_8, \pi' \in \Gamma_8} q^{\frac12 p_+^2 + \frac12 p_-^2} = 
\sum_{\rho \in 2\Gamma_8} q^{\frac14 \rho^2} \sum_{S \in \Gamma_8} q^{S^2} \ + \
\sum_{\rho \in \Gamma_8/2\Gamma_8} \!\!\! q^{\frac14 \rho^2} \sum_{L \in \Gamma_8} q^{(L+\frac{\rho}2)^2} \, .
\ee 
The sums over $S$ and $L$ can be related to $\Theta_8$ functions \cite{Mikhailov:1998si}.
In particular, 
\be
\sum_{S \in \Gamma_8} q^{S^2} =\Theta_8(2\tau) = 1 + 240 \op q^2 + 2160 \op q^4 + 6720 \op q^6 + \cdots \, .
\ee
Putting together the above results in \eqref{z11} leads to
\be
\cz_\rii(\uno, \uno) = \frac1{\bar\eta \eta^{17}}
\sum_{n\in \ZZ, \, \ell \in 2\ZZ} \hspace*{-3mm}
\bar q^{\frac12 p_R^2} \,  q^{\frac12 p_L^2} \left\{  \sum_{\rho \op \in 2\Gamma_8} q^{\frac14 \rho^2}\, 
\Theta_8(2\tau) \ + \ \sum_{\rho \op \in \Gamma_8/2\Gamma_8} \!\!\! q^{\frac14 \rho^2} \sum_{L \in \Gamma_8} q^{(L+\frac{\rho}2)^2} 
\right\} \, .
\label{z11mik}
\ee
The sum over $L$ only depends on the conjugacy classes of $\rho \in \Gamma_8/2\Gamma_8$, denoted $\bar \rho$ \cite{Mikhailov:1998si}.
This will be further elaborated in section \ref{sss:mikproof}.

Substituting \eqref{z1gmik} and \eqref{z11mik} in \eqref{untlat} gives the untwisted lattice sum
\be
\cz_\rii(\uno) = \frac1{\bar\eta \eta^{17}} 
\sum_{n\in \ZZ, \, \ell \in 2\ZZ} \hspace*{-3mm}
\bar q^{\frac12 p_R^2} \,  q^{\frac12 p_L^2}\,  \left\{  \sum_{\rho \op \in 2\Gamma_8} q^{\frac14 \rho^2}
\, \cf_1\left(q, n\!\!\!\!\mod \!2\right) \ + \ \sum_{\rho \op \in \Gamma_8/2\Gamma_8} \!\!\! q^{\frac14 \rho^2}
\cf_2(q,\bar\rho)
\right\} \, .
\label{z1mik}
\ee
The function $\cf_1$ is
\be
\label{f1def}
\cf_1(q, n\!\!\!\mod \!2) = \frac12\left[ \Theta_8(2\tau) +
  e^{i\pi n} \left(\frac{2\eta^3}{\vartheta_2}\right)^{\!\!4} \right] \, .
\ee  
Finally, $\cf_2$ is given by
\be
\label{f2def}
\cf_2(q,\bar\rho) =  \frac12  \sum_{L \in \Gamma_8} q^{(L+\frac{\rho}2)^2} \, .
\ee
The functions $\cf_1$, $\cf_2$, and $\cf_3$ below, match the Mikhailov's generating functions
$F_1$, $F_2$ and $F_3$ given in equations (3.3)-(3.5) in \cite{Mikhailov:1998si} 
up to an overall factor $q/\eta^{24}$. The meaning of these functions will be explained shortly.

\bigskip
\noindent
\underline{Twisted sector}
\medskip

The twisted sector is much simpler. The two terms in $\cz_\rii(g)$ are given in equations \eqref{zg1} and \eqref{zgg}. 
Combining them gives
\be
\cz_\rii(g) = \frac1{\bar\eta \eta^{17}} 
\sum_{n\in \ZZ, \, \ell \in 2\ZZ+1} \hspace*{-5mm}
\bar q^{\frac12 p_R^2} \,  q^{\frac12 p_L^2}\,  \sum_{\rho \op \in \Gamma_8} q^{\frac14 \rho^2}
\, \cf_3\!\left(q, (\tfrac{\rho^2}2 + n)\!\!\!\mod \!2\right) \, .
\label{zgmik}
\ee
The momenta $p_R$ and $p_L$ are again given in \eqref{momp}. The function $\cf_3$ reads
\be
\label{f3def}
\cf_3\!\left(q, (\tfrac{\rho^2}2 + n)\!\!\!\!\mod \!2\right) = \frac12\left[ \left(\frac{\eta^3}{\vartheta_4}\right)^{\!\!4}
-  e^{i\pi (n + \frac{\rho^2}2)} \left(\frac{\eta^3}{\vartheta_3}\right)^{\!\!4} \right] \, .
\ee  
The twisted states constitute class {\bf C} characterized by $\rho \in \Gamma_8$ and $\ell \in 2\ZZ+1$.

\bigskip
\noindent
\underline{Full orbifold}
\medskip

Adding $\cz_\rii(\uno)$ and $\cz_\rii(g)$ yields
\be
\cz_{\rii_{\text{o}}} = \frac1{\bar\eta \eta^{17}} 
\left\{
\sum_{\substack{n\in \ZZ \\ \ell \in 2\ZZ}} \hspace*{-1mm}
\bar q^{\frac12 p_R^2} \,  q^{\frac12 p_L^2}  \hspace*{-1mm}
 \left[  \sum_{\rho \op \in 2\Gamma_8} \hspace*{-2mm} q^{\frac14 \rho^2}
\, \cf_1 \ + \hspace*{-2mm}  \sum_{\rho \op \in \Gamma_8/2\Gamma_8}  \hspace*{-4mm} q^{\frac14 \rho^2} \cf_2  \right] + \hspace*{-3mm}
\sum_{\substack{n\in \ZZ \\ \ell \in 2\ZZ+1}} \hspace*{-2mm}
\bar q^{\frac12 p_R^2} \,  q^{\frac12 p_L^2}\,  \sum_{\rho \op \in \Gamma_8} \hspace*{-1mm} q^{\frac14 \rho^2}
\, \cf_3
\right\} \, .
\label{zorbi}
\ee
The arguments of $\cf_1$, $\cf_2$ and $\cf_3$ are omitted to simplify the expression.
From this result we can read off the content of orbifold states classified according to the possible
domains of $\ell$, $n$ and $\rho$. This information is summarised in Table \ref{abc}.

\begin{table}[h!]\begin{center}
\renewcommand{\arraystretch}{1.5}
\begin{tabular}{c@{\hspace{8mm}}c@{\hspace{8mm}}c@{\hspace{8mm}}c@{\hspace{8mm}}c@{\hspace{8mm}}c}
\hline
class & $\ell$ & $n$ & $\rho$ & sector & generating function   \\
\hline
{\bf A} & $2\ZZ$ & $\ZZ$ & $2\Gamma_8$ & untwisted & $\cf_1(q, n\!\!\!\mod \!2) $  \\ 
{\bf B} & $2\ZZ$ & $\ZZ$ & $\Gamma_8/2\Gamma_8$ & untwisted & $\cf_2(q,\bar\rho) $  \\ 
{\bf C} & $2\ZZ + 1$ & $\ZZ$ & $\Gamma_8$ & twisted & $\cf_3\!\left(q, (\tfrac{\rho^2}2 + n)\!\!\!\mod \!2\right) $  \\ 
 \end{tabular}
\caption{Classes of orbifold states}
  \label{abc}\end{center}\end{table}

\subsubsection{Reading $\rii_{(1)}$ from the partition function}
\label{sss:mikproof}

It turns out that some identities relating the functions $\cf_1$, $\cf_2$ and $\cf_3$ are needed to show that the orbifold states 
lie in the lattice $\rii_{(1)}$. In \cite{Mikhailov:1998si} the relations are proven analytically for the
generating functions $F_c$, $c=1,2,3$, connected to the $\cf_c$ by
\be
\label{mikf}
F_c = \frac{q}{\eta^{24}} \cf_ c \, . 
\ee
The identities can be verified by comparing the $q$-expansions of the $\cf_c$. 
For $\cf_1(q, n\!\!\!\mod \!2)$ and $\cf_3\!\left(q, (\tfrac{\rho^2}2 + n)\!\!\!\mod \!2\right)$ these expansions
are easily found from their definitions.
The expansion of $\cf_2(q,\bar \rho)$ is less direct 
because it depends on $\bar\rho$, which is the conjugacy class
of $\rho$ in $\Gamma_8/2\Gamma_8$. As explained in \cite{Mikhailov:1998si}, if $\frac12 \rho^2$ is odd, then
$\rho$ equals a root modulo $2\Gamma_8$. We denote the conjugacy class by $\Delta_2$. If instead $\frac12 \rho^2$ is even,
then $\rho$ modulo $2\Gamma_8$  is equal to a vector $\varrho \in \Gamma_8$ with $\varrho^2=4$. We denote   
the conjugacy class by $\Delta_4$. 
From the expansions we can check the identities
\begin{subequations}
\label{fidentities}
\begin{align}
\cf_1(q,1) &= \cf_2(q,\Delta_4) =\cf_3(q,0)=8 q + 64 q^2 + 224 q^3 + 512 q^4 + \cdots , \label{idabc}\\
\cf_2(q,\Delta_2) &=\cf_3(q,1)=q^{\frac12}(1 + 28 q + 126 q^2 + 344 q^3 + \cdots)  \, . \label{idbc}
\end{align}
\end{subequations}
The Mikhailov's generating functions $F_c$ verify the same identities because they are equal to the $\cf_c$ up to an overall
factor.

The meaning of the generating functions can be understood by looking at the simple lattice partition function
of the 10-dimensional heterotic string in \eqref{zlatticedef}. In this case there is a generating function
$1/\eta^{16}$.
Now, we know that for each vector in the lattice there is a tower of excited states
created by acting with the oscillators of the $Y^{\hat I}$, $\hat I=1,\cdots, 16$. 
Moreover, the coefficients in the $q$-expansion of $1/\eta^{16}$ precisely count the number of states at each excited level.
The meaning of the Mikhailov's generating functions $F_c$ is completely analogous. The prefactor $q/\eta^{24}$ in the relation
with the $\cf_c$ is well justified. The power of $\eta$ corresponds to the 24 left-moving coordinates, and the power of $q$ just offsets 
the normal ordering constant. In this way their $q$-expansion will be of the form
\be
F(q) = \sum_{N'} d(N', (\ell, n, \rho)) q^{N'} \, ,
\ee
where now $N'$ corresponds to the full left-moving oscillator number. For each state $(\ell, n, \rho)$, the coefficients 
$d(N', (\ell, n, \rho))$ count the number of states with oscillators acting on it and given oscillator number $N'$.
The dependence on $(\ell, n, \rho)$ is necessary because,
as seen in Table \ref{abc}, for each type of state there is an associated generating function.

We are finally ready to state Mikhailov's proof 
that the spectrum of orbifold states can be put into correspondence with the points in the 
lattice $\rii_{(1)}$ displayed in \eqref{mikpoints}. This correspondence is summarised in Table \ref{mik123}. 
For example, the points of type {\bf 1} where $(\ell, n,\rho) \in 2\rii_{(1)}$ can only correlate with points of orbifold class 
{\bf A} which have $\rho \in 2\Gamma_8$, and $\ell \in 2\ZZ$, provided that also $n \in 2\ZZ$.

\begin{table}[h!]\begin{center}
\renewcommand{\arraystretch}{1.5}
\begin{tabular}{cc@{\hspace{6mm}}c@{\hspace{8mm}}l}
\hline
$\rii_{(1)}$ type & $(l,n,\rho)$ & $\frac12\rho^2 + \ell n$  & orbifold class  \\
\hline
{\bf 1} & $2\rii_{(1)}$ & $2\ZZ$ & $\left[{\bf A}, n \in 2\ZZ \right]$  \\[2mm] 
{\bf 2} &  $\rii_{(1)}/2\rii_{(1)}$ & $2\ZZ$ & $\left[{\bf A}, n \in 2\ZZ+1\right]$,  $\left[{\bf B},\frac12 \rho^2 \in 2\ZZ\right]$,
$\left[{\bf C}, \left(\frac12\rho^2 + n\right) \in 2\ZZ\right]$  \\[2mm]
{\bf 3} & $\rii_{(1)}$ & $2\ZZ+1$ & $\left[{\bf B}, \frac12\rho^2 \in 2\ZZ+1\right]$,  
$\left[{\bf C}, \left(\frac12\rho^2 + n\right) \in 2\ZZ+1\right]$ \\
 \end{tabular}
\caption{Points in $\rii_{(1)}$ vs orbifold classes}
\label{mik123}\end{center}\end{table}

For points of type {\bf 2} and {\bf 3} there can be more than one orbifold class as can be understood
by looking at Table \ref{abc}. For these points consistency requires precise identities among the generating
functions. For example, the points of type {\bf 3} must appear with the same generating function whether
they arise in class {\bf B} with $\frac12 \rho^2 \in 2\ZZ+1$, or in class {\bf C}
with $(\frac12\rho^2 + n) \in 2\ZZ+1$.  This means that $\cf_2(q,\Delta_2)$ must be equal to $\cf_3(q,1)$, which
is precisely the identity in \eqref{idbc}. Similarly for the points of type {\bf 2} the functions
$\cf_1(q,1)$, $\cf_2(q,\Delta_4)$ and $\cf_3(q,0)$ must be the same, which is true by virtue of
of the identity \eqref{idabc}. 
For points of type {\bf 1} they just fall in class {\bf A} and occur with generating function $\cf_1(q,0)$.

\subsubsection{T-duality}
\label{sss:tdual}

The previous results can be used to show that the partition function of the $S^1/\ZZ_2$ orbifold is invariant under T-duality.
We consider the simpler  situation with Wilson line $a=0$ in which T-duality is the action
$n \leftrightarrow \ell$,  $R_M \to 1/R_M$.  
The relevant piece of the partition function
is the lattice contribution $\cz_{\rii_{\text{o}}}$ displayed in \eqref{zorbi}. From the corresponding spectrum of
states summarised in Table \ref{abc} it is evident that T-duality mixes twisted and untwisted states as remarked in
\cite{Mikhailov:1998si}. 

To establish T-duality it is enough to show that the quantity between brackets in \eqref{zorbi} is invariant.
The parts with both $n$ and $\ell$ even (odd), arising in the untwisted (twisted) sector, are clearly invariant by themselves.
The remaining question is whether the untwisted sector terms with $n$ odd and $\ell$ even do match
the twisted sector terms with $n$ even and $\ell$ odd. The answer is yes as follows from the equality
\be
\label{tequality}
\sum_{\substack{n\in 2\ZZ+1 \\ \ell \in 2\ZZ}} \hspace*{-1mm}
\bar q^{\frac12 p_R^2} \,  q^{\frac12 p_L^2}  \hspace*{-1mm}
 \left[  \sum_{\rho \op \in 2\Gamma_8} \hspace*{-2mm} q^{\frac14 \rho^2}
\, \cf_1(q,1) \ + \hspace*{-3mm}  \sum_{\rho \op \in \Gamma_8/2\Gamma_8}  \hspace*{-4mm} q^{\frac14 \rho^2} 
\cf_2(q,\bar\rho)  \right] \! =\!\!
\sum_{\substack{n\in 2\ZZ \\ \ell \in 2\ZZ+1}} \hspace*{-2mm}
\bar q^{\frac12 p_R^2} \,  q^{\frac12 p_L^2}\,  \sum_{\rho \op \in \Gamma_8} \hspace*{-1mm} q^{\frac14 \rho^2}
\, \cf_3\!\left(q, \tfrac{\rho^2}2\!\!\!\!\mod \!2\right) \, . 
\ee
In turn this identity can be shown using the properties 
$\cf_1(q,1)=\cf_3(q,0)$, $\cf_2(q,\Delta_4)=\cf_3(q,0)$ and $\cf_2(q,\Delta_2)=\cf_3(q,1)$,
given in \eqref{fidentities}.

\section{World-sheet realisation of gauge symmetries}\label{App:CA}

In this appendix we briefly discuss the Kac-Moody algebras that realize the space-time gauge symmetries
of the CHL theory in 9 dimensions and its toroidal compactifications. 

The space-time $\mre_8\times {\mre}'_8$ gauge symmetry  is realized on the world-sheet   by dimension (1,0) currents $J_1^a\otimes 1$ and $1\otimes J^b_2$, $a,b=1,...,248$,  that obey the OPE
\be
{J}_i^a(z) {J}_i^b(0)\sim \frac{\tilde k_i\delta^{ab}}{z^2}+\frac{i}z f^{ab}{}_c{J}_{i}^c(0)\, ,\quad i=1,2\label{kma}
\ee
at level $k_i=\frac{2\tilde k_i}{\psi_i^2}=1$, where $\tilde k_i=1$, $\psi_i^2=2$ is the norm of the highest root and $f^{ab}{}_c$ are  the structure constants  of the  simply-laced Lie algebra of $\mre_8$. 

The  Sugawara construction induces a representation of the Virasoro algebra with  central charge 
\be
c=\sum_{i=1}^2c_i=\sum_{i=1}^2\frac{ k_i \, {\rm dim}\ G_i }{k_i+{\rm g}_i} \, , \label{cc}
\ee
where ${\rm g}_i$ is the dual Coxeter number of the group $G_i$ (see Table \ref{tab:data}). 
These formulae hold in general for arbitrary products of groups \cite{DiFrancesco:1997nk,Ginsparg:1988ui}. 
For simply laced algebras at level one it follows that $c_i = {\rm rank}\, G_i$. 
In the ten dimensional heterotic string with $G_i=\mre_8$, clearly $c_i=8$ and $c=16$.

\begin{table}[h!]\begin{center}
\renewcommand{\arraystretch}{1.0}
\begin{tabular}{|c|c|c|c|c|c|c|c|c|c|}
\hline
$G$ & $\mra_n$ & $\mrd_n$ & $\mre_6$ & $\mre_7$&$\mre_8$&$\mrb_n$&$\mrc_n$&$\mrf_4$ &$\mrg_2$ \\[3pt]
\hline
${\rm g}$ &  $n+1$  & $2n-2$ &12&18&30&$2n-1$&$n+1$&9&4
  \\ \hline
dim $G $ &  $n(n+2)$ & $n(2n-1)$&78&133&248&$n(2n+1)$&$n(2n+1)$&52&14
 \\ \hline 
 \end{tabular}
\caption{Dual Coxeter number ${\rm g}$ and dimension of the gauge group $G$}
  \label{tab:data}\end{center}\end{table}

The currents of the level $k=1$ untwisted affine Kac Moody algebras  associated with simple
Lie algebras which are simply-laced were constructed using the vertex operators of  the massless gauge bosons of the string spectrum in \cite{Frenkel:1980rn,Segal:1981ap}. 
In the ten dimensional theory, the 248 gauge bosons of each $\mre_8$ comprise the 8 Cartan $\alpha^I_{-1}|0,0\rangle$ or $\alpha^{I+8}_{-1}|0,0\rangle, I=1,...,8$ and the 240 roots $|p^I, p^{I+8}\rangle=|r_1^I, 0\rangle$ or $|0, r_2^{I}\rangle$, with $r_1^I, r_2^I\in\Gamma_8$. Their vertex operators can be written  in terms of the free bosons $Y^I(z)$ and $ {Y'}^{I}(z)$, and the  corresponding currents $J_{1,2}^a$ have the following realisation in the  Cartan basis 
  \bea
H^I_1(z)=i\partial Y^I(z)\, ,\qquad \qquad E_1^{\pm r_1}(z)=c_{r_1} :e^{\pm ir_1\cdot Y(z)}:\, , \\
H^I_2(z)=i\partial {Y'}^I(z)\, ,\qquad \qquad E_2^{\pm r_2}(z)=c_{r_2} :e^{\pm ir_2\cdot Y'(z)}:\, ,
\eea
where   $c_{r_1},c_{r_2}$ are cocycle factors. 
Using the OPEs
\be
\partial Y^I(z)\partial Y^J(0)=-\frac{\delta^{IJ}}{z^2}\, ,\qquad \partial {Y'}^I(z)\partial {Y'}^J(0)=-\frac{\delta^{IJ}}{z^2}\, ,
\ee
the current algebra of $\widehat {\mre_8\times \mre'_8}$ is realized at level $k_1=k_2=\frac{2\tilde k_i}{|\psi_i|^2}=1$, as can be read from
\begin{subequations}\label{alge8e8}
\begin{align}
H^I_i(z)H^J_j(0)&\sim\frac{\delta_{ij}\delta^{IJ}}{z^2}\, ,\\
H_i^I(z) E_j^{\pm r_j}(0)&\sim\frac{\pm r_j^IE_j^{\pm r_j}(0)\delta_{ij}}z\, , \\
E_i^{r_i}(z)E_i^{-r_i}(0)&\sim\frac1{z^2}+\frac{r_i\cdot H_i(0)}{z}\, ,
\end{align}
\end{subequations}

As we have seen, the CHL string in 9 and lower dimensions can be constructed as a $\ZZ_2$ orbifold
involving the outer automorphism that exchanges $\mre_8$ and ${\mre}'_8$.
In 10 dimensions the orbifold by this exchange simply reproduces the original theory.
This can be verified computing the partition function as discussed in Appendix \ref{app:pf},  with $g$ corresponding to the 
action $\mre_8\leftrightarrow \mre'_8$, and using the identities \eqref{fidentities}. 
However, if the exchange of $\mre_8$ and $\mre'_8$ is accompanied by an additional $2\pi$ rotation of the ten dimensional space-time one gets the non-supersymmetric 
$\mre_8$ string \cite{Kawai:1986vd,Forgacs:1988iw}  in which some sectors of the Hilbert space are projected out. As explained in \cite{Forgacs:1988iw}, only  the  products  ${\cal H}_{\rm spF }\otimes{\cal H}_{\rm s}$ and ${\cal H}_{\rm spB } \otimes {\cal H}_{\rm as}$ survive, where ${\cal H}_{\rm spB } ({\cal H}_{\rm spF })$ denotes the Hilbert subspace  of space-time bosons (fermions) which is symmetric (antisymmetric) under the $2\pi$ rotation. Since the  internal Hilbert space ${\cal H}_{\rm int}$
 is an irreducible representation of $\widehat{\mre_8\times \mre'_8}$, its symmetric and antisymmetric subspaces ${\cal H}_{\rm s}$ and ${\cal H}_{\rm as}$ are
not invariant under the full $\mre_8\times \mre'_8$ current algebra, but  are invariant under the algebra of the diagonal currents
  \bea
T^a(z)=J_1^a(z)\otimes 1 + 1\otimes J^a_2(z)\, ,
\eea
since $T^a$ is invariant under the exchange of the two $\mre_8$'s. The diagonal $\hat \mre_8$ is a subalgebra of $\widehat{\mre_8\times \mre'_8}$
 and clearly, the current algebra is realized at level $k_1+k_2=2$.   
 In this case, the central charge obtained from \eqref{cc} is $\frac{31}2$, and the missing $\frac12$ is provided by the coset theory 
 $\frac{(\mre_8\times \mre_8)_{k=1}}{(\mre_8)_{k=2}}$, which is equivalent to the Ising model \cite{Forgacs:1988iw}.

Let us now turn to CHL strings. As we have reviewed in the main text, 
the 9-dimensional theory can be described by a $S^1/\ZZ_2$ orbifold, with $\ZZ_2$ action given by the exchange 
$\mre_8\leftrightarrow\mre'_8$ together with a translation in the compactified direction $x^9\rightarrow x^9+\pi R$. 
For arbitrary values of the compactification radius $R$ and Wilson lines $a$, only the 8 diagonal Cartan  gauge bosons  in the untwisted massless sector survive the orbifold projection, and together with the KK gauge boson of the compactified $x^9$, they account for the 
$\uo^9$ abelian symmetry of the theory, with  generators 
\be
T^I_+=H^I_1+ H^I_2=i\left(\partial Y^I(z)+\partial {Y}^{'I}(z)\right)\qquad {\rm and} \qquad H^9=i\partial X^9(z)\, ,
\ee
where $X^9$ is a free boson.

For arbitrary $R$ and $a=0$ the untwisted states
$\frac1{\sqrt2}\left(|r^I,0\rangle + |0,r^I\rangle\right)$ 
are also massless when $r^I$ is a root of $\mre_8$ and $m=n=0$. Together with the nine Cartan above, they give rise to the  
rank nine gauge symmetry $\mre_8\times \uo$, with $\mre_8$  the diagonal subgroup of the original $\mre_8\times \mre_8$. The corresponding raising and lowering currents are
\be
T_+^{\pm r^I}=E_1^{\pm r^I}+  E_2^{\pm r^I}\, .
\ee
The $\mre_8$ current algebra is realized at level 2, just as in the 10-dimensional $\mre_8$ string theory. 

Lower rank groups and higher level algebras are a hallmark of CHL strings. 
The total central charge $c$ of the Kac-Moody algebra associated to the gauge symmetry also gives useful
information.
In general, in $(10-d)$ dimensions there is a bound
$c \le c_L^{int}$, where the internal piece is $c_L^{int}=16+d$.
This follows because keeping transverse degrees of freedom the total left-moving central charge is $c_L=24$ and the 
world-sheet bosons corresponding to the space-time coordinates contribute $(8-d)$.
It is convenient to write the bound on $c$ as
\be
\label{missingc}
\Delta = 16 + d - c \ge 0 \, .
\ee
We will refer to $\Delta$ as the missing central charge. A consistency condition is that when $\Delta < 1$ it must be equal
to the central charge of a unitary minimal model given by
\be
\label{cminmod}
c_j=1-\frac6{j(j+1)}, \quad   j=3,4, \ldots .
\ee
For instance, in the above $\mre_8\times \uo$ example the Kac-Moody central charge is 
$c=\frac{31}2+1$ and the missing $\Delta=\frac12$ is provided by the $j=3$ minimal
model, i.e. the Ising model which is furthermore equivalent to the coset theory 
$\frac{(\mre_8\times \mre_8)_{k=1}}{(\mre_8)_{k=2}}$.

Continuing with the 9-dimensional CHL string, at the particular radius $R=\sqrt2$ and $a=0$, the states with 
$\ell=\pm1, n=\pm1,\rho=0$  
in the twisted sector  become massless and enhance the $\uo$ of the KK vector  to $\sug(2)$. 
The  vertex operators that create these states involve the left moving  currents
\be 
\label{exe8a1}
\qquad 
H^1(z)=i\partial X^1(z) \, , \qquad E^{\pm }(z)=c_{\pm } \Lambda e^{\pm ip_{L}\cdot X(z)}\, ,
\ee
where
 the fields $X^{\rm a}(z)=e^{\rm a}_i X^i(z)$ with ${\rm a}, i=1,...,d$ have  tangent space indices ${\rm a}$ and standard propagator
\be
\langle X^{\rm a}(z)X^{\rm b}(w)\rangle=-\delta^{\rm{ab}}ln(z-w)\, . \label{prop}
\ee
The momentum in the tangent space is $p_{L{\rm a}}=\hat e_{\rm {a}}{}^ip_{Li}=1$ and $\Lambda$ is a twist field with conformal dimension $h=\frac12$ and OPE
\be
\Lambda(z)\Lambda(0)= \frac1z + {\rm reg}\, ,
\ee 
which  is necessary to build   spin 1 currents \cite{Hamidi:1986vh,Dixon:1986qv}.
From the OPEs
\bea
H(z)H(0)&\sim&\frac{1}{z^2}\, , \\
H(z)E^{\pm }(0)&\sim&\pm\frac{p_{L{\rm a}}E^\pm(0) }z\, ,\\
E^{+}(z)E^{-}(0)&\sim&\frac1{z^2}+\frac{p_{L{\rm a}} H^{\rm a}(0)}z\, ,
\eea
we see that the affine $\sug(2)$ algebra is realized at level $k=\frac{2\tilde k}{p_{L}^2}=2$. 
The central charge of the $\mre_8\times \mra_1$ model at level $k=2$ saturates  $c_L^{int}=17$, as may be verified using the data in Table \ref{tab:data}.

The central charges of all the maximal enhancements listed in Table \ref{tab:CHLd1} can be readily computed.
Except for the $ \mrd_9$ and  $\mre_8 \times \mra_1$ models,
the internal Kac-Moody algebras do not saturate $c_L^{int}=17$.  In some cases, the missing central charge $\Delta$ is provided by 
unitary minimal models, cf. \eqref{cminmod}.
For instance the $\mre_6 \times \mra_3$ at level 2 requires $\Delta=6/7=c_6$. On the other hand,
the $\mra_1 \times \mra_2 \times \mra_6$ current algebra at level 2 leads to $\Delta=\frac{49}{30}$ which could arise 
combining two minimal models with $j=4$ and $j=9$. However in the case $\mrd_5 \times \mra_4$, with $\Delta=8/7$, 
a candidate world-sheet CFT is not obvious. 
It would be interesting to understand if there is a realisation of the missing CFTs in terms of  coset models involving the original and the enhanced gauge groups.

In compactifications of the CHL string to 8 dimensions, the  gauge group is $\uo^{10}$ for generic values of the background fields. 
To analyze maximal enhancement at a special point in moduli space let us choose $E_{11}=2, E_{22}=1, E_{12}=E_{21}=0$ and $a_1=a_2=0$. These moduli actually correspond to starting with the 9-dimensional CHL model with group 
$\mre_8\times \mra_1$ at level $k=2$ discussed above, and further compactifying on a circle of 
radius $R_2=1$.
Following the analysis in section \ref{sec:9-d}, and using equations \eqref{m03D} and \eqref{m04D}, we see that there
are additional untwisted states with $Z^2=4$, having $\rho=0$ and $(\ell^1, \ell^2, n_1,n_2)=\pm(0,2,0,1)$.
These states enhance the $(\mre_8\times\mra_1)_2\times\uo$ gauge symmetry to  $(\mre_8\times\mra_1)_2\times(\mrc_1)_1$. 
Using that $p_L^{\rm a}=\frac1{\sqrt2} l^ie_i{}^{\rm a}=(0,\sqrt2)$,  the vertex operators contain the currents
\be
\label{exe8a1c1}
H^2=i\partial X^2\, , \qquad E^\pm=c^\pm e^{\pm i\sqrt2 X^2}\, ,
\ee
which realize the current algebra of $\mrc_1$ at level $k=1$.

We next consider an example with short and long roots.
Taking $E_{11}=2$, $E_{12}=-2$, $E_{21}=0$, $E_{22}=1$ and $a_1=a_2=0$,  gives gauge symmetry $\mre_8\times \mrc_2$. 
The quantum numbers $(\ell^1,\ell^2,n_1,n_2)$ of the massless states that enhance the $\uo^2$ to $\mrc_2$ are 
$\pm(0,2,-2,1)$, $\pm (2,2,0,1)$, $\pm(1,0,1,0)$ and $\pm(1,2,-1,1)$, and they all have $\rho=0$. The vertex operators contain the currents 
\begin{subequations}
\label{exe8c2}
\begin{align}
E_1^{\pm}(z)&=c_1^\pm \Lambda(z) e^{\pm iX^1(z)}, \hspace*{2cm}
 p^{\rm a}_{1L}=\alpha_1=(1,0)\, ,\\
E_2^{\pm}(z)&=c_2^\pm  e^{\mp i(X^1(z)-X^2(z))}, \qquad \qquad  p^{\rm a}_{2L}=\alpha_2=(-1,1)\, ,\\
E_3^{\pm}(z)&=c_3^\pm \Lambda(z) e^{\pm  iX^2(z)}, \hspace*{1.7cm}
\quad  p^{\rm a}_{3L}=\alpha_3=(0,1)\, ,\\
E_4^{\pm}(z)&=c_4^\pm  e^{\pm  i(X^1(z)+X^2(z))}, \qquad\qquad  p^{\rm a}_{4L}=\alpha_4=(1,1)\, .
\end{align}
\end{subequations}
Together with the Cartan operators $H^1=i\partial X^1$ and $H^2=i\partial X^2$, the current algebra of $\hat \mrc_2$ is realized at level $k=1$ since, $\tilde k=1$ and  the square of the  highest root $\alpha_4$ is 2.

It is straightforward to calculate the central charge of the Kac-Moody algebras of the eight dimensional models listed in Table \ref{tab:CHLd2}. 
As in the nine-dimensional case, they do not saturate $\Delta=0$ in general, but again in most cases one can find combinations of minimal models that account for the missing 
contribution. A consistency check is that when $\Delta < 1$ it is always equal to the central charge of a unitary minimal model.

Finally, let us remark that vertex operators for the twisted states in 
the examples \eqref{exe8a1c1} and \eqref{exe8c2} discussed above do not involve the fields $Y^I, Y^{\prime I}$ explicitly.  
However,  when $\rho$  is non-vanishing, they are expected to be part of the exponentials in the currents.
For instance,  with Wilson line $a=\frac12w_6$ and $R^2=\frac32$ in nine dimensions, the states with quantum numbers 
$(\ell,n,\rho)$ given by $\pm(1, 0, -w_6)$ and $\pm(1,1,0)$ become massless and enhance the gauge group to 
$\mre_7\times \mra_2$, with current algebra realized at level 2. Free field representations of the affine $\sug(3)$ current algebra at  level 2 are known \cite{DiFrancesco:1997nk} as well as of the level 1 non-simply laced 
algebras \cite{Goddard:1986bp, Goddard:1986ts, Bernard:1986rd} involved in the enhanced gauge groups of the eight dimensional theory (see also \cite{Kuwahara:1989xy} for constructions of Kac Moody algebras in terms of free fields). But not all of them can be directly related to the vertex operators of the twisted states of the CHL theory.  
We postpone a detailed analysis of the twisted vertex operators that realize these current algebras to a future publication.

\bibliographystyle{JHEP}
\bibliography{refs-CHL}

\end{document}